\documentclass[a4paper,twocolumn,10pt]{quantumarticle}
\pdfoutput=1

\usepackage[normalem]{ulem}
\usepackage{dsfont}
\usepackage{amsmath}
\usepackage{bm}
\usepackage{amssymb}
\usepackage{graphicx}
\usepackage{braket}
\usepackage{amsthm}
\usepackage{comment}
\usepackage{cancel}
\usepackage{cite}
\usepackage{array}
\usepackage{mathrsfs}
\usepackage{pifont}
\usepackage{multirow}
\usepackage[dvipsnames]{xcolor}

\newcommand{\cmark}{\textcolor{ForestGreen}{\ding{51}}}%
\newcommand{\xmark}{\textcolor{Red}{\ding{55}}}%
\usepackage[colorlinks=true,urlcolor=blue,citecolor=blue,linkcolor=blue]{hyperref}
\usepackage{physics}
\DeclareGraphicsExtensions{.eps,.ps,.pdf}
\def \Tr {\text{Tr}}
\newcommand{\ie}{\hbox{\textit{i.e.}{}}}

\newcommand{\Hs}{\mathcal{H}}
\newcommand{\HSs}{\mathcal{B}}
\newcommand{\BHs}{\mathscr{B}}

\renewcommand{\vec}{\mathbf}
\newcommand{\SU}{\mathrm{SU}}
\newcommand{\SO}{\mathrm{SO}}
\newcommand{\su}{\mathfrak{su}}

\let\oldcdot\cdot
\renewcommand{\cdot}{\boldsymbol{\oldcdot}}

\newcommand{\group}{\mathcal{G}}

\makeatletter
\g@addto@macro\bfseries{\boldmath}
\makeatother

\newtheorem{definition}{Definition}

\newtheorem{lemma}{Lemma}
\newtheorem{proposition}{Proposition}

\begin{document} 

\title{Dynamical decoupling and quantum error correction with $\SU(d)$ symmetries}

\author{Colin Read}
\email{cread@uliege.be}
\affiliation{Institut de Physique Nucléaire, Atomique et de Spectroscopie, CESAM, University of Liège
\\
B-4000 Liège, Belgium}
\author{Eduardo Serrano-Ens\'astiga}
\email{ed.ensastiga@uliege.be}
\affiliation{Institut de Physique Nucléaire, Atomique et de Spectroscopie, CESAM, University of Liège
\\
B-4000 Liège, Belgium}
\author{John Martin}
\email{jmartin@uliege.be}
\affiliation{Institut de Physique Nucléaire, Atomique et de Spectroscopie, CESAM, University of Liège
\\
B-4000 Liège, Belgium}


\begin{abstract}
Dynamical decoupling is a long-established and effective way to suppress unwanted interactions in qubit systems, enabling advances in fields ranging from quantum metrology to quantum computing. For general qudit systems, however, comparable protocols remain rare, mainly because Hamiltonian engineering in higher dimensions lacks the geometric intuition available for qubits. Here we present a general framework for dynamical decoupling in qudit systems, based on Lie group representation theory. By extending the group theory approach to dynamical decoupling, we show how decoupling groups can be systematically identified among the finite subgroups of $\SU(d)$ by analyzing their access to the irreducible components of the operator space. As an application, we construct new pulse sequences for interacting qutrit systems based on finite subgroups of $\SU(3)$, and show how subgroup factorizations and group orientations can be exploited to obtain shorter and more experimentally practical protocols for spin-$1$ systems with large zero-field splitting. We further show that the same symmetry-based framework yields quantum error-correcting codes: whenever a finite subgroup of $\SU(d)$ acts as a decoupling group for the relevant set of operators, the associated one-dimensional symmetry sectors define codespaces satisfying the Knill–Laflamme conditions, thereby unifying dynamical decoupling and quantum error correction in multi-level quantum systems.
\end{abstract}

\maketitle 


\section{Introduction}

Protecting a quantum system from decoherence and dephasing over an extended period of time is essential to maintain the advantage of many quantum technologies, from quantum sensing~\cite{Degen_2017} to quantum information processing~\cite{Nielsen_Chuang_2010,Preskill_2018quantum,NISQAlgo}, compared to their classical counterparts. While quantum error correction codes (QECCs) are widely regarded as the key to achieve arbitrarily long and accurate quantum computations~\cite{lidar2013quantum,Terhal_2015}, their viability relies on a sufficiently low error rate that we are only just beginning to achieve. Furthermore, much lower error rates are necessary to ensure practical implementations in realistic experimental setups. By contrast, many quantum-simulation and quantum-sensing platforms are not naturally compatible with standard QECC architectures. It is therefore of fundamental importance to develop new schemes to mitigate decoherence in these platforms.

\par One such method is \textit{dynamical decoupling} (DD), an open-loop protocol that uses carefully designed pulse sequences to engineer the Hamiltonian of the quantum system of interest. Originally introduced in the context of NMR spectroscopy~\cite{Levitt_2008Book}, this technique is now frequently used in most experimental platforms and for various quantum technologies. In the field of quantum sensing, DD sequences are generally used to increase sensitivity and enable the detection of AC fields in many solid-state sensors, such as NV-centers in diamond~\cite{Taylor_2008,Zhou_2020_metrology,Zhou_2023} or hexagonal Boron Nitride~\cite{Souvik_2025}. DD-based protocols are also commonly used in NV-centers~\cite{Childress_2006,Lukin_2007,Zhao_2012,Lukin_2012,Hanson_2012,Childress_Hanson_2013} and Silicon Carbide~\cite{Widmann_2014} to coherently manipulate the electron spin and neighboring nuclear spins for quantum computing purposes, or in many other relevant platforms, such as superconducting qubits~\cite{Ezzell_2023}, trapped ions~\cite{Biercuk_2009,Biercuk_2009LODD,Pelzer_2024} or neutral atoms~\cite{Cohen_2021,Bluvstein_2022,Chow_2021}, in order to extend the coherence time. 

\par While the most advanced experimental platforms currently being considered for quantum information processing and metrology rely on two-level systems, \ie, qubits, as their building blocks, many theoretical proposals suggest that an architecture based on a register of $d$-level systems, \ie, \textit{qudits}, might lead to more efficient quantum algorithms~\cite{Lanyon_2009,Wang_2020}, simulations~\cite{Neeley_2009,Ringbauer_2024,Popov_2024,vezvaee_2024} or error correction~\cite{Gottesman_2001, Cafaro_2012, Muralidharan_2017,Albert_2020,Gross_2021,Gross_2024,Lim_2023,lim_2025,Uy_2025}. Significant efforts have been made to improve the controllability of the multi-level building blocks across various platforms, such as trapped ions~\cite{Hrmo_2023,Ringbauer_2022}, superconducting qudits~\cite{Goss_2022}, and neutral atoms~\cite{Deutsch_2010,omanakuttan_2021}. Consequently, the development of error mitigation schemes for these systems is of fundamental interest.

\par While the theory of dynamical decoupling for qubit systems is already very mature, its extension to the qudit setting has yet to be fully explored. Decoupling a single qudit is relatively simple. For example, the generalization to qudits of the CPMG sequence, originally introduced to suppress dephasing in a single qubit~\cite{CarrPurcell_1954,Viola_1998}, has been proposed in Refs.\cite{iiyama_2024,Tripathi_2024}. A universal sequence for a single qudit, built on the Heisenberg-Weyl group, has also been proposed in Ref.\cite{Tripathi_2024}. This sequence can be viewed as a generalization of the $XY4$ sequence, which is built on the Klein $4$-group~\cite{Viola_1999,Viola_2003}, or equivalently on the $2$-dihedral group. Fewer results are available regarding the decoupling of interactions between qudits. The first sequences capable of canceling qudit-qudit interactions were proposed in Refs.~\cite{Wocjan_2002,Wocjan_2006, Rotteler_2006, Rötteler_Wocjan_2013}. They require selective addressing of each qudit and a number of pulses that increases linearly with the number of qudits in the register. Similarly, the concatenation of single-qudit sequences proposed in Ref.~\cite{Tripathi_2024} guarantees the decoupling of interactions between qudits, at the cost of a sequence complexity that grows exponentially with the number of qudits.

\par DD sequences that decouple qudit interactions using only global pulses are rare and are known only for very specific systems. Here, “global pulses” means control pulses applied identically and simultaneously to all qudits, so that the sequence complexity does not grow with system size. A short and elegant sequence constructed in Ref.~\cite{Lukin_2017} with linear programming techniques decouples dipole-dipole interactions between two qutrits in the rotating wave approximation. This sequence generalizes the WAHUHA sequence~\cite{Waugh_1968}, widely used in solid-state NMR, to the qutrit setting and requires only a few pulses. A similar sequence was constructed in Ref.~\cite{Lukin_2024}, which cancels dipole-dipole interactions and dephasing in the same system and has been experimentally demonstrated on NV centers. 

\par In this work, we present a systematic approach for constructing DD sequences for the qudit setting. In particular, we build on the abstract group-theoretic framework of~\cite{Viola_1999} and show that finite subgroups of $\SU(d)$ can be used to decouple many-body interactions.
Our framework relies heavily on the decomposition of the relevant operator space into irreducible representations (irreps) of $\SU(d)$~\cite{Hall_2015,Fulton_2013}; we show how one can find symmetries -\ie, finite subgroups of $\SU(d)$- that are \textit{inaccessible} to certain irreps of interest, and how these inaccessible symmetries can be used to construct robust decoupling sequences for all operators that transform as elements of these irreps. We then show how to select the irreps of interest based on the types of interactions to be decoupled. Our framework can be considered a generalization of the general theory of decoupling sequences based on the irreducible tensors of $\SO(3)$, presented in \cite {Llor_1991,Llor_1995,Llor_1995bis,Read_2025}, to the case of $\SU(d)$ or, in fact, any semisimple Lie group. Additionally, we show how our framework can be used to easily identify potential quantum error correction codes based on $\SU(d)$ symmetries.

\par This paper is organized as follows. In Sec.~\ref{Dyn.Dec.}, we first review the necessary concepts related to dynamical decoupling. We then present our theoretical framework based on the notion of \textit{inaccessible symmetries} and explain its connection to \textit{decoupling groups}. We then describe where and how to search for these inaccessible symmetries. Finally, we show how to shorten the resulting sequences by leveraging the Hamiltonian’s inherent symmetries. In Sec.~\ref{ex.SU2}, we first illustrate our framework for interacting spin-$s$ particles under global $\SU(2)$ pulses, recovering the Platonic symmetry groups as decoupling groups~\cite{Llor_1991,Llor_1995,Llor_1995bis,Read_2025}. In Sec.~\ref{ex.SU3}, we then treat interacting qutrits using finite subgroups of $\SU(3)$, recovering known universal sequences~\cite{Tripathi_2024,Read_2025} and identifying longer sequences that decouple arbitrary two- and three-body anisotropic interactions, including $\SU(3)$ generalizations of $\mathrm{TEDD}$\cite{Read_2025}. Next, in Sec.~\ref{sec.NV}, we focus on spin-$1$ systems with large zero-field splitting, and exploit Hamiltonian symmetries to construct shorter DD sequences to cancel disorder and dipole interactions, including analogues of $\mathrm{TEDDY}$\cite{Read_2025facto}. We conclude by showing how group-orientation freedom can simplify implementations and how the construction extends to $\SU(d)$ with $d>3$ in Sec.~\ref{sec.qudit}. Finally, in Sec.~\ref{sec.QEC}, we relate decoupling groups to quantum error correction, showing that a logical subspace with a given symmetry satisfies the Knill–Laflamme conditions when that symmetry is a decoupling group for the relevant set of operators. We then show how, for a given set of dominant physical errors, the framework identifies the appropriate symmetry group and the associated multilevel logical subspace. Several examples are given for spin and qutrit registers. We conclude with a summary and outlook of our work in Sec.~\ref{sec.conc}.

\section{Dynamical decoupling}\label{Dyn.Dec.}
\subsection{Notions of decoupling}

Consider a quantum system ($S$) interacting with a quantum bath ($B$), so that the total Hamiltonian in operator-Schmidt form reads $H=\sum_{\alpha}S_{\alpha}\otimes B_{\alpha}$, where the $S_{\alpha}$'s (resp.\ $B_{\alpha}$'s) act on the system's (resp.\ bath's) Hilbert space $\Hs_S$ (resp.\ $\Hs_B$). The goal is to mitigate these interactions by applying a sequence of equidistant, infinitely short and strong pulses,
\begin{equation}
    \mathrm{Seq}\,\equiv\,-\,P_1\,-\,P_2\,-\,\dots\, -\, P_N.\label{eq.Seq}
\end{equation}
In Eq.~\eqref{eq.Seq}, $P_i$ is a unitary operator in $\HSs(\Hs_S)$ that represents the effect of the $i$-th pulse in the sequence, and the dash $(-)$ represents a free evolution under the Hamiltonian $H$ for a duration $\tau_0$ (the waiting time).
\par The evolution in the toggling frame (interaction picture with respect to the pulse sequence) reduces to a stepwise constant Hamiltonian that is transformed by a sequence of unitary transformations,
\begin{equation}
    H \,\xrightarrow[P_1]{}\, U_2^{\dagger}HU_2\,\xrightarrow[P_2]{}\,\dots \,\xrightarrow[P_{N-1}]{}\,U_N^{\dagger}HU_N\,\xrightarrow[P_N]{}\,\label{Seq.Unitary}
\end{equation}
where the (instantaneous) action of each pulse is represented by an arrow and where we defined the sequence's propagators\footnote{The sequence propagator is defined as a unitary operator of $\HSs(\Hs_S)$ for simplicity, so it is understood that $U_k$ is extended to $U_k\otimes \mathds{1}_B$ in \eqref{Seq.Unitary}.} as 
\begin{equation}
    U_1 = \mathds{1}_S, \quad U_k = P_{k-1}P_{k-2}\dots P_1.
\end{equation}
At time $T=N\tau_0$, the time-evolution operator associated with the time-dependent Hamiltonian~\eqref{Seq.Unitary} can be written in the toggling frame as $\exp (\sum_{n=1}^{\infty}\Omega_n )$ using the Magnus expansion, where $\Omega_n$ denotes the $n$th order term. The expansion converges when $T\norm{H}<\pi$. For $T\norm{H}\ll 1$, the series is well approximated by its first term, so the toggling-frame evolution reduces to a free evolution of duration $T$ under the time-independent Hamiltonian
\begin{equation}
    H_1 = \frac{i\Omega_1}{T}=\frac{1}{N}\sum_{U\in\mathcal{U}}(U^{\dagger}\otimes \mathds{1}_B)H(U\otimes \mathds{1}_B)
\end{equation}
where $\mathcal{U}=\{U_1,\dots,U_N\}$ denotes the propagators defining the frame transformations. To first order, the pulse sequence acts through the quantum map $\Pi_{\mathcal{U}}:\HSs(\Hs_S)\to\HSs(\Hs_S)$,
\begin{equation}
    S\mapsto \Pi_{\mathcal{U}}(S)=\frac{1}{N}\sum_{U\in\mathcal{U}}U^{\dagger}SU ,
\end{equation}
which is applied to each system operator $S_\alpha$ in the Schmidt decomposition.
The goal is therefore to design the pulse sequence such that this quantum operation maps each noise operator $S_\alpha$ to a multiple of the identity operator, or at least to an operator that acts trivially on the relevant subspace or subsystem. In this case, the system is said to be \textit{decoupled} to first order, and the pulse sequence~\eqref{eq.Seq} is called a \textit{dynamical decoupling} sequence.

\subsection{From NMR to group-theoretic decoupling}

The first decoupling sequences were developed in the context of NMR spectroscopy, where the systems to be decoupled were essentially ensembles of spins under coherent dephasing~\cite{Hahn_1950,CarrPurcell_1954}. Many interesting sequences were subsequently developed to suppress more noise mechanisms (\textit{e.g.}, dipole-dipole interactions) and enhance the spectral resolution of NMR experiments, a topic covered in numerous reviews~\cite{Levitt_2007,Levitt_2008Book,Mote_2016}. Later, a more general framework based on abstract group theory was established~\cite{Viola_1999,Zanardi_1999} and dynamical decoupling was introduced into many experiments outside of NMR. Since then, a wide range of methods has been introduced to construct efficient DD sequences, including optimization algorithms~\cite{Lidar_2013,Cappellaro_2022,zhang_2026}, other numerical search methods~\cite{Lukin_2017,Lukin_2020,Zhou_2023,Tyler_2023,Lukin_2024}, graph theory formalisms~\cite{Coote_2025,Minh_2025}, and orthogonal arrays~\cite{Wocjan_2002,Wocjan_2006}, among many others~\cite{Uhrig_2007,Biercuk_2009OFDD,Biercuk_2009LODD,West_2010QDD,Tayler_2025,Xu_2025}. The original approach based on group theory generally performs less well than more optimized approaches, as it relies primarily on the first-order Magnus approximation, whereas optimization-based methods allow for the construction of higher-order sequences for specific Hamiltonians. Nevertheless, sequences derived from this rigorous framework offer significant advantages: they are often more robust against finite-duration and control errors~\cite{Viola_2003}, less dependent on the detailed form of the Hamiltonian~\cite{Read_2025}, and naturally compatible with advanced strategies such as dynamically-corrected gates~\cite {Khodjasteh_2009PRL, Khodjasteh_2009PRA, Khodjasteh_2010, denis_2025}, dynamically-generated decoherence-free subspaces and noiseless subsystems~\cite {Viola_2000,Lidar_2002,Viola_2002,Quiroz_2024}, as well as quantum error correction~\cite {Paz_Silva_2013,Kasatkin_2026}, among others~\cite{Viola_2002,Viola_2005,Santos_2006, Lidar_2008, Quiroz_2012}.
\subsection{Extended group-theoretic framework}
In the remainder of this subsection, we present a slight extension of the group-theoretic framework of Refs.~\cite{Viola_1999,Viola_2003}. The main ingredients remain the interaction subspace—a vector subspace containing the operators to be decoupled—and the \emph{decoupling group}, a finite group of unitary transformations that decouples them. DD sequences are then constructed as Hamiltonian or Eulerian paths on the corresponding Cayley graphs. Our contribution is twofold. First, we use the decomposition of the operator space into irreducible representations of a semisimple Lie group, such as $\SU(d)$, to construct a vector space $V$ that carries a representation of the Lie group and contains the interaction subspace. As explained in Lemma~\ref{lemma.V}, this restricts the search for decoupling groups to a small part of the operator space and further simplifies it, as discussed in Sec.~\ref{sec.ConsV}. Second, we show how to determine systematically whether a finite group is a decoupling group by decomposing $V$ into irreducible representations of that group and checking whether the trivial representation appears. This is explained in detail in Sec.~\ref{sec.InSym}. In Sec.~\ref{sec.rob}, we comment on the robustness of the resulting DD sequences to finite-duration and control errors. Finally, in Sec.~\ref{sec.facto}, we review two simple methods~\cite{Read_2025facto}, compatible with this framework, that exploit symmetries of the interaction subspace to reduce the complexity of the decoupling sequence.

\subsubsection{Mathematical framework}
The mathematical framework relies on standard results from the representation theory of Lie groups and finite groups, available in many textbooks~\cite{Rotman_2012,Fulton_2013,Hall_2015}. In particular, it applies to any semisimple Lie group, since it relies only on the complete reducibility theorem. While we focus on $\SU(d)$, which is directly relevant for qudit systems, the same framework should also allow the design of decoupling protocols based on subgroups of less common Lie groups, such as symplectic or special orthogonal groups.

Consider a representation ($\pi,\Hs_S$) of the semisimple Lie group $\SU(d)$, such that, for all $g\in\SU(d)$, $\pi(g)$ corresponds to a unitary operation on $\Hs_S$ that we can implement experimentally. We want our sequence of pulses to consist of such pulses. It is therefore reasonable to require that the set of propagators of the sequence $\mathcal{U}=\{U_1,\dots,U_N\}$ be the image of a finite subset $\group\subset \SU(d)$ under $\pi$, \ie, $\mathcal{U}= \pi(\group)$. In particular, we can require that the subset $\group$ forms a finite subgroup of $\SU(d)$. In this case, the quantum operation implemented by the sequence will be a projector on a $\group$-invariant subspace\footnote{Note that in many mathematics textbooks, the term “$\group$-invariant subspace” refers to any subspace closed under the elements of $\group$, \ie, one that defines a representation of the group. In this work, we call $\group$-invariant a subspace on which the elements of $\group$ act as the identity.} of $\HSs(\Hs_S)$, which we denote by $\Pi_{\group}$:
\begin{equation}
    S \mapsto \Pi_{\group}(S) = \frac{1}{\abs{\group}}\sum_{g\in\group}\pi(g^{-1})S\pi(g).
    \label{eq.qo}
\end{equation}
This can easily be proved by noting that $\Pi_{\group}^2=\Pi_{\group}$ and $\qty[\pi(g),\Pi_{\group}(S)]=0$ $\forall g\in\group$~\cite{Viola_1999,Zanardi_1999}.

\par In some cases, the $\group$-invariant subspace is trivial, in the sense that it is either the zero subspace\footnote{The zero subspace is the trivial vector space containing only the zero vector; in the present setting, this corresponds to the zero operator.}, or the subspace spanned solely by the identity. In this situation, we say that the symmetry is \textit{inaccessible}. We now define this notion precisely.
\begin{definition}
    \textbf{(inaccessible symmetry)\,}Given a non-trivial representation $(\pi, V)$ of $\SU(d)$, we say that the finite subgroup $\group<\SU(d)$ is an inaccessible symmetry for $V$ if the $\group$-invariant subspace is either the zero subspace or the span of the identity on $V$. 
\end{definition}

\par Inaccessible symmetries are naturally related to dynamical decoupling sequences. We now recall the definitions of \textit{interaction subspace} and \textit{decoupling group}~\cite{Viola_1999}.
\begin{definition}
    \textbf{(interaction subspace)\,}Given a system-bath Hamiltonian of the form $H=\sum_{\alpha}S_{\alpha}\otimes B_{\alpha}$, where the operators $S_{\alpha}$ (resp.\ $B_{\alpha}$) act on the system Hilbert space $\Hs_S$ (resp.\ the bath Hilbert space $\Hs_B$), the vector space spanned by all system operators $S_{\alpha}$ is called the interaction subspace,
    \begin{equation}
        \mathcal{I}_S = \mathrm{span}(\{S_{\alpha}\}_{\alpha}).
    \end{equation}
\end{definition}
\begin{definition}
    \textbf{(decoupling group)\,}Given a representation $(\pi,V)$ of $\SU(d)$ and an interaction subspace $\mathcal{I}_S\subseteq \HSs(V)$, a finite subgroup $\group<\SU(d)$ is called a decoupling group for $\mathcal{I}_S$ if the image of $\mathcal{I}_S$ under $\Pi_{\group}$ is the zero subspace or the span of the identity.
\end{definition}

\par Now that these concepts have been defined, we can move on to a trivial lemma that relates inaccessible symmetries to decoupling groups.

\begin{lemma}
    Let $(\pi,\HSs(\Hs_S))$ be a representation of $\SU(d)$ and $\mathcal{I}_S\subseteq \HSs(\Hs_S)$ an interaction subspace. If a finite subgroup $\group<\SU(d)$ is an inaccessible symmetry for $\HSs(\Hs_S)$, then it is a decoupling group for $\mathcal{I}_S$.
\end{lemma}

This lemma shows that a decoupling group may be constructed from symmetries that are absent from the operator space of the system. It is clear that this condition is sufficient but not necessary for finding a decoupling group; using this construction becomes inefficient as the system size  increases, because the operator space expands, even though $\mathcal{I}_S$ remains unchanged, making more symmetries accessible.  We therefore wish to find a method that ensures the decoupling group depends on the interaction subspace and not on the system size. To do this, we seek a subspace $V$ with $\mathcal{I}_S\subseteq V\subseteq \HSs(\Hs_S)$ that is closed under a finite subgroup $\group$ acting as an inaccessible symmetry of $V$. In other words, we seek a subrepresentation $(\pi,V)$ of the group $\group$. Defining $V$ in this way, we find the following lemma.
\begin{lemma}
Let $(\pi,\HSs(\Hs_S))$ be a representation of $\SU(d)$, and let $\mathcal{I}_S\subseteq \HSs(\Hs_S)$ be an interaction subspace. Let $\group<\SU(d)$ be a finite subgroup, and suppose there exists a subspace $V$ such that $\mathcal{I}_S\subseteq V\subseteq \HSs(\Hs_S)$, with $V$ invariant under the action of $\group$, that is, $\pi(g)^\dagger S\,\pi(g)\in V,\;\forall\,S\in V,\ \forall\, g\in \group$. If $\group$ is an inaccessible symmetry for $V$, then $\group$ is a decoupling group for $\mathcal{I}_S$.
\label{lemma.V}
\end{lemma}
The proof is straightforward; it essentially relies on the closure of $V$ under the group action, which guarantees that the image of $\mathcal{I}_S$ under $\Pi_{\group}$ is confined to the subspace $V$. Thus, the definition of this subspace $V$ allows us to restrict our search for inaccessible symmetries to a corner of the operator space, whose size depends on the interaction subspace, but not on the size of the system's Hilbert space. 

\begin{figure*}[t]
    \centering
    \includegraphics[width=\linewidth]{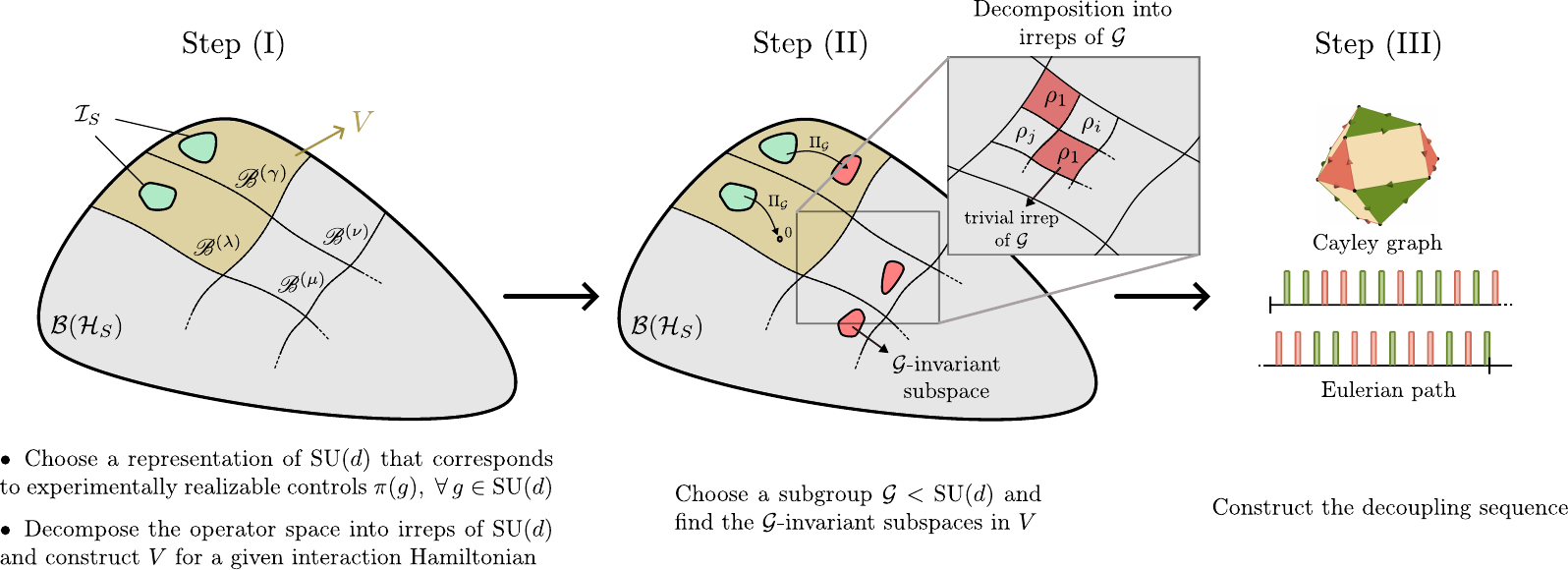}
    \caption{Representation of the conceptual framework. (I) The operator space of the system is decomposed into irreps of $\SU(d)$, which serve as building blocks to construct a vector space $V$ that includes the interaction subspace associated with a given interaction Hamiltonian. (II) A subgroup $\group<\SU(d)$ is then specified, and the $\SU(d)$ irreps are decomposed into irreps of $\group$ in order to find their $\group$-invariant subspace. (III) If the symmetry is inaccessible to the irreps used to construct $V$, then we can construct the Cayley graph from which a decoupling sequence follows.}
    \label{fig:fig1}
\end{figure*}

The last lemma defines the path to constructing DD sequences, which we structure as a three-step procedure.
The first step (I) will consist of decomposing the operator space into irreducible representations (irreps) of $\SU(d)$ and using these building blocks to construct $V$. The second step (II) will consist of decomposing each of these building blocks into irreps of the finite subgroup $\group<\SU(d)$ and looking for the trivial representation. The last step (III) will construct a sequence of pulses from a group through its respective Cayley graph. Below, we explain in detail how to systematically perform these three tasks. The conceptual framework is illustrated in Fig.~\ref{fig:fig1}.

\subsubsection*{Step (I): Constructing \texorpdfstring{$V$}{Lg}}\label{sec.ConsV}

Here, we explain how to construct relevant vector space $V$ containing the interaction space. 
Given a representation $(\pi, \Hs_S)$ of $\SU(d)$ on $\Hs_S$, one naturally obtains a representation on the operator space $\HSs(\Hs_S)$ by 
\begin{equation}
    \pi^{\mathrm{op}} \equiv \pi\otimes \overline{\pi} , \quad \HSs(\Hs_S)\cong\Hs_S\otimes \overline{\Hs_S},
\end{equation}
where $(\overline{\pi}, \overline{\Hs_S})$ denotes the \textit{contragredient}, or \textit{dual}, representation~\cite{Hall_2015}. This representation describes how operators acting on $\Hs_S$ transform under conjugation by elements of $\SU(d)$. Explicitly, for $O\in\HSs(\Hs_S)$ and $g\in\SU(d)$, $\pi^{\mathrm{op}}(g)$ acts on $O$ as 
\begin{equation}
    \pi^{\mathrm{op}}(g)\cdot O = \pi(g)\,O\,\pi(g)^{\dagger}.
    \label{eq.ad}
\end{equation}
Since $(\pi^{\mathrm{op}},\HSs(\Hs_S))$ is a finite-dimensional representation of the semisimple Lie group $\SU(d)$, it decomposes into a direct sum of irreducible representations: 
\begin{equation}
    \HSs(\Hs_S) = \bigoplus_{\mu\in S}\bigoplus_{\alpha\in S_{\mu}}\BHs^{(\mu)}_{\alpha} , \label{eq.Hdecomp}
\end{equation}
where the space of operators has been partitioned into a direct sum of subspaces that (i) are closed under the action of $\pi^{\mathrm{op}}$, \ie, $\forall g\in \SU(d)$, and $\forall \mu,\alpha$,
\begin{equation}
    \pi^{\mathrm{op}}(g)\cdot \BHs^{(\mu)}_{\alpha} \equiv \pi(g)\BHs^{(\mu)}_{\alpha} \pi(g)^{\dagger} = \BHs^{(\mu)}_{\alpha},
\end{equation}
and (ii) do not contain nontrivial subspaces closed under $\pi^{\mathrm{op}}$. 
Accordingly, the linear map $\pi^{\mathrm{op}}$ itself decomposes as
\begin{equation}
    \pi^{\mathrm{op}} = \bigoplus_{\mu\in S}\bigoplus_{\alpha\in S_{\mu}}\pi^{(\mu)}_{\alpha},\label{eq.Pidecomp}
\end{equation}
where $\pi^{(\mu)}_{\alpha}$ describes how $\pi^{\mathrm{op}}$ acts on the subspace $\BHs^{(\mu)}_{\alpha}$. Each irrep $(\pi^{(\mu)}, \BHs^{(\mu)})$ of $\SU(d)$ is entirely characterized by its \textit{highest weight} $\mu$, which is an element of the Cartan subalgebra of $\su(d)$, usually represented as a $(d-1)$-dimensional vector once a basis of the Cartan subalgebra has been chosen. The additional index $\alpha$ distinguishes multiple copies of the same irrep. In Eqs.~\eqref{eq.Hdecomp} and~\eqref{eq.Pidecomp}, $S$ denotes the set of highest weights appearing in the decomposition, and $S_\mu$ is a set of indices that label the copies of the irrep with the highest weight $\mu$.

\par Once $(\pi^{\mathrm{op}}, \HSs(\Hs_S))$ has been decomposed into irreps, one may choose subsets $S'\subseteq S$ and $S'_{\mu}\subset S_{\mu}$ of these irreps and construct a vector space and a linear map
\begin{equation}\label{eq:V}
    V \equiv \bigoplus_{\mu\in S'}\bigoplus_{\alpha\in S'_{\mu}}\BHs^{(\mu)}_{\alpha} , \quad \pi_V \equiv \bigoplus_{\mu\in S'}\bigoplus_{\alpha\in S'_{\mu}}\pi^{(\mu)}_{\alpha} ,
\end{equation}
so that $(\pi_V,V)$ is a representation of $\SU(d)$, and hence is closed under the action of any subgroup of $\SU(d)$. The goal is to select the smallest possible subsets $S'$ and $S'_{\mu}$ such that $\mathcal{I}_S\subseteq V$. One can then search for inaccessible symmetries irrep by irrep. Since irreps with the same highest weight $\mu$ are isomorphic, the existence of an inaccessible symmetry will depend only on the set of highest weights $S'$ appearing in the decomposition.

\subsubsection*{Step (II): Searching for inaccessible symmetries}\label{sec.InSym}

Let us now focus on a specific irreducible representation of $\SU(d)$, $(\pi^{(\mu)}, \BHs^{(\mu)})$, appearing in the decomposition of the vector space $V$ in Eq.~\eqref{eq:V}. The next step is to determine the $\group$-invariant subspace for a given finite group $\group<\SU(d)$. Although $(\pi^{(\mu)}, \BHs^{(\mu)})$ is irreducible as a representation of $\SU(d)$, it is generally reducible when restricted to the subgroup $\group$. We can thus play the same trick as before and decompose it into irreps of $\group$: 
\begin{equation}\label{eq:decompirrepG}
    \BHs^{(\mu)} = \bigoplus_i \rho_i^{\oplus a_i},\quad \pi^{(\mu)}|_{\group}= \bigoplus_i\lambda_i^{\oplus a_i}.
\end{equation}
Here each subspace $\rho_i$, appearing with multiplicity $a_i$, is closed under the action of $\pi^{(\mu)}(g)$ $\forall g\in \group$, and the linear map $\lambda_i$, which is defined on the group $\group$\footnote{But not on all $\SU(d)$. We make this explicit by writing $\pi^{(\mu)}|_{\group}$ in Eq.~\eqref{eq:decompirrepG}.}, describes the action of $\pi^{(\mu)}(g)$ on this specific subspace. The multiplicity $a_i$ gives the number of copies of that irrep in the decomposition.

\par There will be only a finite number of irreps of a finite group, equal to the number of conjugacy classes in the group\footnote{Two elements $x,y\in\group$ are said to be conjugate if there exists $g\in\group$ such that $x=gyg^{-1}$. The conjugacy class of $x$ is the set of all elements conjugate to $x$, namely $\{gxg^{-1}\mid g\in\group\}$.}. All irreps of a finite group are generally listed in what are called the \textit{character tables} of the group, and all that remains is to calculate the multiplicity $a_i$ of each of these irreps. In particular, we are interested in the \textit{trivial irrep} of $\group$ --generally denoted $(\lambda_1,\rho_1)$--, which is the one-dimensional representation on which every element of the group acts as the identity. Any element defined on such an irrep is thus $\group$-invariant, which means that the $\group$-invariant subspace in $\BHs^{(\mu)}$ is nothing but the vector space $\rho_1^{\oplus a_1}$. We thus conclude that $\group$ will be inaccessible to the considered $\SU(d)$ irrep $\BHs^{(\mu)}$ if and only if the multiplicity $a_1$ of the trivial irrep of $\group$ is equal to zero. We must now calculate this multiplicity.

\par The multiplicity $a_i$ is given by the inner product between the \textit{characters} of the representations $(\pi^{(\mu)},\BHs^{(\mu)})$ and $(\lambda_i,\rho_i)$, that is
\begin{equation}
    a_i = (\chi_{(\mu)},\chi_i) = \frac{1}{\abs{\group}}\sum_{g\in\group}\overline{\chi_{(\mu)}(g)}\chi_i(g).
    \label{Eq.char.multi}
\end{equation}
Here, $\chi_{(\mu)}$ and $\chi_i$ denote the corresponding characters, i.e., complex-valued functions of the conjugacy classes of the group. The irreducible characters $\{\chi_i(g)|\,g\in\group\}$ are listed in the character table of $\group$. For the trivial irrep, the character is equal to one for every element of the group. To calculate the characters of $(\pi^{(\mu)}, \BHs^{(\mu)})$, we must use Weyl's character formula, which reads (using the notations of Ref.~\cite{Hall_2015}) 
\begin{equation}
    \chi_{(\mu)}(e^{H}) = \frac{\sum_{w\in W}\mathrm{det}(w)\, e^{\expval{w\cdot(\mu + \delta), H}}}{\sum_{w\in W}\mathrm{det}(w)\, e^{\expval{w\cdot\delta, H}}}
\label{Eq.Weyl.character}
\end{equation}
where $\delta$ is half the sum of the positive roots of the algebra; $W$ is the Weyl group; $\mu$ is the highest weight; $\expval{X,Y}=\Tr[X^*Y]$ is the inner product on the Cartan subalgebra; and $H$ is an element of the Cartan subalgebra. Although the formula gives only the character of diagonal elements of $\SU(d)$, one can find any other character by exploiting the invariance of the character under conjugation by an element of $\SU(d)$, \ie,
\begin{equation}
    \chi_{(\mu)}(X) = \chi_{(\mu)}(YXY^{\dagger})\quad\forall Y,X\in \SU(d).
\end{equation}

\par The results of this section are summarized in the following proposition.
\begin{proposition}
    Let $(\pi^{(\mu)},\BHs^{(\mu)})$ be an irreducible representation of $\SU(d)$, and let $\group<\SU(d)$ be a finite subgroup. Then, $\group$ is inaccessible for $\BHs^{(\mu)}$ if and only if the trivial irrep of $\group$ does not appear in the decomposition \eqref{eq:decompirrepG} of $(\pi^{(\mu)},\BHs^{(\mu)})$, that is, its multiplicity
    \begin{equation}
        a_1 = \frac{1}{\abs{\group}}\sum_{g\in\group} \chi_{(\mu)}(g)
    \end{equation}
    is equal to zero, where $\chi_{(\mu)}(g)$ is given by the Weyl character formula \eqref{Eq.Weyl.character}.
\end{proposition}

\subsubsection*{Step (III): Constructing pulse sequences}
Lastly, suppose we find a subspace $V$ with an inaccessible symmetry $\group$. In this case, we have a decoupling group that defines the quantum operation to be applied to implement the decoupling. However, we still need to construct the actual pulse sequence that implements this quantum operation. A useful method~\cite{Viola_2003} for doing this is to use what is called the \textit{Cayley graph} of the group $\group$ with respect to a generating set $\Gamma$, which we will denote by $C(\group,\Gamma)$. In such a graph, a vertex is assigned to each element of the group. We then choose a generating set, which is a subset of elements $\Gamma=\{\gamma_{\lambda}\}_{\lambda} \subset \group$ that generates the entire group by multiplying them together. Directed edges are then added to the graph, connecting vertices $i\to j$ if an element $\gamma_{\lambda}\in\Gamma$ satisfies $g_j=\gamma_{\lambda}g_i$. Colors can be assigned to the different generators and the edges colored accordingly to distinguish them. The overall result is a directed graph where each vertex has $\abs{\Gamma}$ incoming and outgoing edges (one of each color).

\par We can then find an Eulerian circuit (which always exists), which visits each edge exactly once, or a Hamiltonian circuit (if it exists), which visits each vertex exactly once. We then specify a starting point, and we are left with a sequence of edges that corresponds to a sequence of pulses. This sequence implements the desired quantum operation and consists of $\abs{\Gamma}$ distinct pulses, for a total of $\abs{\Gamma}\abs{\group}$ pulses (for the Eulerian path) or $\abs{\group}$ pulses (for the Hamiltonian path). Each pulse is, by construction, a unitary operator that represents an element of the generating set $\Gamma$. 

\subsection{Eulerian path and robustness to finite-duration and control errors} \label{sec.rob}

The group theory framework presented above relies on the ideal pulse approximation, according to which each pulse is considered error-free,  infinitely fast, and strong, such that its duration is much shorter than the interval between two successive pulses. If this approximation does not hold, two types of errors should be taken into account: control errors, due to imperfect pulses, and finite-duration errors, caused by the finite duration of the pulses. We now quantify these errors for a single pulse of propagator $P(t)$, which implements the unitary pulse $P=P(\tau)$ in a duration $\tau$, and explain how these imperfections can be addressed. 

\par \textit{Finite-duration errors.} During the implementation of the pulse, the unwanted Hamiltonian $H$ remains active and contributes to the pulse errors. These are finite-duration errors, quantified by moving to the rotating frame with respect to the pulse and performing a first-order Magnus expansion~\cite{Viola_2003}. The noisy pulse will therefore be given by 
\begin{equation}
    P_{\mathrm{noisy}} \approx P e^{-i\tau H_P}
\end{equation}
where 
\begin{equation}
    H_P = \frac{1}{\tau}\int_0^{\tau} dt\,P^{\dagger}(t)HP(t)
\end{equation}
is the finite-duration error Hamiltonian. 

\par \textit{Control errors.} Errors introduced by the imperfect pulse can be modeled by an error Hamiltonian (which may be time-dependent) $H_{\mathrm{err}}^P(t)$ acting during the duration of the pulse. In the rotating frame with respect to the pulse, this contributes to the first-order error as follows:
\begin{equation}
    P_{\mathrm{noisy}} \approx P e^{-i\tau H_P' }
\end{equation}
where 
\begin{equation}
\label{Eq.HPprime}
    H_P' = \frac{1}{\tau}\int_0^{\tau} dt\,P^{\dagger}(t)\big(H + H_{\mathrm{err}}^P(t)\big)P(t)
\end{equation}
includes both the finite-duration error Hamiltonian and the effective control error.

\par In order to deal with these errors, one can follow an Eulerian path on the Cayley graph $C(\group,\Gamma)$ of the group $\group$ with respect to the generating set $\Gamma=\qty{\gamma_{\lambda}}_{\lambda=1}^{\abs{\Gamma}}$, as discovered in Ref.~\cite{Viola_2003}. Each pulse of this sequence will correspond to an element of the generating set, $P\in \pi(\Gamma)$ where $\pi$ is the linear map defining the relevant representation of $\SU(d)$ acting on the system's Hilbert space, and we can thus label the different pulses used in the sequence $P_{\lambda} = \pi(\gamma_{\lambda})$. Similarly, we label the propagator $P_{\lambda}(t)$, such that $P_{\lambda}(\tau) = P_{\lambda}$. In general, we can assume that there is no waiting time between pulses\footnote{We can simply consider a finite waiting time $\tau_0$ and a pulse duration $\tau-\tau_0$ by choosing a pulse propagator $P_{\lambda}(t)=\mathds{1}_S$ for $t\in[0,\tau_0]$.}.

\par If the control errors are \textit{systematic}--that is, the same error Hamiltonian $H_{\mathrm{err}}^{P_{\lambda}}(t)$ acts whenever a pulse $P_{\lambda}$ is implemented--, then choosing an Eulerian path ensures that the quantum operation undergone by the Hamiltonian in the toggling frame is given by~\cite{Viola_2003} 
\begin{equation}\begin{aligned}
    H &\mapsto \frac{1}{\abs{\Gamma}}\sum_{\lambda =1}^{\abs{\Gamma}}\frac{1}{\abs{\group}}\sum_{g\in\group} \pi(g)^{\dagger} H_{P_{\lambda}}'\pi(g) \\ 
    &= \frac{1}{\abs{\Gamma}}\sum_{\lambda =1}^{\abs{\Gamma}}\Pi_{\group}\qty(H_{P_{\lambda}}')
\end{aligned}
\end{equation}
where $H_{P_{\lambda}}'$, as defined in Eq.~\eqref{Eq.HPprime},
is the error Hamiltonian of the pulse $P_{\lambda}$. In conclusion, the error Hamiltonian corresponding to each pulse $P_{\lambda}$, $\lambda=1,\dots, \abs{\Gamma}$, used in the sequence will be symmetrized independently. If $\group$ is a decoupling group for these operators, then robustness against finite-duration and control errors is guaranteed.

\par If $\group<\SU(d)$ is an inaccessible symmetry for a certain vector space $V$ that is closed under $\SU(d)$ and contains both the unwanted Hamiltonian $H$ and the control error Hamiltonian $H_{\mathrm{err}}^{P_{\lambda}}(t)$ $\forall t$, then robustness to finite-duration errors is guaranteed provided the pulse propagator $P_{\lambda}(t)$ remains within the representation of $\SU(d)$ at all times. That is, there must exist a smooth path $\gamma_{\lambda}(t)$ such that $\gamma_{\lambda}(0)=\mathds{1}_{d\times d}$, $\gamma_{\lambda}(\tau)=\gamma_{\lambda}$ and satisfying $\gamma_{\lambda}(t)\in \SU(d)$ and $P_{\lambda}(t)=\pi(\gamma_{\lambda}(t))$ $\forall t$. Equivalently, the generating Hamiltonian of the pulse must lie in the $\mathfrak{su}(d)$ algebra at every time. Since $V$ is closed under $\SU(d)$, the unitary evolution generated during the pulse keeps the transformed Hamiltonian inside $V$, and $\group$ therefore also decouples the error Hamiltonian~\eqref{Eq.HPprime}. By contrast, if the pulse does not satisfy this condition,  the transformed Hamiltonian may leave $V$ during the pulse, and robustness is generally lost.

\par
\begin{figure}[t]
    \centering
    \includegraphics[width=\linewidth]{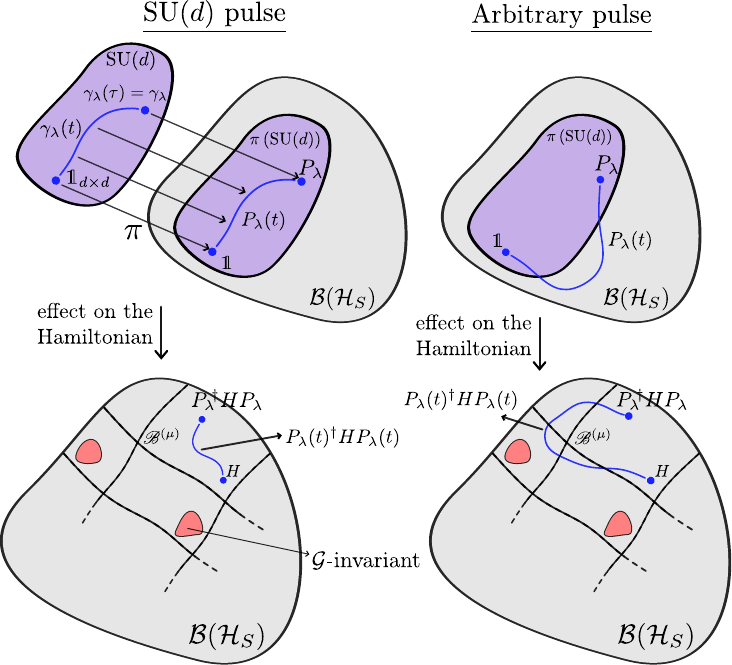}
    \caption{Effect of pulse designs on robustness. Left: when the pulse $P_\lambda(t)$ is generated by a smooth path in $\SU(d)$, the transformed Hamiltonian stays within the subspace $V$ which is closed under $\SU(d)$ for all times, thus ensuring that $\group$ remains an inaccessible symmetry for the relevant error Hamiltonian. Right: if the pulse trajectory leaves $\SU(d)$, the transformed Hamiltonian may leave $V$, so $\group$ no longer decouples the pulse errors.} 
    \label{fig:fig2}
\end{figure}

 This condition on the design of the pulses is sketched in Fig.~\ref{fig:fig2}. This design is not very restrictive and leaves plenty of degrees of freedom for optimizing the pulses.

\subsection{Reducing sequence complexity through group factorization}
\label{sec.facto}
Consider the projector $\Pi_{\group}$ defined in Eq.~\eqref{eq.qo} on the $\group$-invariant subspace of some vector space. To apply this symmetrization to the Hamiltonian of our system, we need a pulse sequence whose minimum length is equal to the order of the group. However, when the element $O$ of the vector space on which this projector acts has certain symmetries, it is possible to reduce the complexity of this protocol. The first step in understanding this mechanism is to write $\group$ as the product of one of its subgroups $K<\group$ with a set of representatives of its right- or left-cosets, \ie, 
\begin{equation}
    \group = K\,S_R = S_L\,K
\end{equation}
for a certain set of group elements $S_R$ and $S_L$ of order $\abs{S_R}=\abs{S_L}=\abs{\group}/\abs{K}$. Note that these sets need not necessarily be subgroups. Based on this decomposition, there are two ways to construct decoupling groups. The first method exploits the symmetries of $O$ under the coset representatives $S_R$, allowing $K$ itself to serve as the decoupling group (see e.g.~Ref.~\cite{Read_2025facto}). The second exploits the symmetries of $O$ under $K$~; when $K$ is normal, the relevant reduced structure is the quotient group $\group/K$ that serves as the decoupling group (see Ref.~\cite{Viola_2003}).

\par \textit{First method.} Assume that the set of representatives of the right-cosets is a symmetry group of the operator $O \in \HSs (\Hs)$. Then,
\begin{equation}\begin{aligned}
    \Pi_{K}(O) &= \Pi_{K}\qty(\frac{1}{\abs{S_R}}\sum_{s\in S_R}\pi^{\mathrm{op}}(s) \cdot O)\\
    &= \frac{1}{\abs{K}}\sum_{k\in K}\frac{1}{\abs{S_R}}\sum_{s\in S_R}\pi^{\mathrm{op}}(k) \cdot \left( \pi^{\mathrm{op}}(s) \cdot O \right) \\ 
    &= \frac{1}{\abs{\group}}\sum_{g\in \group}\pi^{\mathrm{op}}(g) \cdot O \equiv\Pi_{\group}(O).
\end{aligned}\end{equation}
Thus, the symmetries of $O$ allow the projection onto the $\group$-invariant subspace to be implemented using only the subgroup $K$. $K$ can thus be used as a decoupling group, which yields a pulse sequence of minimum length $\abs{K}$. In Ref.~\cite{Read_2025facto}, such factorizations of $\SO(3)$ point groups into products of subgroups were used to construct DD sequences for Hamiltonians with different $\SU(2)$ symmetries. There, the Hamiltonian symmetries were identified through the Majorana representation, which provides an intuitive framework to design short sequences with interesting decoupling properties.

\par \textit{Second method.} In this case, we require that $K$ be a symmetry subgroup of $O$ and that $K$ be a normal subgroup of $\group$, denoted by $K\triangleleft \group$. Then, the left and right cosets of $K$ in $\group$ coincide, so $\group = KS = SK$, and we may simply speak of cosets. Although the set $S$ of coset representatives need not itself form a group, 
 the set of cosets of $K$ in $\group$ does when $K$ is normal; this group is the \textit{quotient group} $\group/K$, with multiplication
$\group/K\times \group/K \to \group/K$ defined by 
\begin{equation}
    (s_1\,K) \times (s_2\,K) = s_1s_2\,K. 
\end{equation}
Its generators are therefore cosets $\Gamma = \qty{s_{\gamma}\, K}_{\gamma=1}^{\abs{\Gamma}}$, represented by a choice of elements $\Gamma'=\qty{s_{\gamma}}_{\gamma=1}^{\abs{\Gamma}}\subset S$. A Hamiltonian or Eulerian path on the Cayley graph of $\group$ thus gives a sequence of cosets of $\Gamma$, which can be implemented by replacing each coset $s_\gamma K$ with a chosen representative $s_\gamma$. This yields a sequence of implementable pulses realizing the projection
\begin{equation}\begin{aligned}
    \Pi_{\group/K}(O) &= \frac{1}{\abs{\group/K}}\sum_{s\in S}\pi^{\mathrm{op}}(s) \cdot \qty(\frac{1}{\abs{K}}\sum_{k\in K}\pi^{\mathrm{op}}(k) \cdot O) \\ 
    &= \frac{1}{\abs{\group}}\sum_{g\in \group}\pi^{\mathrm{op}}(g) \cdot O\equiv\Pi_{\group}(O).
\end{aligned}\end{equation}
Designing a sequence of pulses on the Cayley graph of a quotient group was first considered in Ref.~\cite{Viola_2003}.

\par As explained in Ref.~\cite{Read_2025facto}, leveraging the symmetries of the interaction subspace to reduce the complexity of the sequence typically comes at the expense of robustness to finite-duration errors. This is because the symmetry of the Hamiltonian is generally broken by the unitary transformation in the finite-duration error Hamiltonian~\eqref{Eq.HPprime}. However, robustness to control errors is often still guaranteed. For each sequence designed in Sec.~\ref{sec.NV}, where the two methods above are used, we will explain to what extent robustness has been lost.

\section{Recovering \texorpdfstring{$\SU(2)$}{Lg} decoupling of interacting spins} \label{ex.SU2}
In this section, we focus on the case of $\SU(2)$, relevant for decoupling of interacting spin ensembles and large-spin qudits. In this specific case, our framework is closely related to the irreducible tensor formalism developed in Refs.~\cite{Llor_1991, Llor_1995, Llor_1995bis} for homonuclear decoupling in zero-field NMR. More recently, unaware of these contributions, we developed a general theory similar to this framework but based on the Majorana representation~\cite{Read_2025,Read_2025facto}, which identifies inaccessible symmetries through three-dimensional geometric figures that facilitate physical intuition, robustness analysis, and yield additional results. We also used the general theory of Eulerian sequences~\cite{Viola_2003} to improve the robustness of the sequences. In this section, we show that those results can be recovered in terms of pure representation theory, by searching for the trivial representation of finite groups.

\subsection{Representations of \texorpdfstring{$\SU(2)$}{Lg}}

\par The standard two-dimensional representation of $\SU(2)$ corresponds to the highest weight\footnote{Formally, the notion of highest weight is defined relative to a Cartan–Weyl basis of the Lie algebra. However, since the subsequent discussion does not require an explicit specification of the remaining basis elements, we omit them for brevity and clarity.}
\begin{equation}
    \mu_{\mathrm{std}} = \begin{pmatrix}
        \frac{1}{2}&0\\
        0&-\frac{1}{2}
    \end{pmatrix}=\frac{1}{2}\sigma_z .
\end{equation}
In physics, this representation describes the action of an element of $\SU(2)$ on a single spin-$1/2$. Any other $\SU(2)$-irrep can be obtained as a symmetric power of this irrep. In other words, the irrep of highest weight $\mu_{\mathrm{std}}$ is a \textit{fundamental representation}, in the sense that any $\SU(2)$-irrep has a highest weight that can be written in the form $N\mu_{\mathrm{std}}$ for a non-negative integer $N$. This integer uniquely specifies the irrep and can be encoded in a one-dimensional vector $\vec{d} = (N)$, whose entries are called \textit{Dynkin coefficients}. For groups of higher rank, such as $\SU(d)$ with $d>2$, the corresponding Dynkin vector has additional components. Thus, the vector $\vec{d}$ labels the $\SU(2)$-irrep of highest weight $\mu = N\mu_{\mathrm{std}}$, which corresponds exactly to the spin-$N/2$ irrep; its elements transform under $\SU(2)$ as spin-$N/2$ states. Note that in the case of $\SU(2)$, it is customary to describe an irrep using half the Dynkin coefficient; thus, we write $\mu \cong j$ to denote the spin-$j$ irrep with Dynkin coefficient $(2j)$ and highest weight $\mu = j\sigma_z$.

\par We now examine two physical systems to which we apply our theoretical framework to find decoupling groups: we will begin with the case of a single spin-$j$ as a warm up, and then move on to the case of a spin-$s$ ensemble.

\subsection{A single spin-\texorpdfstring{$j$}{Lg}}

We first consider the case of a single spin-$j$. The Hilbert space $\Hs^{(j)}$ of such a spin naturally defines an irrep of $\SU(2)$, with highest weight $\mu = j\sigma_z\cong j$ and Dynkin coefficient $(2j)$, via the linear map $\pi^{(j)}:\SU(2)\to \HSs(\Hs^{(j)}):$
\begin{equation}
    U=e^{-i\theta \hat{n}\cdot \vec{\sigma}/2}\mapsto \pi^{(j)}(U) = e^{-i\theta \hat{n}\cdot \vec{J}},
\end{equation}
where $\vec{J}=\qty(J_x,\,J_y,\,J_z)^T$ is the vector of angular momentum operators for a spin-$j$, with each operator in
\begin{equation}
    \HSs(\Hs^{(j)}) = \Hs^{(j)}\otimes \overline{\Hs^{(j)}}.
\end{equation}
In $\SU(2)$, the dual of an irrep is isomorphic to the original irrep, and a tensor product of two irreps can be decomposed into irreps using the rules of addition of angular momenta, \ie,
\begin{equation}
    \BHs^{(L_1)}\otimes \BHs^{(L_2)} \cong \BHs^{(L_{\mathrm{min}})}\oplus  \BHs^{(L_{\mathrm{min}}+1)} \oplus\dots\oplus \BHs^{(L_{\mathrm{max}})}
\end{equation}
with $L_{\mathrm{min}}=\abs{L_2-L_1}$ and $L_{\mathrm{max}}=L_1+L_2$, and we find
\begin{equation}
    \HSs(\Hs^{(j)}) \cong \bigoplus_{L=0}^{2j}\BHs^{(L)}
\end{equation}
where only integer values of $L$ appear in the decomposition. Note that the spin-$0$ irrep $\BHs^{(0)}$ is the trivial irrep spanned by the identity operator. \textit{Universal decoupling}\footnote{A sequence is said to achieve universal decoupling of a quantum system if it decouples an arbitrary operator acting on that system.} of a large spin $j$ would therefore require finding inaccessible symmetries in the irreps of highest weights $\mu\cong k$, $\forall k\in\{1,2,\dots,2j\}$. These irreps are also irreps of $\SO(3)$, so that one can search for symmetry groups associated with rotational symmetry groups in $\mathbb{R}^3$ (\textit{e.g.}, $n$-dihedral, tetrahedral, octahedral, and icosahedral point groups). The Weyl character formula~\eqref{Eq.Weyl.character} reduces in this case to a simple formula that depends only on the rotation angle $\theta$ of each element of the group, 
\begin{equation}
    \chi_{(L)}(\theta) = \frac{\sin\qty((2L+1)\theta/2)}{\sin(\theta/2)} .
\end{equation}
Having established this, we can compute the multiplicity $a_1$ of the trivial irrep $\rho_1$ for each value of $L$ for a given subgroup $\group$ of $\SU(2)$.
\begin{figure}[h]
    \centering
    \includegraphics[width=0.8\linewidth]{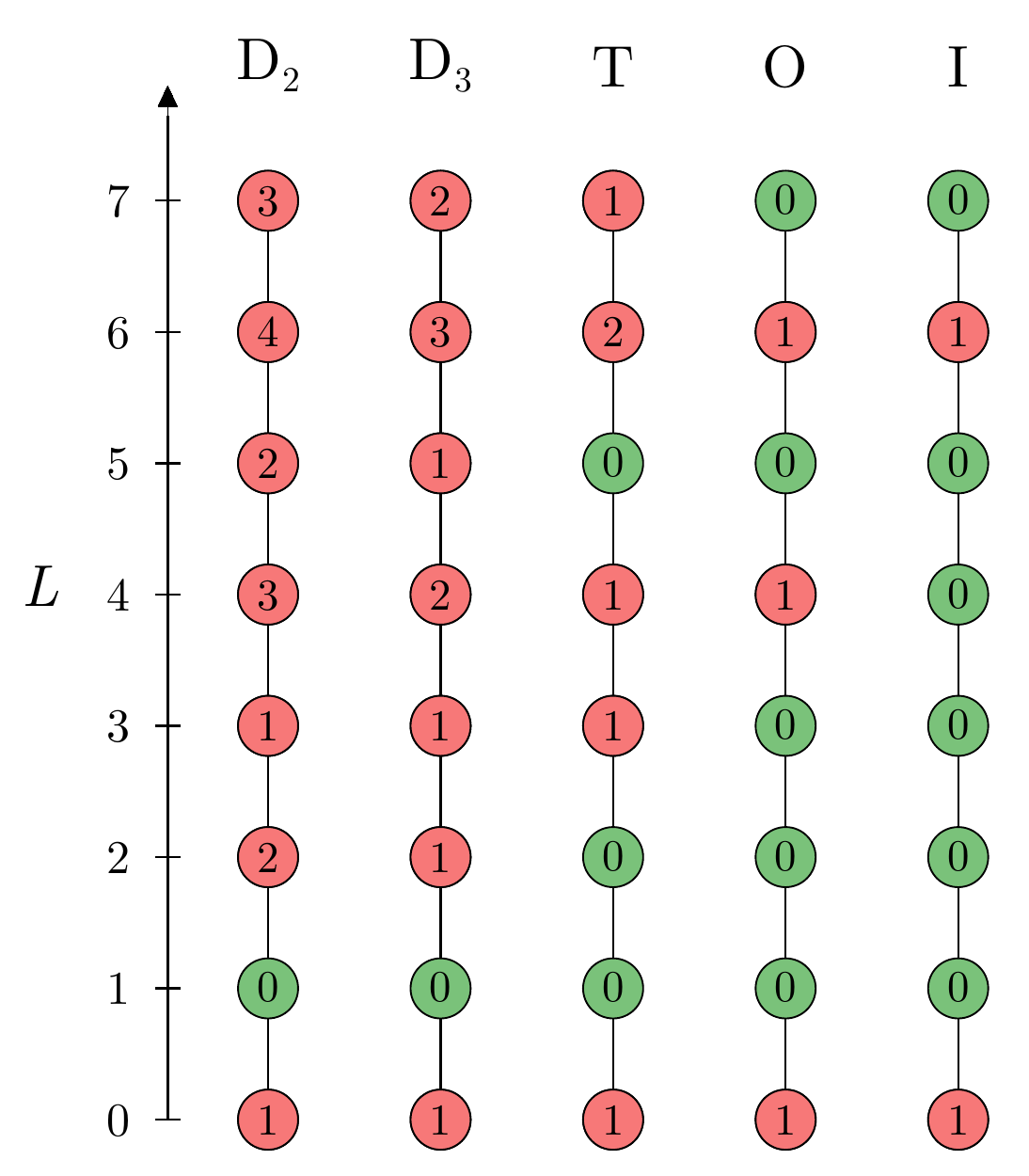}
    \caption{Multiplicity of the trivial spin-$0$ irrep $\BHs^{(0)}$ in the decomposition of $\BHs^{(L)}$ into irreps of the point groups $\group\in\{\mathrm{D}_2,\mathrm{D}_3,\mathrm{T},\mathrm{O},\mathrm{I}\}$ which are subgroups of $\SO(3)$. Entries shown in green correspond to zero multiplicity, meaning that the associated symmetry is inaccessible, while entries shown in red correspond to nonzero multiplicity, with the corresponding value indicated explicitly.}
    \label{fig:su2}
\end{figure}
\par The resulting multiplicities are shown in Fig.~\ref{fig:su2} for several point groups $\group$. The spin-$L$ irreps are labeled by their value of $L$ along a single axis. Each circle represents an $\SU(2)$-irrep and is colored green when the corresponding symmetry is inaccessible, and red when it is accessible. The number displayed in the circle indicates the multiplicity of the trivial irrep in the decomposition \eqref{eq:decompirrepG}. These results show that universal decoupling can be achieved using polyhedral point groups for spin $j<3$. For $j=3$ and beyond, however, even the icosahedral group—being the largest exceptional subgroup of $\SO(3)$—becomes an accessible symmetry.

\par 
If the interaction subspace is not the whole space of operators $\HSs (\Hs^{(j)})$, it is possible to reduce the decoupling group. For example, if the errors are linear in the angular momentum operator ---\textit{e.g.}, dissipation ($J_-$) and dephasing ($J_z$), or more generally depolarization---, one can construct a vector space $V \cong \BHs^{(1)}$ that includes all these operators, and we see that $\mathrm{D}_2$ is a decoupling group regardless of the quantum number $j$. In the presence of higher-rank error operators, such as with optical pumping ($J_+J_-$) or squeezing ($J_z^2$), the relevant representation space becomes $V \cong \BHs^{(0)}\oplus \BHs^{(1)}\oplus \BHs^{(2)}$. In this case, the tetrahedral point group $\mathrm{T}$ must be used to achieve decoupling. The addition of rank-$3$ (resp.\ rank-$4$ and $5$) errors would require the use of the octahedral (resp.\ icosahedral) point groups. The results mentioned in this and the previous paragraph were also obtained in Sec.~V of Ref.~\cite{Read_2025}. We have recovered all these results here.

\subsection{Ensemble of interacting spins}

The second case of particular interest is an ensemble of $N$ interacting spin-$s$, to which we want to apply only global rotations (global $\SU(2)$ pulses on the ensemble). The corresponding representation of $\SU(2)$ on this tensor product Hilbert space $\Hs_S = \bigotimes_{i=1}^N \Hs^{(s)}_i$ is defined by 
\begin{equation}
    \pi:\,U=e^{-i\theta \hat{n}\cdot \vec{\sigma}/2}\mapsto \pi(U) = \bigotimes_{i=1}^Ne^{-i\theta \hat{n}\cdot \vec{S}^{(i)}}
    \label{Eq.ten.prod}
\end{equation}
where $\vec{S}^{(i)}$ is the spin-$s$ operator for the $i$th spin of the ensemble. The decomposition of the space of operators is, in this case, more involved, and we obtain 
\begin{equation}
    \HSs(\Hs_S) = \bigoplus_{L=0}^{2Ns}\bigoplus_{\alpha}\BHs^{(L)}_{\alpha}
\end{equation}
where, once again, only irreps of integer spin appear, and the different irreps can appear with multiplicities greater than one, hence the need for the additional index $\alpha$. Furthermore, with the exception of the $\BHs^{(0)}$-irrep spanned by the identity operator, the trivial irreps appearing in the decomposition contain non-trivial operators that are invariant under $\SU(2)$---\textit{e.g.}, exchange interaction ($\vec{S}^{(i)}\cdot\vec{S}^{(j)}$) or the rank-3 invariant $\vec{S}^{(i)}\cdot(\vec{S}^{(j)}\times \vec{S}^{(k)})$. It may seem more complicated to construct the subspace $V$ for a given interaction subspace $\mathcal{I}_S$, but we can nevertheless determine quite easily which highest weights will appear in $V$ depending on the types of interaction. 
\par For example, on-site disorder linear in the spin operators (terms proportional to $\hat{n}_i\cdot \vec{S}^{(i)}$ for some $\hat{n}_i$), or dephasing and dissipation, will be included in a vector space $V$ containing only irreps of the type $\BHs^{(1)}$. On the other hand, a disorder quadratic in the spin operator, \textit{e.g.}, due to crystal strain inhomogeneities in NV-centers ($S_z^2$), will be included in $V$ if we add certain $\BHs^{(2)}$ irreps and the trivial irrep $\BHs^{(0)}$ which contains the span of the global identity.
\par Various spin-spin interactions can be included this way. For example, dipole-dipole interactions ($\propto 3\hat{n}\cdot\vec{S}^{(i)}\hat{n}\cdot \vec{S}^{(j)} - \vec{S}^{(i)}\cdot\vec{S}^{(j)}$) will involve purely $\BHs^{(2)}$ irreps, or, in general, any two-body interaction $I_{ij}$ between the $i$th and $j$th spins satisfying $\mathrm{Tr}\qty[I_{ij}(\vec{S}^{(i)}\cdot\vec{S}^{(j)})]=0$---anisotropic interactions---will be included in $\BHs^{(1)}$ and $\BHs^{(2)}$ irreps. 
\par Once the types of interactions have been listed and the corresponding irreps identified, we refer to Fig.~\ref{fig:su2} to find an inaccessible symmetry. In conclusion, we recover the results presented in Sec.~VI of Ref.~\cite{Read_2025}.
\subsection{Exploiting representation freedom to enhance dynamical decoupling}
\label{SubSection.diff.irrep}
In this subsection, we show that our framework introduces an additional degree of freedom that can be used to improve dynamical decoupling: the choice of the $\SU(2)$ representation itself. This freedom is not merely a different mathematical description of the same control problem. Instead, changing the representation alters the decomposition of the interaction space into $\SU(2)$ irreps and, consequently, changes the interaction terms that can be associated with inaccessible symmetries. Consequently, by selecting a representation suited to the Hamiltonian to be decoupled, one can obtain new DD sequences and, in some cases, decouple interactions that would not be removable with the standard choice of representation.

To illustrate this idea, we consider an ensemble of two interacting spin-$s$ particles. In the previous subsection, we used the canonical tensor-product representation of $\SU(2)$ defined in Eq.~\eqref{Eq.ten.prod}, in which the same transformation is applied to both subsystems, that is
\begin{equation}
    \pi(U) = e^{-i \theta  \hat{n}\cdot \vec{S}^{(1)}} e^{-i \theta  \hat{n}\cdot \vec{S}^{(2)}} .
\end{equation}
We now show that adopting a different representation for one of the subsystems leads to a different organization of the same interaction space into irreducible sectors. This additional flexibility can be exploited to enlarge the set of accessible decoupling strategies.

Let us restrict our attention to interaction Hamiltonians that are linear combinations of bilinear spin operators, which belong to the space $\mathcal{V} = \mathrm{span}(\{ S_{\alpha}^{(1)} S_{\beta}^{(2)} \}_{\alpha,\beta}) \subset \HSs(\Hs_S)$. With respect to the canonical tensor-product representation, this space decomposes into irreducible subspaces as $ \mathcal{V} \cong \BHs^{(0)}\oplus \BHs^{(1)}\oplus \BHs^{(2)}$ where
\begin{equation}
\begin{aligned}
    & \BHs^{(0)} =  
    \mathrm{span} \Big( \Big\{
\mathbf{S}^{(1)} \cdot \mathbf{S}^{(2)}
\Big\} \Big) ,
    \\
&    \BHs^{(1)} = 
\mathrm{span} \Big( \Big\{
\sum_{\beta,\gamma} \epsilon_{\alpha \beta \gamma} S_{\beta}^{(1)} S_{\gamma}^{(2)}
\Big\}_{\alpha=x,y,z} \Big) ,
\\
&     \BHs^{(2)} = \mathrm{span} \Big( \Big\{
S_+^{(1)} S_+^{(2)} , S_+^{(1)} S_z^{(2)} + S_z^{(1)} S_+^{(2)} , 
\\
& \qquad \qquad \qquad \;\; \mathbf{S}^{(1)}\cdot \mathbf{S}^{(2)} - 3 S_z^{(1)} S_z^{(2)} ,
\\
& \qquad \qquad \qquad \;\; S_-^{(1)} S_z^{(2)} + S_z^{(1)} S_-^{(2)}
, S_-^{(1)} S_-^{(2)} 
\Big\} \Big) .
\end{aligned}
\end{equation}
As established in the previous subsections, operators in $\BHs^{(1)}$ and $\BHs^{(2)}$ will be decoupled by an inaccessible symmetry of the $1$-irrep and $2$-irrep, respectively, for example, $\mathrm{D}_2$ and $\mathrm{T}$. In contrast, the isotropic component $\BHs^{(0)}$ cannot be removed by any DD sequence built solely from rotations.

Let us now consider the same system and the same interaction, but choose a different $\SU(2)$ representation, namely
\begin{equation}
    \widetilde{\pi}(U) = e^{-i \theta  \hat{n}\cdot \vec{S}^{(1)}} \left( e^{-i \theta  \hat{n}\cdot \vec{S}^{(2)}} \right)^* ,
    \label{Eq.Wide.Pi}
\end{equation}
that is, the second subsystem transforms according to the dual representation. In practical terms, the representation is such that whenever a rotation by an angle $\theta$ about $\hat{n}=(n_x,n_y,n_z)$ is performed on the first subsystem, a rotation by the same angle but about the reflected axis $(-n_x,n_y,-n_z)$ is performed on the second subsystem. The space $\mathcal{V}$ still decomposes as $\mathcal{V} \cong \widetilde{\BHs}^{(0)}\oplus \widetilde{\BHs}^{(1)}\oplus \widetilde{\BHs}^{(2)} $ but the basis of each irrep is now different. In particular, the new decomposition is obtained from the previous one by replacing $S_{\pm}^{(2)} \rightarrow - S_{\mp}^{(2)}$ and $S_z^{(2)} \rightarrow -S_z^{(2)}$ in the expression of the operators
, or equivalently, $S_{x/z}^{(2)} \rightarrow -S_{x/z}^{(2)}$ and $S_{y}^{(2)}$ remains invariant. Thus,
\begin{equation}
    \begin{aligned}
          & \widetilde{\BHs}^{(0)} = 
\mathrm{span} \left( \left\{ S_{z}^{(1)} S_{z}^{(2)} + \frac{1}{2} \left( 
    S_{+}^{(1)} S_{+}^{(2)} +     S_{-}^{(1)} S_{-}^{(2)}
    \right) \right\} \right) ,
    \\
&    \widetilde{\BHs}^{(1)} = 
\mathrm{span} \Big( \Big\{
S_z^{(1)}S_-^{(2)} - S_+^{(1)}S_z^{(2)} 
, 
S_z^{(1)}S_+^{(2)} - S_-^{(1)}S_z^{(2)}, 
\\
& \qquad \qquad \qquad \quad
S_-^{(1)}S_-^{(2)} - S_+^{(1)}S_+^{(2)} 
\Big\} \Big) ,
\\
&    \widetilde{\BHs}^{(2)} = 
\mathrm{span} \Big( \Big\{
S_+^{(1)}S_-^{(2)} , S_+^{(1)} S_z^{(2)} + S_z^{(1)} S_-^{(2)} , 
\\
& \qquad \qquad \qquad \;\;  S_+^{(1)}S_+^{(2)} + S_-^{(1)} S_-^{(2)} - 4 S_z^{(1)} S_z^{(2)} ,
\\
& \qquad \qquad \qquad \;\;  S_-^{(1)} S_z^{(2)} + S_z^{(1)} S_+^{(2)} , S_-^{(1)} S_+^{(2)}
\Big\} \Big) .
    \end{aligned}
\end{equation} 
This simple example already shows the advantage of the additional freedom provided by our framework: by changing the representation, the same physical Hamiltonian may be assigned to a different irrep sector and therefore becomes compatible with different inaccessible symmetries and different DD protocols. For example, consider the interaction Hamiltonian $H_S= S_+^{(1)}S_+^{(2)}+S_-^{(1)}S_-^{(2)}-4 S_z^{(1)}S_z^{(2)}$. In the canonical representation, this Hamiltonian has components in $V = \BHs^{(0)} \oplus \BHs^{(2)}$, so it contains an isotropic contribution and therefore cannot be fully decoupled using $\pi(U)$ pulses alone. By contrast, in the $\widetilde{\pi}$ representation, one has $V = \widetilde{\BHs}^{(2)}$, so the same Hamiltonian lies entirely in a single $2$-irrep sector and can therefore be decoupled using, for instance, the inaccessible symmetry $\mathrm{T}$. As a second example, consider $H_S = i(S_-^{(1)}S_-^{(2)} - S_+^{(1)}S_+^{(2)}) $. In the canonical representation, this Hamiltonian belongs to $\BHs^{(2)}$, and is therefore decoupled by a TEDD sequence. However, in the representation $\widetilde{\pi}$ it belongs to $\widetilde{\BHs}^{(1)}$, so one may instead use an Eulerian DD sequence based on $\mathrm{D}_2$, implemented with $\SU(2)$ pulses through $\widetilde{\pi}$ as in~\eqref{Eq.Wide.Pi}. 

\par
\begin{figure}[t]
    \centering
    \includegraphics[width=0.95\linewidth]{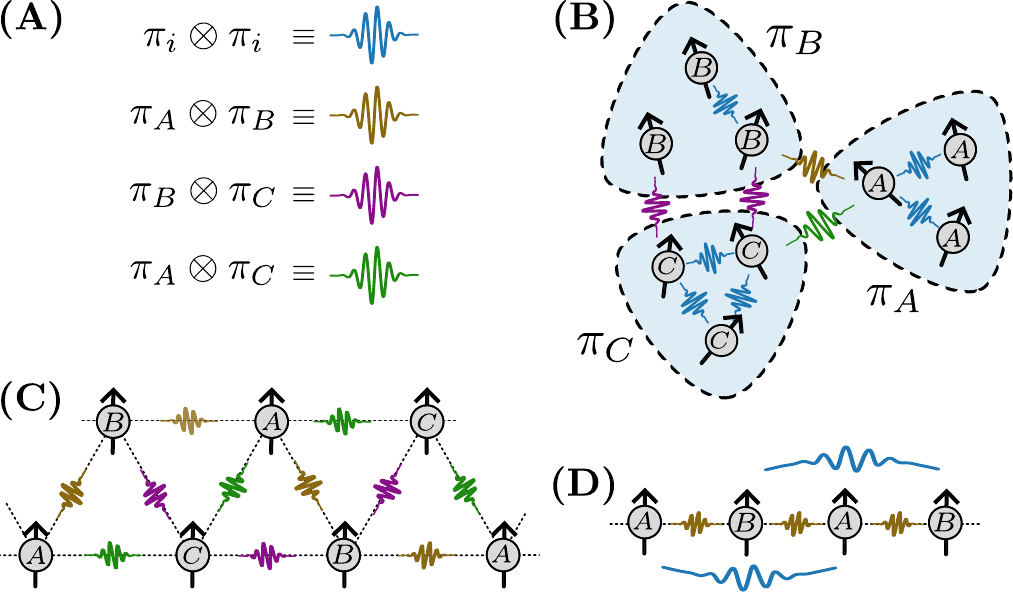}
    \caption{Sketch illustrating the role of local group representations in engineering interactions. A) For each pair of subsystems, the tensor product of the corresponding representations has its own isotropic interaction sector. B) Each spin is assigned a local representation $\pi_{\alpha}$, with $\alpha=A,B,\dots$. A DD sequence will decouple all interaction terms except the isotropic components. For a Hamiltonian with nearest-neighbor interactions, the geometry and the choice of representation for each spin determine the engineered interaction (see the examples C and D).}
    \label{fig:Dif.rep}
\end{figure}

 This simple example shows that the irrep decomposition of the interaction space is not determined solely by the Hamiltonian, but also by the chosen $\SU(2)$ representation. As a result, the inaccessible symmetries, and therefore the available DD sequences, may vary from one representation to another. More generally, consider an ensemble of $N$ spin-$s$ particles partitioned into several subsystems, and define the global representation as the tensor product $\tilde{\pi}\equiv \bigotimes_{i=1}^N\pi_{\alpha_i}^{(i)}$, $\alpha_i\in\{A,B,C,\dots\}$, where different local representations may be assigned to different spins, namely:
\begin{equation}
\begin{aligned}
    \pi_A(U)={}& e^{-i\theta\hat{n}\cdot \vec{S}},\\ 
    \pi_B(U)={}&e^{-i\theta\hat{n}\cdot (R_{\hat{n}_B}(\theta_B)\vec{S})},\\
    \vdots{}&\label{gen.Rep}
\end{aligned}
\end{equation}
where $R_{\hat n_B}(\theta_B)\in \SO(3)$ denotes the rotation matrix corresponding to a rotation by angle $\theta_B$ about the axis $\hat n_B$.
In this setting, a generic two-body Hamiltonian will be symmetrized differently depending on the representations assigned to the two subsystems involved (see Fig.~\ref{fig:Dif.rep}A,B for an illustration). This makes the choice of representation a practical tool for engineering interaction Hamiltonians in spin systems. For instance, in a system with a fixed spatial geometry and only nearest-neighbor couplings, an appropriate assignment of representations can be used to design the effective interactions between different spins; see, e.g., Fig.~\ref{fig:Dif.rep}C. 
A particularly interesting case is that of a spin chain, as shown in Fig.~\ref{fig:Dif.rep}D, with nearest-neighbor exchange interactions of the form $\vec{S}^{(1)}\cdot \vec{S}^{(2)}$. Choosing the local representations $\pi_A$ and $\pi_B$ in Eq.~\eqref{gen.Rep} with $\theta_B=2\pi/3$ makes this interaction anisotropic, so that it can be decoupled by a $\mathrm{TEDD}$ sequence. Moreover, if one further chooses $\hat{z}\cdot \hat{n}_B=\frac{1}{\sqrt{3}}$ in Eq.~\eqref{gen.Rep}, then the long-range secular dipole-dipole interaction is also anisotropic under the mixed representation $\pi_{A}\otimes \pi_B$. Since it is likewise anisotropic under $\pi_{i}\otimes \pi_i$, $i\in\{A,B\}$, the corresponding $\mathrm{TEDD}$ sequence cancels dipolar interactions between any pair of spins in the chain.

\section{Application: \texorpdfstring{$\SU(3)$}{Lg} decoupling of interacting qutrits}\label{ex.SU3}

\par We now consider an ensemble of interacting qutrits that we aim to decouple using only global $\SU(3)$ pulses. Examples of such sequences include the $9$-pulse sequence constructed on the Heisenberg-Weyl group~\cite{Tripathi_2024}, which cancels any arbitrary disorder, and the $3X_3$ sequence~\cite{Tripathi_2024}, which decouples disorder under the rotating wave approximation. These sequences generalize the $XY4$ and $\mathrm{CPMG}$ sequences, respectively. In contrast, there are few sequences that decouple qutrit-qutrit interactions. A generalized WAHUHA sequence has been constructed to suppress the dipole-dipole interaction in the rotating wave approximation~\cite{Lukin_2017}, and a sequence was presented in Ref.~\cite{Lukin_2024} that also cancels disorder in the same approximation, while providing some robustness to certain finite-duration and control errors.

\par Developing new decoupling sequences for such systems would be of great interest for quantum sensing or quantum simulation applications using NV-centers~\cite{Taylor_2008,Zhou_2020_metrology,Zhou_2023,Lukin_2024} and hexagonal Boron Nitride~\cite{Gottscholl_2021, Tarkanyi_2025, Sasaki_2023,Souvik_2025,Mahajan_2025}, where the large zero-field splitting of the spin-$1$ electronic ground state breaks the $\SU(2)$ symmetry of the spin system. $\SU(3)$ pulses can be easily implemented in these systems by applying a MW field in resonance with the accessible ESR transitions. Similarly, due to their large quadrupole moments, nuclear spin-$1$ isotopes ($^2$H, $^{14}$N and $^6$Li) may be treated as natural qutrits where $\SU(3)$ pulses can be implemented using a RF field in resonance with the accessible NMR transitions. As such, new decoupling sequences for these systems could prove useful in the context of quadrupolar NMR spectroscopy~\cite{Jerschow_2005,Ashbrook_2014}.

\subsection{Constructing \texorpdfstring{$V$}{Lg}}
\label{Subsec.V.SU3}

We now consider an ensemble of interacting qutrits ($N$ identical $3$-level quantum systems). The action of $\SU(3)$ on the system Hilbert space $\Hs_S$ is defined by the tensor-product representation $(\pi,\Hs_S)$ with 
\begin{equation}
    \pi = \bigotimes_{i=1}^N \pi^{(\mu_{\mathrm{std}})}_i, \quad \Hs_S = \bigotimes_{i=1}^N \BHs^{(\mu_{\mathrm{std}})}_{i}
\end{equation}
where $(\pi^{(\mu_{\mathrm{std})}},\BHs^{(\mu_{\mathrm{std}})})$ denotes the standard representation of $\SU(3)$, that is, the three-dimensional representation describing the action of $\SU(3)$ on the Hilbert space of a single qutrit. 

For $\SU(3)$, the Cartan algebra is two-dimensional; accordingly, each highest weight is represented by a two-component vector in a given basis. In particular, there exist two distinguished highest weights, $\mu_1$ and $\mu_2$, called the \emph{fundamental weights}, such that all other highest weights can be written uniquely as a non-negative integer linear combination of them:
\begin{equation}
    \mu_1 = \begin{pmatrix}
        \frac{2}{3} & 0 & 0\\ 
        0 & -\frac{1}{3} & 0 \\ 
        0 & 0 & -\frac{1}{3}
    \end{pmatrix},\quad \mu_2 = \begin{pmatrix}
        \frac{1}{3}& 0 & 0\\ 
        0 & \frac{1}{3} & 0 \\ 
        0 & 0 & -\frac{2}{3}
    \end{pmatrix}.
\end{equation}
Thus, any irrep of $\SU(3)$ with highest weight $\mu$ may be labeled by its Dynkin coefficients $\vec{d} = (d_1,d_2)$, defined by the expansion of its highest weight as $\mu = d_1\mu_1+d_2\mu_2$. Accordingly, we now use the notation $(\pi^{\mu},\BHs^{(\mu)})\equiv(d_1,d_2)$ to distinguish the different irreps. 

\par
\begin{figure*}[t]
    \centering
    \includegraphics[width=0.9\linewidth]{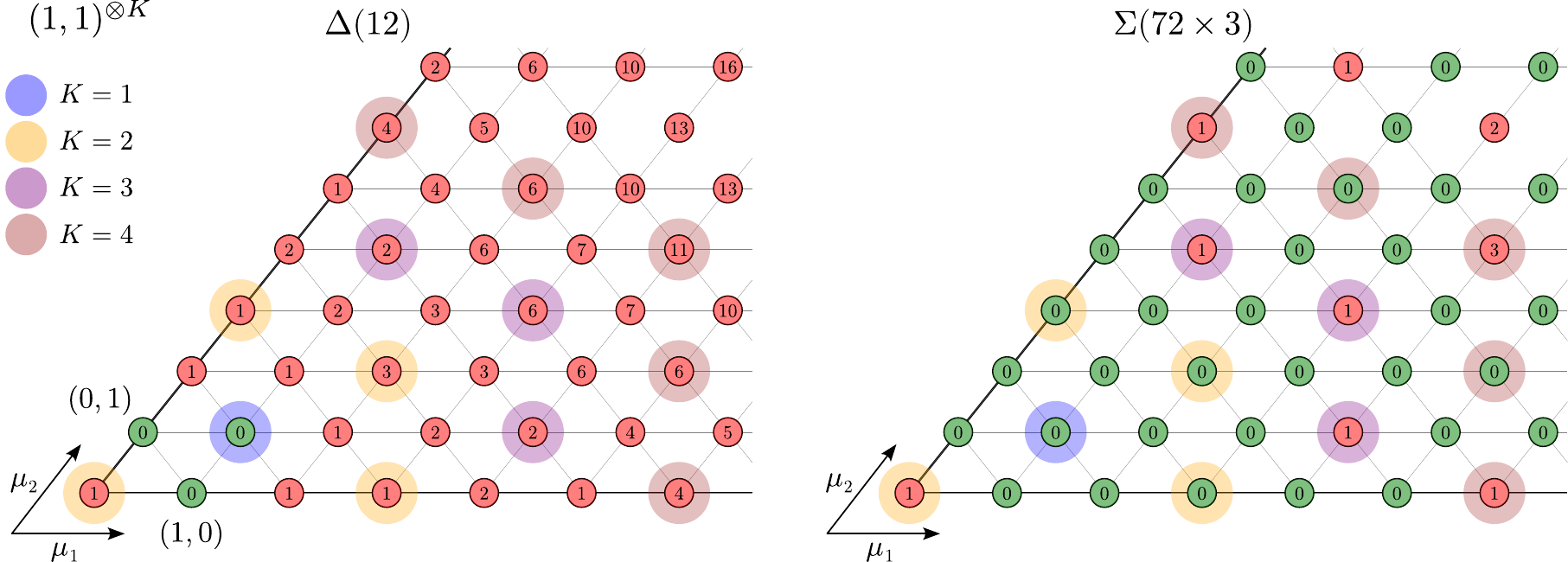}
    \caption{Fundamental Weyl chamber of $\SU(3)$, including its boundary. Each circle represents an irrep of $\SU(3)$, labeled by its Dynkin coefficients, with the displayed number giving the multiplicity of the trivial irrep. Green circles correspond to inaccessible symmetries, whereas red circles correspond to accessible symmetries. The irreps appearing in the decomposition of $(1,1)^{\otimes K}$ are surrounded by a colored halo, whose color indicates the smallest value of $K$ for which the irrep appears.}
    \label{fig:su3}
\end{figure*}

 The standard representation of $\SU(3)$ has Dynkin coefficients $(1,0)$, while its dual has Dynkin coefficients $(0,1)$ (they are precisely the two fundamental representations). More generally, the dual of the irrep with Dynkin coefficients $(p,q)$ is the irrep with Dynkin coefficients $(q,p)$, $(p,q)\cong \overline{(q,p)}$. In principle, the decomposition of $\HSs(\Hs_S)$ in irreps is highly nontrivial, since it generally involves many irreps with different Dynkin coefficients. However, as in the $\SU(2)$ case, it is sufficient to isolate only those irreps that are relevant for the dominant unwanted interactions. To this end, we first consider the operator space of a single qutrit, that is, $\HSs(\Hs_S)$ for $N=1$. Its decomposition into $\SU(3)$ irreps reads 
\begin{equation}
    (1,0)\otimes \overline{(1,0)} \cong (1,0)\otimes (0,1) \cong (0,0)\oplus (1,1)
\end{equation}
where $(0,0)$ is the trivial irrep (of dimension $1$) containing the identity operator and $(1,1)$ is the adjoint representation (of dimension $3^2-1=8$), containing the traceless operators acting on a single qutrit---which can be written as linear combinations of the Gell-Mann matrices. If the dominant errors are single-qutrit errors, such as dissipation, dephasing, and disorder, a vector space $V$ can be constructed that contains only $(1,1)$ irreps. For general $K$-body interactions, the relevant irreps can be identified from the decomposition of the tensor product $(1,1)^{\otimes K}$, which can be done, for example, using Young tableaux. For $K=2,3,$ and $4$, one obtains
\begin{equation}
\begin{aligned}
(1,1)^{\otimes 2} &= (0,0)\oplus (1,1)^{\oplus 2}\oplus (0,3)\oplus (3,0)\oplus (2,2),\\
(1,1)^{\otimes 3} &\to \{(0,0),(1,1),(0,3),(3,0),(2,2),(4,1),\\
                  &\qquad(1,4),(3,3)\},\\
(1,1)^{\otimes 4} &\to \{(0,0),(1,1),(0,3),(3,0),(2,2),(4,1),\\
                  &\qquad\{(1,4),(3,3),(6,0),(0,6),(5,2),(2,5),\\
                  &\qquad(4,4)\}.
\end{aligned}
\end{equation}
Here, the arrow $\to$ indicates that the full decomposition contains the irreps listed on the right-hand side, possibly with multiplicity greater than one. These irreducible representations can then be represented as points on the two-dimensional lattice of Dynkin labels shown in Fig.~\ref{fig:su3}, in contrast with the one-dimensional labeling used for $\SU(2)$ in Fig.~\ref{fig:su2}.

\subsection{Search for inaccessible symmetries}

We consider the finite groups of $\SU(3)$ presented in Ref.~\cite{Grimus_2010}, which we list in Tab.~\ref{tab:su3subgroups} for convenience. The generators are written using the same notations as in Ref.~\cite{Grimus_2010}, defining the constants $\xi_n = e^{2\pi i/n}$, $\nu_1 = \frac{-1+\sqrt{5}}{2}$, $\nu_2 = \frac{-1-\sqrt{5}}{2}$ and the generating matrices
\begin{widetext}
\begin{equation}
\begin{aligned}
    &A(n) = \begin{pmatrix}
        1 & 0 & 0\\ 
        0 & \xi_{n} & 0 \\ 
        0 & 0 & \xi_{n}^{-1} 
    \end{pmatrix} , \quad 
    A = A(2) , \quad
    B = \begin{pmatrix}
        0 & 0 & -1\\ 
        0 & -1 & 0 \\ 
        -1 & 0 & 0
    \end{pmatrix} ,\quad
    C = A(3) , \quad
    D = \begin{pmatrix}
         \xi_{9}^2 & 0 & 0\\ 
        0 & \xi_{9}^2 & 0 \\ 
        0 & 0 & \xi_{9}^2\xi_{3}
    \end{pmatrix} , \quad
    \\
    & E = \begin{pmatrix}
        0 & 1 & 0\\ 
        0 & 0 & 1 \\ 
        1 & 0 & 0
    \end{pmatrix} , \quad  F = \begin{pmatrix}
        -1 & 0 & 0 \\ 
        0 & 0 & -\xi_3 \\ 
        0 & -\xi_3^2 & 0
    \end{pmatrix} , \quad V = \frac{1}{\sqrt{3}i}\begin{pmatrix}
        1 & 1 & 1\\ 
        1 & \xi_3 & \xi_3^2\\ 
        1 & \xi_3^2 & \xi_3
    \end{pmatrix} , \quad
    W = \frac{1}{2}\begin{pmatrix}
        -1 & \nu_2 & \nu_1 \\ 
        \nu_2 & \nu_1 & -1 \\ 
        \nu_1 & -1 & \nu_2
    \end{pmatrix} ,
   \\
    &  X = \frac{1}{\sqrt{3}i}\begin{pmatrix}
        1 & 1 & \xi_3^2 \\ 
        1 & \xi_3 & \xi_3 \\ 
        \xi_3 & 1 & \xi_3 
    \end{pmatrix} , \quad Y = \begin{pmatrix}
        \xi_7 & 0 & 0 \\ 
        0 & \xi_7^2 & 0 \\ 
        0 & 0 & \xi_7^4
    \end{pmatrix} , \quad Z = \frac{i}{\sqrt{7}} \begin{pmatrix}
        \xi_7^4 - \xi_7^3 & \xi_7^2 - \xi_7^5 & \xi_7 - \xi_7^6 \\ 
        \xi_7^2 - \xi_7^5 & \xi_7 - \xi_7^6 & \xi_7^4 - \xi_7^3 \\ 
        \xi_7 - \xi_7^6 & \xi_7^4 - \xi_7^3 & \xi_7^2 - \xi_7^5
    \end{pmatrix} .
\end{aligned}\label{mats.}
\end{equation}
\end{widetext}
These finite subgroups comprise two infinite dihedral-like families, denoted $\Delta(3n^2)$ and $\Delta(6n^2)$, together with six exceptional groups, denoted $\Sigma(n)$ or $\Sigma(n\times 3)$. Further information on their structure, such as generators, character tables, and chains of subgroups, can be found in Refs.~\cite{Luhn_2007,Luhn_2009} for the dihedral-like subgroups and in Refs.~\cite{Luhn_2007Sigma,Grimus_2010} for the exceptional groups. 

The special notation $\Sigma(n\times 3)$ indicates that the center of $\SU(3)$, here denoted $\mathbb{Z}_3$, is a subgroup of $\Sigma(n\times 3)$. However, in a general representation of $\SU(3)$, the set of representatives of the quotient $\Sigma(n\times 3)/\mathbb{Z}_3$ does not necessarily form a group (see the discussion below). By contrast, the notation $\Sigma(n)$ indicates that the corresponding group does not include $\mathbb{Z}_3$ as a subgroup.

\begin{table}[t]
    \renewcommand{\arraystretch}{1.7}
    \setlength{\tabcolsep}{10pt}
    \centering
    \begin{tabular}{c|c|c}
       Group  & Order & Generating set(s)  \\ \hline 
        $\Delta(3n^2)$ & $3n^2$ & $\qty{A(n),E}$\\ 
        $\Delta(6n^2)$ & $6n^2$ &$\qty{A(n),E,B}$\\
        $\Sigma(60)$ & $60$ & $\qty{E,AW}$, $\qty{EA,W}$\\ 
        $\Sigma(168)$ & $168$ & $\qty{Y,Z}$\\ 
        $\Sigma(36\times 3)$ & $108$ & $\qty{C,V}$, $\qty{E,V}$\\ 
        $\Sigma(72\times 3)$ & $216$ & $\qty{V,X}$\\ 
        $\Sigma(216\times 3)$ & $648$ & $\qty{V,D}$\\ 
        $\Sigma(360\times 3)$ & $1080$ & $\qty{A,E,W,F}$\\ 
    \end{tabular}
    \caption{Selected finite subgroups of $\SU(3)$ considered in this work, together with their order and one or several examples of generating sets. Here, $n$ is a positive integer parametrizing the two infinite families $\Delta(3n^2)$ and $\Delta(6n^2)$, with $n\geq 2$ and $n\geq 1$, respectively. Table adapted from Ref.~\cite{Grimus_2010}. }
    \label{tab:su3subgroups}
\end{table}

\par An analogous situation occurs for $\SU(2)$, where the binary polyhedral groups are the lifts to $\SU(2)$ of the ordinary polyhedral subgroups of $\mathrm{SO}(3)$, obtained by adjoining the center of $\SU(2)$, namely $\pm \mathds{1}_{2\times 2}$. The ordinary polyhedral groups themselves are not subgroups of $\SU(2)$, whereas their binary counterparts are. Nevertheless, in the context of dynamical decoupling, it is sufficient to work with the ordinary polyhedral groups since only integer-spin irreps appear in the relevant decomposition. This follows from the fact that the representation of $\SU(2)$ acting on the space of operators~\eqref{eq.ad} is defined by the tensor product of $\pi$ with its contragredient representation. As a result, every element of the center of $\SU(2)$ will have the same representation, that is $\pi^{\mathrm{op}}(-\mathds{1}_{2\times 2}) = \pi^{\mathrm{op}}(\mathds{1}_{2\times 2})$ because the two minus signs cancel each other. Consequently, the irreps of $\SU(2)$ that appear in the decomposition are in fact also irreps of the quotient $\SU(2)/\qty{\pm \mathds{1}_{2\times 2}}\cong \SO(3)$, and therefore only integer-spin irreducible representations can appear.

\par The same mechanism arises for $\SU(3)$. 
In this case, the center has three elements 
$\mathbb{Z}_3=\qty{\xi_3^{n}\mathds{1}_{3\times 3}\,:\,n=0,1,2}$ 
that differ to the identity up to an overall phase factor. 
For representations of $\SU(3)$ acting on operator spaces, the representation is of the form $\pi\otimes \overline{\pi}$, so these phase factors cancel between the two tensor factors. Consequently, every element of the center is represented by the same element. It follows that, in such representations (and hence in all irreps appearing in the decomposition), the sets of representatives of the quotient groups $\Sigma(n\times 3)/\mathbb{Z}_3$ are indeed finite groups. In particular, if a finite subgroup $\group$ contains the center of $\SU(3)$, then it can be reduced to the quotient group $\group/\mathbb{Z}_3$, which is a finite group in the representations of interest and can then be used to construct a DD sequence.
\par Having identified a selected set of relevant groups, we can calculate the multiplicity $a_1$ of the trivial irrep for each subgroup and for each $\SU(3)$-irrep. Our results are shown for the subgroups $\Delta(12)$ (from the $\Delta(3n^2)$ family) and $\Sigma(72\times 3)$. In each panel, we represent the closure of the open fundamental Weyl chamber, with each circle corresponding to an irrep positioned on the lattice according to its Dynkin coefficients. Green circles indicate that the corresponding symmetry is inaccessible, whereas red circles indicate that it is accessible; the number in each circle gives the multiplicity of the trivial irrep. The irreps relevant to $K$-body interactions are surrounded by a wider colored circle, whose color indicates the smallest value of $K$ for which the irrep appears in the decomposition of $(1,1)^{\otimes K}$. For example, the orange circles mark the irreps that first appear in the decomposition $(1,1)^{\otimes 2}$, namely $(3,0)$, $(0,3)$, $(2,2)$, and $(0,0)$.

\begin{table*}[t]
    \renewcommand{\arraystretch}{1.7}
    \setlength{\tabcolsep}{10pt}
    \centering
    \begin{tabular}{|c|c|c||c|cc|cc|ccc|}
       \hline\multicolumn{2}{|c|}{\multirow{2}{*}{Group}}  & \multirow{2}{*}{Order} & \multicolumn{1}{c|}{$K=1$} & \multicolumn{2}{c|}{$K=2$} & \multicolumn{2}{c|}{$K=3$} & \multicolumn{3}{c|}{$K=4$}\\\cline{4-11}
      \multicolumn{2}{|c|}{} & & (1,1) & (3,0) & (2,2) & (4,1)  & (3,3) & (6,0) & (5,2) & (4,4)\\ \hline \hline 
       \multirow{3}{*}{$\Delta(3n^2)$} & $n=2$ & 12 & \cmark &\xmark&\xmark &\xmark&\xmark &\xmark&\xmark&\xmark\\ 
       & $n=3$ & 27* &\cmark &\xmark&\xmark &\xmark&\xmark &\xmark&\xmark&\xmark\\ 
       & $n=4$ & 48 &\cmark &\xmark&\xmark &\cmark&\xmark &\xmark&\xmark&\xmark\\ \hline 
       \multirow{3}{*}{$\Delta(6n^2)$} & $n=1$ & 6 &\xmark &\xmark&\xmark &\xmark&\xmark &\xmark&\xmark&\xmark\\ 
        & $n=2$ & 24 &\cmark &\cmark&\xmark &\xmark&\xmark &\xmark&\xmark&\xmark\\ 
       & $n=3$ & 54* &\cmark &\cmark&\xmark &\xmark&\xmark &\xmark&\xmark&\xmark\\ \hline 
       \multicolumn{2}{|c|}{$\Sigma(60)$} & 60 & \cmark&\cmark&\xmark &\cmark&\xmark &\xmark&\xmark&\xmark\\\hline 
       \multicolumn{2}{|c|}{$\Sigma(168)$} & 168 &\cmark &\cmark&\cmark &\cmark&\xmark &\xmark&\cmark&\xmark\\\hline 
       \multicolumn{2}{|c|}{$\Sigma(36\times 3)$} & 108* & \cmark&\cmark&\xmark &\xmark&\xmark &\xmark&\xmark&\xmark\\\hline 
       \multicolumn{2}{|c|}{$\Sigma(72\times 3)$} & 216* &\cmark&\cmark&\cmark &\xmark&\xmark &\xmark&\cmark&\xmark\\\hline 
       \multicolumn{2}{|c|}{$\Sigma(216\times 3)$} & 648* &\cmark &\cmark&\cmark &\cmark&\xmark &\cmark&\cmark&\xmark\\\hline 
      \multicolumn{2}{|c|}{$\Sigma(360\times 3)$} & 1080* &\cmark &\cmark&\cmark &\cmark&\cmark &\xmark&\cmark&\xmark\\\hline 
    \end{tabular}

    \caption{Summary of the accessibility results for all exceptional finite subgroups of $\SU(3)$ and for the smallest dihedral-like subgroups. For each irrep, specified by its Dynkin coefficients, and each subgroup, the table indicates whether the corresponding  symmetry is inaccessible (\cmark) or accessible (\xmark). The order of each group is given, where an asterisk indicates that the group contains the center of $\SU(3)$. In that case, for representations acting on operator spaces, one may pass to the quotient by the center, thereby obtaining an equivalent decoupling sequence with one third as many pulses.}
    \label{tab:su3}
\end{table*}

\par We summarize our results in Table~\ref{tab:su3}. In the left-hand side of this table, each subgroup is indicated by its order, and for dihedral series, we specify which series the group belongs to. An asterisk next to the order indicates that the group contains the center of $\SU(3)$ and that it can indeed be divided by three. On the right, each column corresponds to an irrep of $\SU(3)$ that appears in the decomposition of $(1,1)^{\otimes K}$ for several values of $K$. For each group and for each irrep, we assign a green checkmark (\cmark) if the symmetry is inaccessible and a red cross (\xmark) if it is accessible. This table indicates which group can serve as a decoupling group for arbitrary and anisotropic $K$-body interactions within a qutrit ensemble. We review a few interesting observations below. 
\begin{center}
    \textbf{Subgroup $\Delta(12)$}
\end{center}
The smallest subgroup of $\SU(3)$ that is inaccessible to $(1,1)$ is $\Delta(12)$, of order 12. This group may serve as a decoupling group, leading to a 12-pulse sequence when following a Hamiltonian path and a 24-pulse sequence when following an Eulerian path. However, this sequence is not new. Indeed, $\Delta(12)\cong T$ where $T\in\SO(3)$ is the tetrahedral group. In previous work~\cite{Read_2025}, it was proven that $T$ is a universal decoupling group for a single spin-$1$ particle, and thus for a single qutrit. 

\begin{center}
    \textbf{Subgroup $\Delta(27)$}
\end{center}

The subgroup $\Delta(27)$ is also a valid decoupling group for a single qutrit, but it contains 27 elements. However, it contains the center of $\SU(3)$, and a pulse sequence can then be constructed using the representatives of the quotient group $\Delta(27)/\mathbb{Z}_3$. This sequence has only 9 elements, making it the smallest universal decoupling group for a single qutrit, leading to a 9-pulse sequence for the Hamiltonian path and 18-pulse sequence for the Eulerian path. It turns out that this sequence can be identified as the Heisenberg-Weyl sequence presented in Ref.~\cite{Tripathi_2024}, which generalizes the XY4 sequence providing universal decoupling of a single qubit~\cite{Viola_1999}.

    \begin{center}
    \textbf{Subgroups $\Sigma(168)$ and $\Sigma(72\times 3)$}
\end{center}

The smallest subgroup capable of decoupling arbitrary and anisotropic two-qutrit interactions is $\Sigma(168)$, which has 168 elements. On the other hand, $\Sigma(72\times 3)$ achieves similar decoupling properties but contains the center of $\SU(3)$. We can therefore use the quotient group and thus construct a sequence using $\Sigma(72\times 3)/\mathbb{Z}_3$, which has 72 elements. The corresponding pulse sequence is then composed of 72 pulses for the Hamiltonian path and 144 pulses for the Eulerian path. The resulting decoupling sequences can be seen as a generalization of the $\mathrm{TEDD}$ sequence~\cite{Read_2025}, which provides similar decoupling properties for interacting spin systems and is built on the tetrahedral point group.
    
    \begin{center}
    \textbf{Subgroup $\Sigma(360\times 3)$}
\end{center}

Only the group $\Sigma(360\times 3)$ achieves the decoupling of arbitrary and anisotropic three-body interactions between qutrits. Since it contains $\mathbb{Z}_3$, the quotient group can be used. However, the sequence involves at least 360 pulses, making it difficult to implement in any experimental setup. The resulting sequence can be seen as a generalization of the $\mathrm{OEDD}$ sequence, built on the octahedral point group, which decouples three-body interactions in qubit ensembles.

Among the finite subgroups of $\SU(3)$ that we have explored, none of the corresponding DD sequences decouples four- and five-body interactions between qutrits. This is a notable difference from the qubit case, where the icosahedral point group decouples such many-body interactions.

\subsection{Benchmark in interacting qutrit ensembles}

We now illustrate the decoupling performance of the above sequences in an ensemble of interacting qutrits. Consider $N$ qutrits governed by the Hamiltonian 
\begin{equation}
    H = \Delta \sum_{i} \delta_{i}\,\hat{n}_i\cdot \bm{\lambda}^{(i)} + \Gamma\sum_{i<j} \gamma_{ij}\,\bm{\lambda}^{(i)}\cdot \hat{M}_{ij} \bm{\lambda}^{(j)},\label{eq.Ham.SU3}
\end{equation}
where the first term describes random on-site disorder and the second term arbitrary pairwise interactions. Here, $\bm{\lambda}^{(i)}= (\lambda_1^{(i)},\lambda_2^{(i)},\dots,\lambda_8^{(i)})^T$ is the vector of Gell-Mann matrices acting on the $i$th qutrit, $\hat{n}_i$ is a random unit vector in eight dimensions, and $\hat{M}_{ij}$ is a random $8\times 8$ matrix with unit Frobenius norm. The coefficients $\delta_i$ and $\gamma_{ij}$ are sampled independently from the uniform distribution on $[-1/2,1/2]$. We focus on anisotropic interactions ($\Tr[\hat{M}_{ij}]=0$), since the isotropic components cannot be decoupled using only global pulses. Finally, we generate 1000 Hamiltonians of the form~\eqref{eq.Ham.SU3} with $\Delta$ and $\Gamma$ varying over the whole parameter range.
\begin{figure*}[t]
    \centering
    \includegraphics[width=\linewidth]{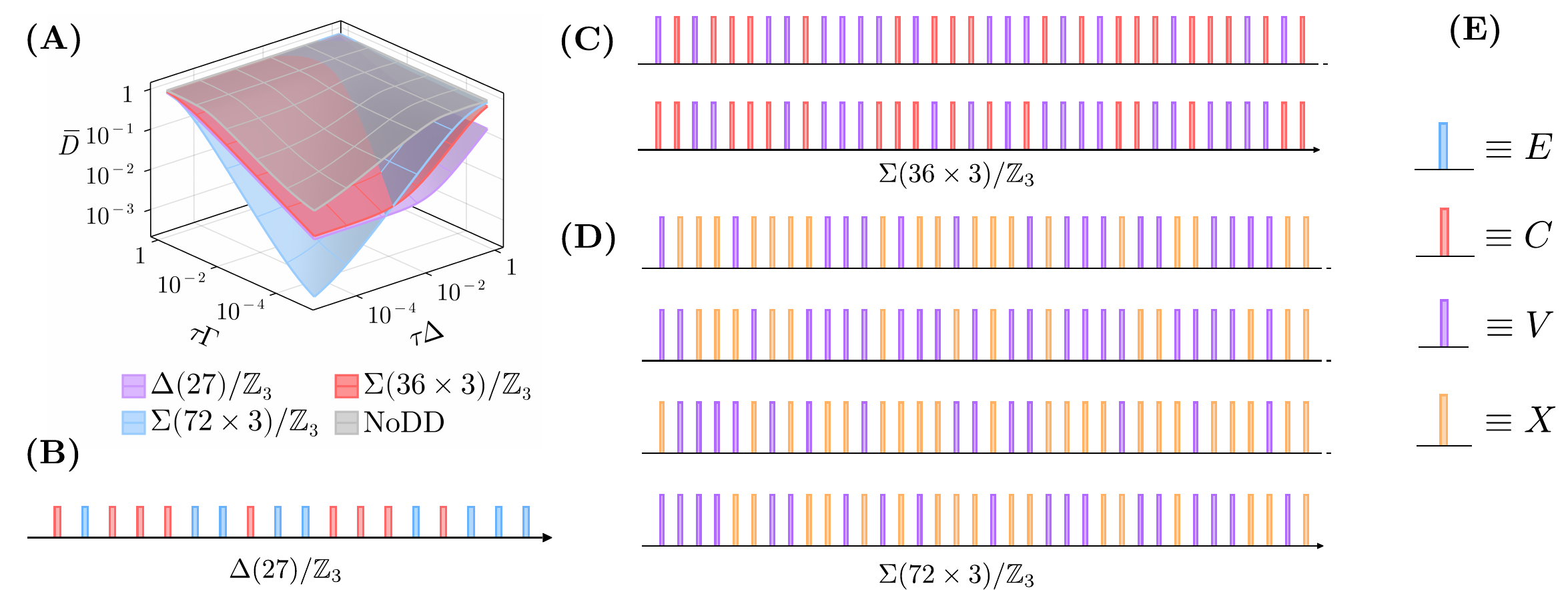}
    \caption{(A) Average distance $\bar{D}$ in the $(\tau \Delta,\tau\Gamma)$ parameter space for an ensemble of three qutrits, averaged over 1000 randomly generated Hamiltonians (see main text). The "NoDD" curve corresponds to a free evolution of the system under the noise Hamiltonian for a duration equal to that of the shortest sequence, $\Delta(27)/\mathbb{Z}_3$. (B,C,D) Eulerian sequences used in the simulations. (E) Pulses used in the sequences, as defined in Eq.~\eqref{mats.}.}
    \label{fig:fig5}
\end{figure*}
\par We generate random Eulerian paths on the Cayley graphs of the quotient groups $\Delta(27)/\mathbb{Z}_3$, $\Sigma(36\times 3)/\mathbb{Z}_3$, and $\Sigma(72\times 3)/\mathbb{Z}_3$, choosing the generators $\qty{C,E}$, $\qty{V,C}$ and $\qty{V,X}$, respectively. Because each group has two generators, the resulting DD sequences consist of $18$, $72$, and $144$ pulses. We consider that each sequence is composed of ideal pulses separated by a free-evolution duration $\tau$. To quantify performance, we calculate the Hilbert-Schmidt distance between the system time-evolution operator $U$ and the identity
\begin{equation}
    D(U, \mathds{1}) = \sqrt{1 - \frac{\abs{\Tr\qty[U]}}{d_S}},\label{eq.dist}
\end{equation}
where $d_S=3^N$ is the dimension of the system’s Hilbert space, over a wide range of parameters $(\tau \Delta, \tau \Gamma)$. The results are plotted in Fig.~\ref{fig:fig5}~A, where the scaling of the distance metric as a function of $\tau \Delta$ and $\tau \Gamma$ reveals first-order decoupling of the disorder term only for $\Delta(27)/\mathbb{Z}_3$ and $\Sigma(36\times 3)/\mathbb{Z}_3$, and of the full Hamiltonian for $\Sigma(72\times 3)/\mathbb{Z}_3$. The Eulerian sequences used in the simulations are shown in Fig.~\ref{fig:fig5}B-D, where each pulse is represented by a colored rectangle (see Fig.~\ref{fig:fig5}E).

\subsection{A case study: interacting spin-$1$ systems with a large zero-field splitting}\label{sec.NV}

We will now examine a more specific case: an ensemble of spins-$1$ (\emph{i.e.}, qutrits) with large zero-field and Zeeman splitting (along the same axis), with dipole-dipole interactions. This example is particularly relevant for quantum sensing platforms based on solid-state defects where the electronic ground state is a spin-$1$, such as ensembles of NV centers or hexagonal Boron Nitride. We will begin with a description of the Hamiltonian, and then show how two methods can be used to reduce the complexity of the pulse sequences found in this section. In the first method, the symmetries in the Hamiltonian are leveraged to design smaller sequences on the subgroups of inaccessible symmetries, as presented in Sec.~\ref{sec.facto}. In the second method, degrees of freedom in the orientation of the decoupling group are leveraged to simplify the form of the pulses in the sequence.

\subsubsection{The Hamiltonian and its decoupling}

Consider the case where the system consists of a set of interacting spin-$1$ particles with dipole-dipole coupling given by 
\begin{equation}
    H_{\mathrm{dd}} = \sum_{i,j}\frac{J_0}{r_{ij}^3}\qty[3\qty(\hat{r}_{ij}\cdot \vec{S}^{(i)})\qty(\hat{r}_{ij}\cdot \vec{S}^{(j)}) - \vec{S}^{(i)}\cdot \vec{S}^{(j)}]
\end{equation}
where $J_0$ is a constant and $\vec{r}_{ij} = r_{ij}\hat{r}_{ij}$ is the vector connecting the position of the $i$-th spin to the $j$-th spin. Additionally, disorder in the system may contribute through the Hamiltonian 
\begin{equation}
    H_{\mathrm{dis}} = \sum_i\delta_i \hat{n}_i\cdot \vec{S}^{(i)},
\end{equation}
where each spin experiences a small field of intensity $\delta_i$ and direction $\hat{n}_i$ due to its local environment. 

\par In the case where the internal Hamiltonian is given by $H_0 = \omega_0 \sum_iS_z^{(i)}$, one traditionally moves to the rotating frame where the energy non-conserving terms of the dipolar interaction are eliminated (RWA or secular approximation). The Hamiltonian then becomes 
\begin{equation}
    H_{\mathrm{dd}}^{\mathrm{sec},\SU(2)} = \sum_{i,j}J_{ij}\qty[3\,S_z^{(i)}S_z^{(j)} - \vec{S}^{(i)}\cdot \vec{S}^{(j)}] , \label{eq.DipSU2}
\end{equation}
with $J_{ij} = J_0(1-3\cos^2(\theta_{ij}))/r_{ij}^3$ and $\cos(\theta_{ij}) = \hat{z}\cdot \hat{r}_{ij}$. This interaction is anisotropic in $\SU(2)$ and can thus be suppressed by a dynamical decoupling sequence that performs a global rotation of the ensemble. However, in cases where strong anharmonicity appears in the spin energy levels, \textit{e.g.}, due to a large zero-field splitting in NV-centers or quadrupolar NMR, the internal Hamiltonian is given by 
\begin{equation}
    H_0 =  \omega_0\sum_i S_z^{(i)} + \chi \sum_i\qty(S_z^{(i)})^2,
\end{equation}
and the dipole-dipole Hamiltonian in the rotating frame differs from~\eqref{eq.DipSU2} and is now given by~\cite{Lukin_2017} 
\begin{multline}
    H_{\mathrm{dd}}^{\mathrm{sec},\SU(3)} = -\sum_{i,j}\frac{J_{ij}}{4}\bigg[\lambda_1^{(i)}\lambda_1^{(j)} +\lambda_2^{(i)}\lambda_2^{(j)} + \lambda_6^{(i)}\lambda_6^{(j)}\\+\lambda_7^{(i)}\lambda_7^{(j)}
    -(\lambda_3^{(i)}+\sqrt{3}\lambda_8^{(i)})(\lambda_3^{(j)}+\sqrt{3}\lambda_8^{(j)})\bigg]\label{eq.dipsu3}
\end{multline}
where the explicit expressions for $\lambda_{\alpha}$ are given in Ref.~\cite{Lukin_2017}. This Hamiltonian is no longer anisotropic under $\SU(2)$ and therefore cannot be entirely suppressed by global $\SU(2)$ pulses. However, it is anisotropic under $\SU(3)$ and can therefore be suppressed by our $\Sigma(72\times 3)/\mathbb{Z}_3$ and $\Sigma(168)$ sequences ($72$ and $168$ pulses, respectively). In both cases, the disorder Hamiltonian, in the rotating frame, is given by 
\begin{equation}
    H_{\mathrm{dis}}^{\mathrm{RWA}} = \sum_i\delta_i n_{i,z}S_z^{(i)},\label{dis}
\end{equation}
and can also be decoupled by these sequences.

\subsubsection{Shortening sequences by group factorization}\label{method1}

As explained in Sec.~\ref{sec.facto}, an inaccessible symmetry $\group$ may admit a factorization $\group=K\,S_R$, where $K<\group$ is a subgroup and $S_R\subset\group$ is a set of right representatives. If the elements of $S_R$ are symmetries of the Hamiltonian to be decoupled, then $K$ itself can be used as a decoupling group, reducing the number of pulses by a factor of $\abs{S_R}$. If $S_R$ is not a symmetry of the Hamiltonian, one may instead conjugate the whole group inside $\SU(3)$, that is, apply a unitary transformation 
\begin{equation}
    \group \mapsto U^{\dagger}\group U,\quad U\in\SU(3)
\end{equation}
and look for a unitary $U$ such that the transformed set of representatives $U^{\dagger}S_RU$ leaves the Hamiltonian invariant. When such a $U$ exists, the transformed subgroup $U^{\dagger}KU$ can be used as a decoupling group. This process can be thought of as a change of \textit{orientation} of the finite group inside $\SU(3)$, with the goal of maximizing the number of group elements that coincide with the Hamiltonian symmetries.

\par In the present case, the interaction Hamiltonian~\eqref{eq.dipsu3} and the disorder Hamiltonian~\eqref{dis} are both invariant under conjugation by arbitrary diagonal $\SU(3)$ matrices. Moreover, the interaction Hamiltonian is also invariant under conjugation by the matrix $B$ in Eq.~\eqref{mats.}, which swaps $\lambda_{1,2}$ with $\lambda_{6,7}$ and changes the sign of $(\lambda_3+\sqrt{3}\lambda_8)$. We now introduce several sequences constructed from subgroups of $\Sigma(72\times 3)$ and $\Sigma(168)$ that take advantage of these symmetries.

\begin{center}
    \textbf{Subgroup $\Sigma(36\times 3)\triangleleft \Sigma(72\times 3)$}
\end{center}
In this case, we exploit the factorization~\cite{Grimus_2010} 
\begin{equation}
    \Sigma(72\times  3) = \Sigma(36\times 3) \qty{\mathds{1},X}
\end{equation}
and choose an orientation of $\Sigma(36\times 3)$ by selecting a matrix $U$ that diagonalizes $X$. This choice is not unique; multiplying $U$ on the right by any element of the Weyl group of $\SU(3)$, which amounts to permuting its columns, gives another valid orientation of $\Sigma(36\times 3)$. An Eulerian pulse sequence based on this construction is shown in Fig.~\ref{fig:fig5}, where the pulses $C$ and $V$ are replaced by $U^{\dagger}CU$ and $U^{\dagger}VU$, respectively. The corresponding generating Hamiltonians of these pulses are given in Appendix~\ref{ap.36}, where we also compare the performance of the different orientations. In particular, we find that one specific orientation decouples the disorder Hamiltonian more effectively, which is especially relevant for NV-centers, where disorder is often the dominant error mechanism~\cite{Zhou_2023, Zhou_2020_metrology, Lukin_2024}. This highlights that optimizing the orientation of the decoupling group can improve the decoupling performance once higher-order Magnus terms are taken into account.

\par Because $\Sigma(36\times 3)$ is an inaccessible symmetry for the irrep $(1,1)$, this sequence decouples arbitrary disorder. Its Eulerian structure makes  it robust against finite-duration errors caused by the disorder term and against essentially any systematic error in the pulses whose error Hamiltonian is a linear combination of the Gell-Mann matrices. It is, however, not robust against finite-duration errors caused by the dipolar Hamiltonian, since these involve two-body terms with no particular symmetry. As noted in Appendix~\ref{ap.36}, it might still be possible to design the sequence's pulses (\textit{e.g.}, using optimal control algorithms) in such a way that these error terms acquire enough symmetry to recover robustness.

\begin{center}
    \textbf{Subgroup $\Delta(24)< \Sigma(168)$}
\end{center}

In this case, we exploit the factorization
\begin{equation}
    \Sigma(168) = \Delta(24) \qty{Y^n\,|\, n=0,\dots 6}
\end{equation}
where $\Delta(24)$, a member of the $\Delta(6n^2)$ family, is isomorphic to the octahedral point group. Although the matrix $Y$ in Eq.~\eqref{mats.} is already diagonal, identifying the corresponding $\Delta(24)$ subgroup inside $\Sigma(168)$ is nontrivial. This task was worked out in Ref.~\cite{King_2009}, which allows one to construct a generating set for the relevant orientation of $\Delta(24)$. From this, one obtains an Eulerian sequence of $48$ pulses. Further details on the construction of these sequences are given in Appendix~\ref{ap.24}.

\par As in the previous case, the sequence is robust against arbitrary systematic control errors and against finite-duration errors generated by the disorder term, but not against finite-duration errors generated by the dipolar interaction.

\begin{center}
    \textbf{Subgroup $\Delta(54)\triangleleft \Sigma(72\times 3)$}
\end{center}

This case differs significantly from the two previous ones. Our goal is to exploit the Hamiltonian symmetries associated with $\Delta(54)$ so that the quotient group $\Sigma(72\times 3)/\Delta(54)$, which is isomorphic to the Klein four-group $K_4\cong\mathbb{Z}_2\times\mathbb{Z}_2$~\cite{Grimus_2010}, can serve as a decoupling group, following the method of Sec.~\ref{sec.facto}. Unfortunately, neither the disorder Hamiltonian nor the dipolar Hamiltonian has enough symmetries. To overcome this, we use a multisymmetrization procedure~\cite{Read_2025facto}. In this approach, two sequences are nested: an inner sequence performs the first symmetrization, while an outer sequence symmetrizes the remaining part of the Hamiltonian. Concretely, each waiting interval of the outer sequence is replaced by one full cycle of the inner sequence, so that the total sequence length is the product of the lengths of the two layers. In the protocol that we describe below, the inner layer serves two purposes: it suppresses the disorder and imposes additional symmetry on the dipolar Hamiltonian. The outer layer then exploits these induced symmetries to complete the decoupling.

\par We take as inner layer the length-$3$ sequence generated on the cyclic group of order $3$ generated by the unitary $E$, which is composed of three successive applications of $E$, \ie,
\begin{equation}
    \mathcal{S}_{\mathrm{inner}} \equiv -\,E\,-\,E\,-\,E.
\end{equation}
This sequence suppresses the disorder Hamiltonian because the disorder is invariant under the diagonal matrices $\mathrm{diag}(1,-1,-1)$ and $\mathrm{diag}(-1,-1,1)$, which generate the Klein four-group $K_4$. Together with the cyclic group generated by $E$, these symmetries form the tetrahedral group $\Delta(12)$, which is a decoupling group for $(1,1)$.
The dipolar Hamiltonian $H_{\mathrm{dd}}^{\mathrm{RWA},\SU(3)}$ is not directly suppressed by the inner layer. Rather, the sequence uses its symmetries under $B$ and the diagonal matrix $C$ to implement a $\Delta(54)$ symmetrization. The outer layer can therefore be designed on the Cayley graph of the quotient group $\Sigma(72\times 3)/\Delta(54)\cong K_4$. Using a Hamiltonian path, we obtain the $4$-pulse outer sequence
\begin{equation}
    \mathcal{S}_{\mathrm{outer}} \equiv -\,X\,-\,V\,-\,X\,-\,V,
\end{equation}
so that the full nested sequence contains only 12 pulses. A similar construction can be obtained from $\Sigma(72\times 3)/\Delta(27)\cong Q_8$~\cite{Grimus_2010}. More details are given in Appendix~\ref{ap.54}.

\par It is important to note that these nested sequences lack the robustness usually associated with group-based constructions. Even when both the inner and outer layers are chosen to be Eulerian, the resulting protocol is not robust against finite-duration pulses, including those arising from the disorder term, nor against control errors. 
These limitations diminish the practical relevance of the nested sequences. Nevertheless, we believe that their construction is of theoretical interest, as it illustrates how different subgroup chains can be combined to generate more exotic decoupling protocols.

\subsubsection{Pulse simplification for experimental implementation}\label{method2}

So far, we have shown that the remaining freedom in the orientation of the group can be used to design sequences on suitable subgroups, thereby reducing the length of the protocol. This leads to several sequences with a moderate number of pulses, each built from two distinct unitary pulses corresponding to the generators of the relevant subgroup. Unfortunately, the generating Hamiltonians of these pulses are dense in the $3\times 3$ matrix representation. A direct implementation would therefore require simultaneous control of all three transitions between the energy levels, which is generally forbidden by selection rules. For example, the double quantum transition $\ket{1}\leftrightarrow\ket{-1}$ is not accessible in ESR or NMR. In practice, such pulses must therefore be synthesized effectively using only the allowed transitions, for instance through quantum optimal control, Floquet engineering, or composite-pulse techniques~\cite{Lindon_2023,lopez_2026}. This may result in longer protocols and more complicated pulse shapes.

\par Rather than choosing the orientation of the group $\group$ to exploit the symmetries of the interaction Hamiltonian, one may instead use this freedom to simplify the pulse implementation itself. More precisely, we may seek an orientation of the group, obtained by conjugation in $\SU(3)$, such that the generating Hamiltonian of every pulse used in the sequence acts trivially on the $\ket{1}\leftrightarrow\ket{-1}$ transition. In the $\{\ket{1},\ket{0},\ket{-1}\}$ basis, this is equivalent to requiring that their $[1,3]$ and $[3,1]$ matrix elements be zero. Such Hamiltonians take the form
\begin{equation}
    H = \varphi \begin{pmatrix}
        h_1 & h_2 & 0 \\
        h_2^* & -h_1+h_3 & h_4\\
        0 & h_4^* & -h_3
    \end{pmatrix}\label{eq.Hdb}
\end{equation}
where $\varphi$, $h_1$, and $h_3$ are real parameters, while $h_2$ and $h_4$ are complex parameters. Hamiltonians of this form can be generated by operating two oscillating magnetic fields~\cite{Rugar_2014}, $B_{a}(t)$ and $B_{b}(t)$, slightly detuned from the $\ket{1}\leftrightarrow\ket{0}$ and $\ket{-1}\leftrightarrow\ket{0}$ transitions, respectively:
\begin{equation}
    \vec{B}_{a,b}(t) = B_{a,b}\cos(\omega_{a,b}t+\phi_{a,b})\hat{x} .\label{eq.magnfield}
\end{equation} 
These $\SU(3)$ pulses are called \textit{double-driving pulses}. The detunings of the driving frequencies from the transition frequencies generate the diagonal terms in~\eqref{eq.Hdb}, while the amplitudes and phases determine the complex coefficients $h_2$ and $h_4$. In the special case $B_a=B_b$, $\omega_a=\omega_{10}$, and $\omega_b=\omega_{-10}$, where $\omega_{\pm10}$ are the frequencies of the $\ket{\pm1}\leftrightarrow\ket{0}$ transitions, one recovers the balanced double-driving pulses introduced in Ref.~\cite{Lukin_2024} in the context of dynamical decoupling.
\begin{figure}[t]
    \centering
    \includegraphics[width=\linewidth]{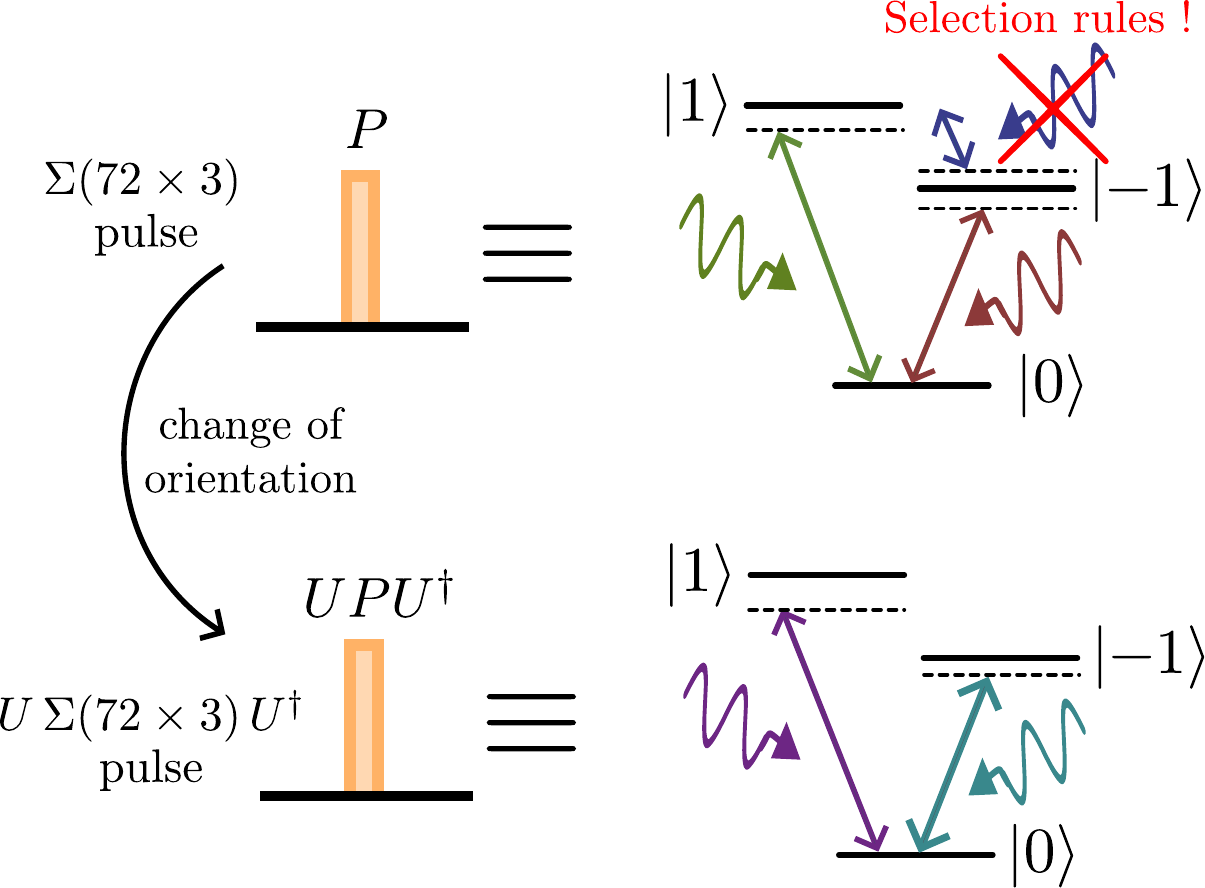}
    \caption{Selection rules may prohibit certain transitions and thus limit the set of pulses that can be generated experimentally. Group conjugation allows the pulse generators to be modified accordingly and the sequence to be adapted to these constraints.
    }
    \label{fig:or.p}
\end{figure}

\par The main idea is illustrated in Fig.~\ref{fig:or.p}. Explicit orientations of the group $\Sigma(72\times 3)$ satisfying these constraints are given in Appendix~\ref{ap.Orient}.

\section{\texorpdfstring{$\SU(d)$}{Lg} decoupling}\label{sec.qudit}

\subsection{Arbitrary $d$}
The results of Sec.~\ref{ex.SU3} generalize to qudit ensembles, in which global $\SU(d)$ pulses are applied to suppress anisotropic interactions and disorder. In this setting,  irreps are labeled by their Dynkin coefficients, which we write as a $(d-1)$ dimensional vector $\vec{d} = (d_1,d_2,\dots , d_{d-1})$. The fundamental representation is characterized by the Dynkin labels $(1,\bm{0}_{d-2})$, where $\bm{0}_{d-2}$ denotes a sequence of $d-2$ zeros, while its dual is given by $(\bm{0}_{d-2},1)$. The operator space of a single qudit then decomposes as
\begin{equation}
    \begin{aligned}
        (1,\bm{0}_{d-2})\otimes \overline{(1,\bm{0}_{d-2})} &= (1,\bm{0}_{d-2})\otimes(\bm{0}_{d-2},1)\\ 
        &= (\bm{0}_{d-1})\oplus (1,\bm{0}_{d-3},1) ,
    \end{aligned}
\end{equation}
where the trivial irrep $(\bm{0}_{d-1})$ corresponds to the one-dimensional subspace spanned by the identity operator. A decoupling group for a single qudit may therefore be obtained  by identifying an inaccessible symmetry for the nontrivial irrep $(1,\bm{0}_{d-3},1)$.

\par To decouple qudit-qudit interactions, one should decompose the tensor product $(1,\bm{0}_{d-3},1)^{\otimes 2}$ into irreps, which can be done systematically for any value of $d$ using Young's tableaux. For $d=4$, we obtain 
\begin{equation}
    \begin{aligned}
        (1,0,1)^{\otimes 2}={}&(0,0,0)\oplus (1,0,1)^{\oplus 2}\oplus (0,2,0)\\
        &\oplus (2,1,0)\oplus (0,1,2)\oplus (2,0,2).
    \end{aligned}
\end{equation}
For $d>4$, we find the closed-form formula (proof in Appendix~\ref{ap:YT})
\begin{equation}
    \begin{aligned}
        (1,\vec{0}_{d-3},1)^{\otimes 2}={}&(\vec{0}_{d-1})\oplus (1,\vec{0}_{d-3},1)^{\oplus 2}\\
        &\oplus (0,1,\vec{0}_{d-5},1,0) \oplus (2,\vec{0}_{d-4},1,0)\\
        &\oplus (0,1,\vec{0}_{d-4},2)\oplus (2,\vec{0}_{d-3},2).
    \end{aligned}\label{eq:quditqudit}
\end{equation}

\subsection{$d=4$}
\par For $\SU(4)$, several exceptional point groups and dihedral-like subgroups are listed in Ref.~\cite{Amihay_2001} together with their corresponding matrix generators. Unfortunately, although many of these finite subgroups are inaccessible to $(1,0,1)$, none of the smallest finite groups considered in this work are inaccessible to $(2,0,2)$. A few subgroups of larger order were found to satisfy the relevant condition for ququarts-ququarts decoupling, for example, the group XI, of order $7!$. However, their large size makes them highly impractical for experimental implementation. It remains possible that suitable groups of more manageable order exist but were not listed in Ref.~\cite{Amihay_2001} or included as subgroups of the larger groups identified there. For instance, additional results on the classification of finite groups of $\SU(4)$ are given in~\cite{Teixeira_2025}. Beyond $\SU(4)$, there does not appear to be such a comprehensive classification of finite subgroups, comparable to the Refs.~\cite{Grimus_2010,Amihay_2001}, likely because these Lie groups are less commonly used. Nevertheless, $\SU(d)$ gauge theories have been the subject of in-depth studies in quantum field theory~\cite{Lucini_2001,Luigi_2002, Lucini_2004, MartinLu_2004}, so that related work in this field may contain useful results on finite subgroups for $d>4$.

\section{Link with quantum error correction}\label{sec.QEC}
\label{Sec7.QECC}
Current quantum-information platforms remain noisy and limited to relatively small system sizes because of both technical and fundamental challenges, including decoherence and dephasing. Although dynamical decoupling can mitigate some of these errors, it is not sufficient by itself to guarantee arbitrarily long and accurate computation, \ie, fault-tolerant quantum computation. In principle, fault tolerance is achieved through \textit{quantum error correction}~\cite{Nielsen_Chuang_2010,lidar2013quantum}. In this approach, a \textit{logical} qubit (or qudit) is encoded into a larger physical system, so that the quantum information is distributed across a higher-dimensional Hilbert space. Computation is then performed using logical gates acting on the encoded degrees of freedom. When errors affect the underlying physical system, the encoding makes it possible to detect and correct them, thereby protecting the quantum information stored in the logical state. We refer to such encodings as \textit{quantum error-correcting codes} (QECCs).

\par A remarkably broad range of QECCs has been developed over the past decades~\cite{ErrorCorrectionZoo}. In the earliest proposals, a logical qubit was encoded into a large register of physical qubits~\cite{Shor_1995,Shor_1996,Gottesman_1997,gottesman_1998}. Since then, many more exotic QECCs have been proposed, in which the logical system is encoded, for example, in an oscillator~\cite{Gottesman_2001,Knill_2006,Fluhmann_2019}, in the rotational states of a molecule~\cite{Albert_2020,Albert_2024}, or in several spin systems~\cite{Gross_2021,OmanakuttanAndGross_2023,Omanakuttan_2023,Gross_2024,Omanakuttan_2024,Teixeira_2025bis,franke_2026}. 

\par We now show that a large class of codes can be constructed from symmetry groups of $\SU(d)$, and that the choice of an appropriate one for a given physical platform can be made within the same general framework developed above. The central idea is as follows: a subspace consisting of states with a given symmetry satisfies the Knill-Laflamme condition whenever that symmetry defines a decoupling group for the relevant set of error operators. Consequently, constructing a codespace reduces to identifying a symmetry that is accessible in the physical Hilbert space but inaccessible to the relevant set of irreps. We focus on two cases of particular interest: first, the case where a logical qudit is encoded into the collective spin subspace (symmetric subspace) of an ensemble of spin-$s$ particles; and second, the case where a logical qudit is encoded into a register of qutrits.

\subsection{Relation between decoupling groups and QECCs}

Consider a physical system whose dominant errors are described by a set of operators $\mathcal{E}$, and let $\Hs_C$ be a candidate $d$-dimensional codespace spanned by a set of orthogonal states $\qty{\ket{\psi_i}}_{i=1}^{d}$. A QECC for $\mathcal{E}$ is a subspace $\Hs_C$ for which the encoded quantum information can be protected against the errors in $\mathcal{E}$, in the sense that these errors can be detected and, when the stronger conditions below are satisfied, corrected by a suitable recovery operation. This particular encoding can \emph{detect} any error $E_k\in\mathcal{E}$ provided that
\begin{equation}
\label{Eq.QECC.detect}
    \bra{\psi_i} E_k\ket{\psi_j} = \delta_{ij}C_k\quad \forall k,i,j
\end{equation}
where $C_k$ is a constant that may depend on $E_k$. Similarly, the code can \emph{correct} the errors in $\mathcal{E}$ if the stronger conditions are satisfied~\cite{PhysRevA.55.900},
\begin{equation}
     \bra{\psi_i} E_p^{\dagger}E_k\ket{\psi_j} = \delta_{ij}C_{kp}\quad \forall k,p,i,j,
\end{equation}
where $C_{kp}$ is a constant that depends on $E_k$ and $E_p$. These relations are the \emph{Knill-Laflamme (KL) conditions} that any codespace should fulfill to be a QECC. Equivalently, the correction condition may be written as
\begin{equation}
    \bra{\psi_i}F_k\ket{\psi_j} = \delta_{ij}C_{k}'\quad \forall k,i,j
\end{equation}
where each $F_k$ belongs to the set $\mathcal{E}^2 \equiv \qty{E_p^{\dagger}E_q}_{p,q}$, and $C_k'$ depends on $F_k$. In other words, the KL conditions require that every operator in the span of $\mathcal{E}^2$ acts on the codespace as a scalar multiple of the identity. Hence, if the condition is verified on a basis of $\mathrm{span}(\mathcal{E}^2)$, it automatically holds for every operator in that subspace.

\par Consider now a pure state $\ket{\psi}\in \mathcal{H}$ that is invariant under a finite subgroup $\group<\SU(d)$, \ie,
\begin{equation}
    \pi(g)\ket{\psi} = e^{i\theta(g)}\ket{\psi}
\end{equation}
for every $g \in \group$, where $\pi$ denotes the linear map defining the $\SU(d)$ representation on $\mathcal{H}$ and $\theta(g)$ is a real global phase depending on $g$. Then, for any operator $F$ acting on $\mathcal{H}$, its expectation value in the state $\ket{\psi}$ satisfies
\begin{equation}\begin{aligned}
    \bra{\psi}F\ket{\psi} &= \frac{1}{\abs{\group}}\sum_{g\in\group}\bra{\psi}\pi(g)^{\dagger}F\pi(g)\ket{\psi}\\
    &= \bra{\psi}\Pi_{\group}(F)\ket{\psi}
    \end{aligned}\label{eq.Psisym}
\end{equation}
where $\Pi_{\group}$ is the projector onto the $\group$-invariant subspace defined in Eq.~\eqref{eq.qo}. Now let $\qty{\ket{\psi_{i}}}$ be a set of orthogonal states spanning a subspace $\Hs_C \subseteq \mathcal{H}$, and suppose that all of them possess the same symmetry, \ie,
\begin{equation}
    \pi(g)\ket{\psi_i} = e^{i\theta(g)}\ket{\psi_i} \quad \forall i
\end{equation}
with the same global phase factor $e^{i\theta(g)}$ for each basis state. It follows that every vector in $\Hs_C$, being a linear combination of the $\ket{\psi_{i}}$, has the same symmetry and therefore also satisfies the relation~\eqref{eq.Psisym}. In particular, if $\group$ is a decoupling group for $\mathcal{E}^2$, then for every $F_k \in \mathcal{E}^2$ and every pair of states $\ket{\phi_i},\ket{\psi_j}\in\mathcal{H}_C$, one has  
\begin{equation}\begin{aligned}
        \bra{\phi_i}F_k\ket{\psi_j} &=  \bra{\phi_i}\Pi_{\group}(F_k)\ket{\psi_j} \\ 
        &= c_k \langle\phi_i|\psi_j\rangle \quad \forall i,j,k,
\end{aligned}
\end{equation}
where in the second line we used the decoupling condition $\Pi_{\group}(F_k)\propto \mathds{1}$. Therefore, the restriction of every operator $F_k \in \mathcal{E}^2$ to $\mathcal{H}_C$ is proportional to the identity, and the Knill–Laflamme conditions are automatically satisfied. Hence, any subspace spanned by states sharing the same symmetry defines a QECC whenever that symmetry group is a decoupling group for the error algebra $\mathcal{E}^2$. This is summarized by the following lemma.
\begin{lemma}\label{lemmaQECC}
    Let $(\pi,\Hs_S)$ be a representation of $\SU(d)$ on the Hilbert space $\Hs_S$ of a physical system. Let $\mathcal{E}$ denote a set of error operators, and let $\mathcal{E}^2=\{E_p^\dagger E_q\}_{p,q}$. Let $V_{\mathcal{E}}\supseteq \mathcal{E}$ and $V_{\mathcal{E}^2}\supseteq \mathcal{E}^2$ be subspaces closed under $\SU(d)$.
    \par Let $\group<\SU(d)$ be a subgroup of $\SU(d)$, and let $\Hs_C = \mathrm{span}(\qty{\ket{\psi_i}}_{i=1}^k)$ be a $k$-dimensional subspace spanned by orthogonal states satisfying 
    \begin{equation}
        \pi(g)\ket{\psi_i} = e^{i\theta(g)}\ket{\psi_i}\quad \forall i,\,\forall g\in\group
    \end{equation}
    for some phase $\theta(g)$ independent of $i$.
    Then, the following statements hold true:
    \begin{enumerate}
         \item If $\group$ is a decoupling group for $\mathcal{E}$, then $\Hs_C$ is a quantum error-detecting code for $\mathcal{E}$.
         \item If $\group$ is a decoupling group for $\mathcal{E}^2$, then $\Hs_C$ satisfies the Knill-Laflamme conditions for $\mathcal{E}$, and hence is a quantum error-correcting code. 
       \item If $\group$ is an inaccessible symmetry for $V_{\mathcal{E}}$, then $\Hs_C$ is a quantum error-detecting code for $\mathcal{E}$.
        \item If $\group$ is an inaccessible symmetry for $V_{\mathcal{E}^2}$, then $\Hs_C$ satisfies the Knill-Laflamme conditions for $\mathcal{E}$, and hence is a quantum error-correcting code. 
    \end{enumerate}
\end{lemma}

With the above lemma, we realize that many results obtained in the context of DD can be used in the context of QECC to choose the symmetry of the codespace. In particular, choosing a symmetry $\group$ that is inaccessible to $V_{\mathcal{E}^2}$ guarantees that the KL condition is satisfied, and factorizing this symmetry as presented in Sec.~\ref{sec.facto} allows us to exploit the symmetries of $\mathcal{E}^2$ (\textit{e.g.}, in the case of biased noise) to construct codespaces with smaller and less restrictive symmetries.

\subsection{Constructing subspaces with \texorpdfstring{$\group$-symmetry}{Lg}}

In this section, we explain how to construct subspaces with the symmetry of a finite group $\group$. Let $(\pi,\Hs_S)$ be a representation of $\SU(d)$ on the physical Hilbert space $\Hs_S$. Restricting this representation to a finite subgroup $\group<\SU(d)$, the Hilbert space decomposes into irreps of $\group$ as
\begin{equation}\label{irrepdecompQECC}
    \Hs_S = \bigoplus_i\rho_i^{\oplus a_i},\quad \pi = \bigoplus_i\lambda_i^{\oplus a_i}.
\end{equation}
where $\rho_i$ denotes an irrep of $\group$, $\lambda_i$ its corresponding representation map, and $a_i$ its multiplicity in $\Hs_S$. The multiplicities are given by the character inner product, $a_i = (\chi_{\Hs_S},\chi_{i})$ as in Eq.~\eqref{Eq.char.multi}, where $\chi_{\Hs_S}$ is the character of $\Hs_S$ restricted to $\group$, and $\chi_i$ is the character of $\rho_i$. The characters $\chi_i$ are obtained from the character table of $\group$.

\par Among the irreps of $\group$, some may be one-dimensional; in those representations, each group element acts simply by multiplication with a phase factor. Let $\rho_1,\dots,\rho_l$ denote the distinct (non-isomorphic) one-dimensional irreps appearing in $\Hs_S$ as in Eq.~\eqref{irrepdecompQECC}. Then the decomposition may be written as
\begin{equation}
    \Hs_S = \overbrace{\rho_1^{\oplus a_1}\oplus \rho_2^{\oplus a_2}\oplus\dots\oplus \rho_l^{\oplus a_l}}^{\mathrm{1D- irreps}}\oplus \rho_{l+1}^{\oplus a_{l+1}}\oplus\cdots.
\end{equation}
For any vector $\ket{\psi}\in \rho_j$, with $j=1,\dots,l$, the action of $\group$ is given by
\begin{equation}
    \pi(g)\ket{\psi}=\lambda_j(g)\ket{\psi} = e^{i\theta_j(g)}\ket{\psi}\quad \forall g\in\group,
\end{equation}
since every one-dimensional irrep acts by a phase. It follows that every vector in the multiplicity space $\rho_j^{\oplus a_j}$ transforms according to the same character $\lambda_j$. Therefore, each subspace $\rho_j^{\oplus a_j}$ is a natural candidate codespace: all of its states share the same symmetry under $\group$.

\par The projector onto the $a_j$-dimensional subspace $\rho_j^{\oplus a_j}$ is given by~\cite{Fulton_2013}
\begin{equation}
    P(\rho_j^{\oplus a_j}) = \frac{1}{|\group|}\sum_{g\in\group}\overline{\chi_j(g)} \pi(g)
\end{equation}
where $\chi_j$ is the character of the one-dimensional irrep $\rho_j$. A basis for the corresponding codespace can be obtained by diagonalizing this projector. Any one-dimensional irrep $\rho_j$ can be used in this construction; it need not be the trivial irrep. What matters is that all code states transform by the same character of $\group$, so that group actions contribute only a phase, which cancels in the relevant expectation values. Consequently, the codespace may be supported on several copies of the same irrep, but not on a direct sum of non-isomorphic one-dimensional irreps. This is because if $\ket{\psi} \in \rho_i$ and $\ket{\phi} \in \rho_j\ncong\rho_i$, then a linear superposition $\ket{\eta} = \alpha\ket{\psi}+\beta \ket{\phi}$ will not, in general, have the symmetry of $\group$. Indeed, 
\begin{equation}\begin{aligned}
    \pi(g) \ket{\eta} &= \alpha\lambda_i(g)\ket{\psi}+\beta\lambda_j(g)\ket{\phi}\\& = \alpha\, e^{i\theta_i(g)}\ket{\psi}+\beta \, e^{i\theta_j(g)}\ket{\phi}
\end{aligned}\end{equation}
which is not proportional to $\ket{\eta}$ in general, since $\theta_i(g)\neq\theta_j(g)$ for non-isomorphic one-dimensional irreps. Therefore, only multiple copies of the same one-dimensional irrep define a subspace in which every state has the same $\group$-symmetry.

\subsection{Connections to recent works in QEC}

Many QECCs in the literature have been constructed using closely related representation-theory ideas. In particular, Refs.~\cite{Gross_2021,OmanakuttanAndGross_2023,Teixeira_2023} introduced codespaces for spin systems based on multiplicity spaces associated with several copies of two-dimensional irreducible representations of finite rotational symmetry groups. Efforts have been made to generalize these results to $\SU(3)$~\cite{herbert_2023} and $\SU(d)$~\cite{Uy_2025}. Especially relevant developments in this direction were presented  in Refs.~\cite{Teixeira_2024, Teixeira_2026}. Compared with the construction considered here, an important advantage of encoding information within a multi-dimensional irrep of a finite group is that the group elements themselves may act nontrivially on the codespace and can therefore be used as logical gates.

\par Very recently, Ref.~\cite{bradshaw_2026} introduced a general framework based on representation theory, closely aligned with the one developed here, in which $\group$-invariant subspaces are also identified as natural codespaces, along with general procedures for syndrome extraction and the implementation of logic gates. Those results apply directly to any QECC constructed within our framework. In particular, when the code is supported on the trivial irrep, its logical operations may be chosen from the normalizer\footnote{For a subgroup $K<T$, the normalizer of $K$ in $T$ is the set of elements $\qty{t\in T\,|\, tKt^{-1}=K}$.} of the group $\group$ in $\SU(d)$, denoted $N_{\mathrm{SU}(d)}(\group)$. Indeed, elements of $N_{\mathrm{SU}(d)}(\group)$ will map the subspace $\rho_1^{\oplus a_1}$, spanned by all copies of the trivial irrep $\rho_1$ of $\group$, to itself. To see this, let $t\in N_{\mathrm{SU}(d)}(\group)$ and $\ket{\psi}\in \rho_1^{\oplus a_1}$. Since $\ket{\psi}$ is invariant under $\group$, the state $\pi(t)\ket{\psi}$ is also invariant under $\group$, and therefore still belongs to $\rho_1^{\oplus a_1}$. Consequently, the elements of $N_{\mathrm{SU}(d)}(\group)$ that are not in  $\group$---those that are representatives of the quotient group $N_{\mathrm{SU}(d)}/\group$---will act as a non-trivial set of logical gates.

\subsection{Application to spin-\texorpdfstring{$s$ ensembles}{Lg}}

As a first application, we construct QECCs for an ensemble of $N$ spins-$s$ particles that can correct disorder and two-body interactions. The relevant symmetry group is $\SU(2)$, since each individual spin transforms under the spin-$s$ representation of $\SU(2)$, and the physical errors considered are naturally organized according to their transformation properties under spatial rotations. Restricting to the collective subspace of the ensemble, the $N$-body system behaves as an effective single spin with total angular momentum $j=Ns$, so the problem reduces to studying finite subgroups of $\SU(2)$ (or, for integer $j$, equivalently of $\SO(3)$) acting on this effective spin-$j$ Hilbert space\footnote{While for $N$ spin-$s$ the symmetric subspace contains different blocks transforming according to different irreps of $\SU(2)$, we can focus on the "top block" corresponding to the spin-$j=Ns$ irrep, which always appears~\cite{Chryssomalakos_2021}.}. The logical qubit states are then encoded as two orthogonal states inside a symmetry sector of this collective-spin space, that is, inside a two-dimensional subspace whose vectors all transform according to the same one-dimensional irrep of the chosen finite subgroup. For a given error model, the corresponding symmetry can be identified systematically, and we present several explicit examples below. For simplicity, we focus on the encoding of a logical qubit, although the same construction applies more generally to logical qudits.

\par The codespace will thus be realized within the collective subspace of the spin system, which is isomorphic to the Hilbert space of a single spin with quantum number $j=Ns$. To construct such codes, one should in principle consider binary polyhedral groups. However, for half-integer values of $j$, the multiplicity of every one-dimensional irrep of the binary $2$-dihedral group ($2\mathrm{D}_2$) is zero. Since each binary Platonic group contains $2\mathrm{D}_2$ as a subgroup, it follows that their one-dimensional irreps also occur with zero multiplicity in half-integer spin representations. Therefore, only integer values of $j$ can yield the one-dimensional symmetry sectors relevant to our construction.
For integer $j$, these representations are also representations of $\SO(3)$, so the ordinary polyhedral character tables may be used instead of the binary ones; these are standard and can be found in textbooks on representation theory~\cite{Rotman_2012,Fulton_2013}. It is worth noting that the present construction produces codes in integer-spin $j$ sectors, whereas recent spin-based QEC schemes~\cite{Gross_2021,OmanakuttanAndGross_2023,Gross_2024,Omanakuttan_2024} have focused instead on half-integer values of $j$.

\begin{figure*}[t]
    \centering
    \includegraphics[width=\linewidth]{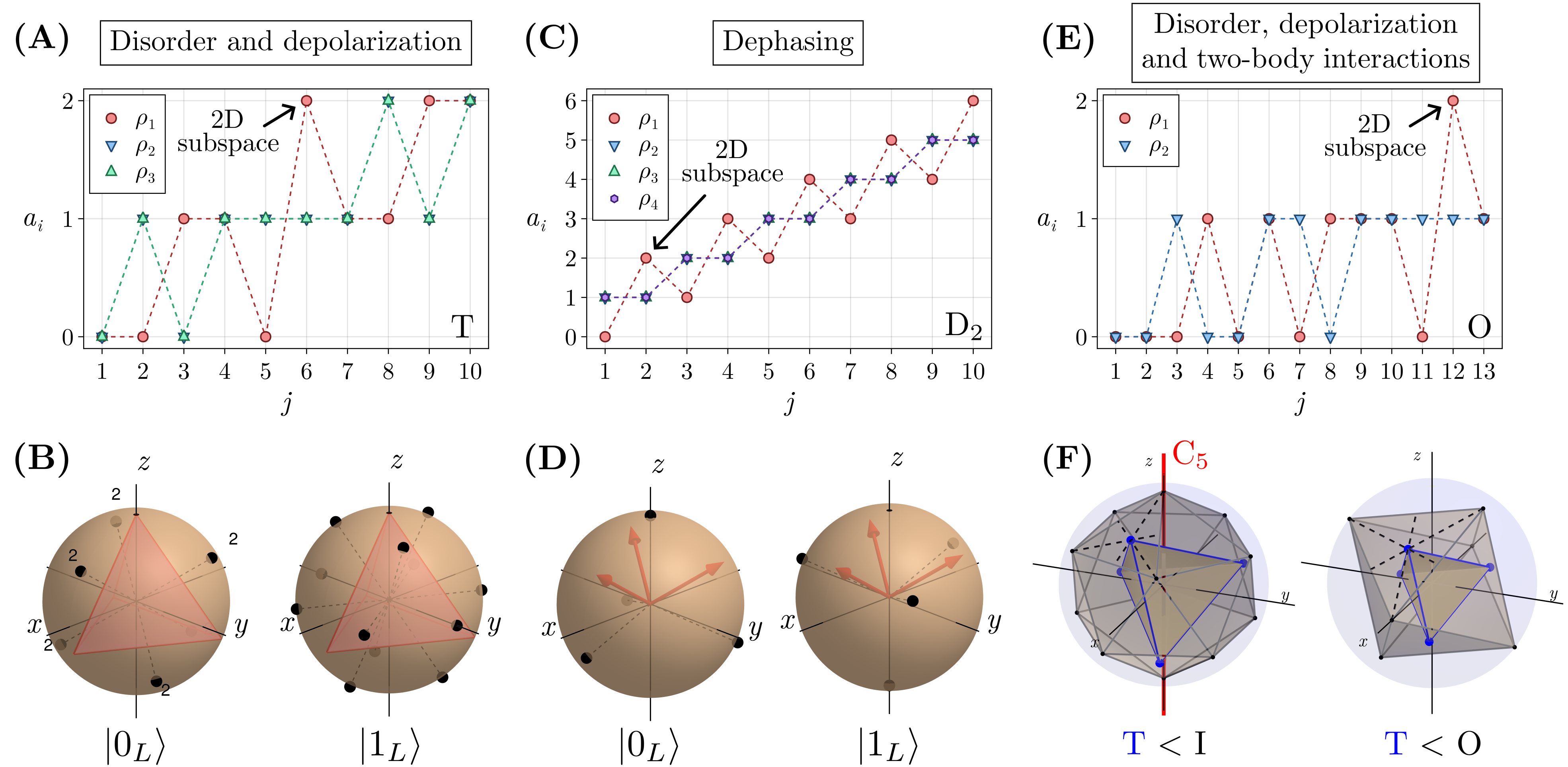}
    \caption{(A,C,E) Multiplicities of the one-dimensional irreps of the finite point groups $\mathcal{G}\in\{\mathrm{T},\mathrm{D}_2,\mathrm{O}\}$ in the spin-$j$ irrep, shown as a function of $j$. The first two-dimensional subspace relevant in QEC is indicated by an arrow. (B,D) Logical states $\ket{0_L}$ and $\ket{1_L}$ spanning the corresponding two-dimensional subspace. (F) Specific orientation of the octahedral point group and its tetrahedral subgroup, as explained in the main text. The $\mathrm{C}_5$ symmetry axis of the icosahedral point group is highlighted in red.}
    \label{fig:codespaces_spin}
\end{figure*}

\subsubsection{Correcting arbitrary disorder and depolarization}\label{arb.dis}

\par In this first example, we take $\mathcal{E}$ to consist of one-body error operators that are linear in the spin-$s$ operators, such as $S_z$ and $S_{\pm}$. These single-spin errors, to first order, model imperfections arising from disorder in the ensemble, as well as dephasing and, more generally, depolarization. In this case, the $\SU(2)$-invariant space $V_{\mathcal{E}}\supseteq \mathcal{E}$ to consider in Lemma~\ref{lemmaQECC} is composed of spin-$1$ irreps. By contrast, $V_{\mathcal{E}^2}\supseteq \mathcal{E}^2$ generally contains spin-$0$, spin-$1$, and spin-$2$ irreps, since $\mathcal{E}^2$ is generated by products of rank-$1$ spin operators, which transform inside $1\otimes 1=0\oplus 1\oplus 2$, and rank-1 spin operators. In particular, $\mathcal{E}^2$ includes bilinear combinations of spin operators, such as $S_z^{(i)}\otimes S_+^{(j)}$ or $S_z^{(i)}\otimes S_z^{(j)}$, as well as quadratic operators such as $S_z^2$ and $S_zS_+$ and linear operators such as $S_{z}$ and $S_{\pm}$. Referring to Fig.~\ref{fig:su2}, we then identify suitable inaccessible symmetries. A $2$-dihedral symmetry ($\mathrm{D}_2$) is sufficient for error detection, while tetrahedral symmetry ($\mathrm{T}$) yields error correction. Note also that spin-$0$ irreps in $V_{\mathcal{E}^2}$, corresponding to operators that are invariant under $\SU(2)$, such as $\vec{S}^{(i)}\cdot \vec{S}^{(j)}$, automatically satisfy the Knill–Laflamme condition on the collective-spin subspace, since that subspace is an eigenspace of such invariants. Indeed, for any $\ket{\psi}$ and $\ket{\phi}$ in the collective-spin subspace,
\begin{equation}
    \bra{\psi}\vec{S}^{(i)}\cdot \vec{S}^{(j)}\ket{\phi} = c_{ij} \bra{\psi}\ket{\phi}\quad\forall i,j.
\end{equation}

\par The multiplicities of the three one-dimensional irreps of $\mathrm{T}$ (listed in Table~\ref{tab:char.T}) are shown in Fig.~\ref{fig:codespaces_spin}A as a function of $j=Ns$. The first two-dimensional codespace appears for $j=6$, in the sector corresponding to the trivial irrep $\rho_1$. In Fig.~\ref{fig:codespaces_spin}B, we plot the Majorana constellation~\cite{Majorana_1932} of two orthogonal states of this subspace, chosen as the logical states $\ket{0_L}$ and $\ket{1_L}$. For any state in this codespace, the Majorana constellation consists of a tetrahedral arrangement in which each vertex is replaced by three stars forming an equilateral triangle. Moving within the subspace corresponds to continuously changing the size of these triangles and rotating each of them about the axis joining the center of the sphere to the barycenter of the corresponding triangle. In particular, one can choose a logical basis such that both $\ket{0_L}$ and $\ket{1_L}$ exhibit octahedral symmetry. Using this basis, we find that some elements of the octahedral point group perform a logical Pauli-$Z$ gate (see Appendix~\ref{Ap.T}). This codespace can be realized with $N=6/s$ spin-$s$ particles, provided $N$ is an integer; for example, with $12$ spin-$\tfrac12$ particles or $6$ spin-$1$ particles.

\subsubsection{Correcting dephasing}\label{deph}

We now consider the same system in the presence of dephasing only, so that $\mathcal{E}$ consists of operators proportional to $S_z$. As before, $V_{\mathcal{E}^2}$ contains spin-$1$ and spin-$2$ irreps, so tetrahedral symmetry remains sufficient for error correction. In this case, however, every operator in $\mathcal{E}^2$ is invariant under rotations about the $z$-axis (\textit{e.g.}, $S_z^2$, $S_z^{(i)}\otimes S_z^{(j)}$, $S_z$). This additional axial symmetry can be exploited using the group factorization $\mathrm{T} = \mathrm{D}_2 \mathrm{C}_3$, where $\mathrm{C}_3$ denotes a $2\pi/3$ rotation symmetry of the tetrahedron and $\mathrm{D}_2$ its 2-dihedral symmetry. By orienting the tetrahedron so that one of its $\mathrm{C}_3$ axes coincides with $z$ axis, the corresponding orientation of $\mathrm{D}_2$ becomes a decoupling group for $\mathcal{E}^2$. In axis-angle notation, this $\mathrm{D}_2$ subgroup is generated by the rotations $((\sqrt{2},0,1)/\sqrt{3},\pi)$ and $((-1/\sqrt{6},1/\sqrt{2},1/\sqrt{3}),\pi)$ in axis-angle notation, and coincides with the subgroup used in Ref.~\cite{Read_2025facto} to construct the $\mathrm{TEDDY}$ decoupling sequence. A code can therefore be constructed on a 2-dimensional subspace associated with one of the four one-dimensional irreps of $\mathrm{D}_2$, as long as $\mathrm{D}_2$ is properly oriented. Moreover, since $\mathrm{D}_2\triangleleft \mathrm{T}$, the elements of $\mathrm{T}$ outside $\mathrm{D}_2$---namely, the rotations of angle $2\pi/3$ and $4\pi/3$--- act as nontrivial logical gates.

\par Figure~\ref{fig:codespaces_spin}C shows the multiplicities of the one-dimensional irreps of $\mathrm{D}_2$ as a function of $j=Ns$ for integer values of $j$. The first $2$-dimensional codespace appears for $j=2$. In Fig.~\ref{fig:codespaces_spin}D, we plot the Majorana constellations of two orthogonal states spanning this subspace, chosen as the logical states. One can choose the logical basis so that these two states correspond to two tetrahedra with antipodal symmetry, and a rotation of angle $2\pi/3$ around the $z$ axis can be seen to add a relative phase $2\pi/3$ between the logical zero and one (see Appendix~\ref{Ap.D}). This codespace exists for collective spin $j=Ns=2$. Therefore, the allowed realizations are four spin-$\tfrac12$ particles, two spin-$1$ particles, or one spin-$2$ particle.

\subsubsection{Correcting disorder, depolarization and two-body interactions}\label{twobdy}

We now consider a more general error model in which $\mathcal{E}$ contains both single-spin disorder and depolarization terms, such as $S_z^{(i)}$ and $S_{\pm}^{(i)}$, together with linear two-body interactions, for example, the dipolar coupling
\begin{equation}
    D_{ij}(\hat{r}_{ij}) \propto 3(\hat{r}_{ij}\cdot\vec{S}^{(i)})(\hat{r}_{ij}\cdot\vec{S}^{(j)}) - \vec{S}^{(i)}\cdot\vec{S}^{(j)}.
\end{equation}
In the most general case, the invariant space $V_{\mathcal{E}^2}$ contains irreps of spin $L=1,\dots 4$. Figure~\ref{fig:su2} shows that the icosahedral symmetry is required for error correction. As before, scalar contributions may also appear in $\mathcal E^2$, including rank-$3$ and rank-$4$ rotational invariants such as $\vec{S}^{(i)}\cdot(\vec{S}^{(j)}\times \vec{S}^{(k)})$ and $(\vec{S}^{(i)}\cdot \vec{S}^{(j)})(\vec{S}^{(k)}\cdot \vec{S}^{(l)})$. Since the collective spin subspace is an irreducible $\SU(2)$ sector, these invariant operators act on it as scalar multiples of the identity and therefore automatically satisfy the KL condition. We now discuss two situations in which the required symmetry can be reduced by exploiting group factorization.

\par First, within the rotating-wave approximation, only operators invariant under rotations about the $z$-axis appear. In that case, one can use the factorization $\mathrm{I}=\mathrm{T}\mathrm{C}_5$. A suitable rotation to the logical states $\ket{0_L}$ and $\ket{1_L}$ shown in Fig.~\ref{fig:codespaces_spin}B is then sufficient to obtain a code with the required symmetry for error correction. This specific orientation of $\mathrm{T}$ is represented in Fig.~\ref{fig:codespaces_spin}F, and explicit generators of the group for this orientation are given in Ref.~\cite{Read_2025facto}. The first relevant codespace therefore occurs for a spin $j=6$.

\par Suppose now that, in addition to the previous errors, we also wish to correct dissipation errors generated by $S_{\pm}$. Including these operators enlarges $\mathcal E$ and breaks the rotational symmetry about the $z$-axis for the operators of $\mathcal{E}^2$ that transform according to a spin-$L=1,2,3$ irrep. The situation is different for the spin-$4$ sector: the only operators transforming as spin-$4$ arise from products of spin-spin interaction terms, for example, $D_{ij}(\hat{r}_{ij})\otimes D_{kl}(\hat{r}_{kl})$ or $\qty(D_{ij}(\hat{r}_{ij}))^2$. If we only wish to correct these interactions within the rotating-wave approximation (that is, when $\hat{r}_{ij} = \hat{z}$ $\forall i,j$), then the only operators that transform according to the spin-4 irrep will have a rotational symmetry around $z$ that we can use. We can thus use a specific orientation of the octahedral point group---to ensure that spin-$L=1$, $2$ and $3$ irreps are decoupled---such that its tetrahedral subgroup has the orientation described in Fig.~\ref{fig:codespaces_spin}F, leveraging again the factorization $\mathrm{I}=\mathrm{T}\mathrm{C}_5$. The first relevant codespace appears at $j = 12$ (see Fig.~\ref{fig:codespaces_spin}E) which can be realized using $N=12/s$ particles with spin $s\in\{\tfrac{1}{2}, 1, \tfrac{3}{2}, 2, 3, 4, 6, 12\}$.

\subsection{Application to qutrit ensembles}

We now consider a register of $N$ qutrits on which we wish to encode a logical qubit. Since arbitrary two-qutrit error correction would require a symmetry inaccessible to every nontrivial irrep contained in $V_{\mathcal E^2}$, and since the decomposition of $(1,1)^{\otimes 4}$ includes the irrep $(4,4)$ for which no such inaccessible symmetry exists, our construction cannot yield a QECC correcting arbitrary two-qutrit errors. For arbitrary single-qutrit errors, however, the relevant space for error correction is only $(1,1)^{\otimes 2}$, because the basic error operators transform in the $(1,1)$ irrep. In that case, a codespace with symmetry $\Sigma(168)$ or $\Sigma(72\times 3)$ is sufficient. In the special case of pure dephasing, all operators in $\mathcal{E}^2$ are invariant under unitary transformation by diagonal matrices, and the required codespace symmetry can be reduced to $\Sigma(36\times 3)$ or $\Delta(24)$, using the specific orientations described in Sec.~\ref{method1}.

\par
\begin{figure}[h]
    \centering
    \includegraphics[width=\linewidth]{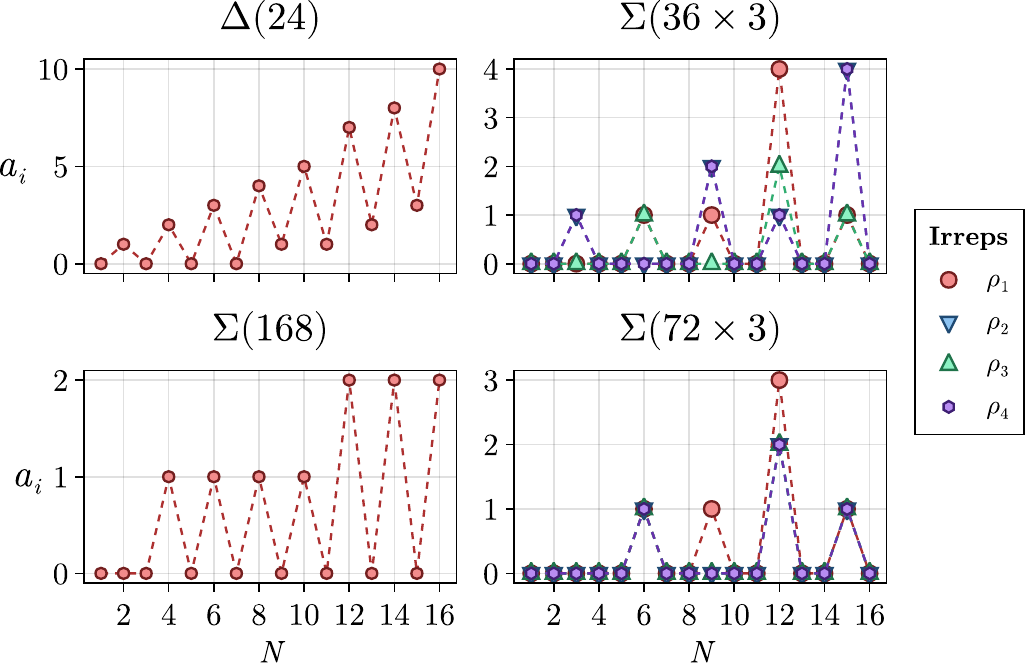}
    \caption{Multiplicities of the one-dimensional irreps of several finite groups in the symmetric subspace of $N$ qutrits, which is isomorphic to the $\SU(3)$ irrep $(N,0)$, shown as a function of $N$.}
    \label{fig:codespace3}
\end{figure}

 To construct a codespace, we restrict to the symmetric subspace of $N$ qutrits, which is isomorphic to the irrep of $\SU(3)$ with highest weight $(N,0)$. On this subspace, any $\SU(3)$-invariant operator acts as a scalar; examples include isotropic two-qutrit interactions of the form $\sum_{\mu=1}^8\lambda_{\mu}^{(i)}\otimes \lambda_{\mu}^{(j)}$. Figure~\ref{fig:codespace3} shows the multiplicities of the one-dimensional irreps of the subgroups $\Sigma(36\times 3)$, $\Sigma(72\times 3)$, $\Delta(24)$ and $\Sigma(168)$ as a function of the number $N$ of qutrits. 

\par The smallest QECC capable of detecting and correcting dephasing errors requires four qutrits and is associated with the octahedral group $\Delta(24)$. Using instead the symmetry $\Sigma(36\times 3)$, one finds a six-qutrit QECC with similar properties. In this case, the codespace admits a basis in which the logical states $\ket{0_L}$ and $\ket{1_L}$ are each individually symmetric under $\Sigma(72\times 3)$ (see Appendix~\ref{ap:Sigmas}). Since $\Sigma(36\times 3)\triangleleft \Sigma(72\times 3)$, the elements of $\Sigma(72\times 3)/\Sigma(36\times 3)$ correspond to non-trivial logical gates in this six-qutrit QECC. For example, the element $X\in\Sigma(72\times 3)$ (but $X\not\in\Sigma(36\times 3)$) that was diagonalized to identify the relevant orientation of $\Sigma(36\times 3)$ can be used to implement a Pauli-$Z$ gate. Since it is diagonal, it commutes with the set of errors and can be implemented fault-tolerantly.

\par For the correction of arbitrary single-qutrit errors using either $\Sigma(168)$ or $\Sigma(72\times 3)$ symmetry, at least twelve qutrits are necessary. In the case of $\Sigma(72\times 3)$, the two-dimensional logical subspace can be constructed from any of the four one-dimensional irreps. However, choosing the trivial irrep $\rho_1$ yields a three-dimensional multiplicity space, thereby allowing the encoding of a logical qutrit rather than only a logical qubit. Moreover, one can choose a basis of the codespace $\rho_1^{\oplus 3}$ such that the three logical basis states are each individually invariant under $\Sigma(216\times 3)$ (see Appendix~\ref{ap:Sigmas}). As before, the elements of $\Sigma(216\times 3)$ that are not in $\Sigma(72\times 3)\triangleleft \Sigma(216\times 3)$ then provide non-trivial logical gates on this three-dimensional codespace.

\section{Discussion and conclusion}\label{sec.conc}

In this work, we have developed a general theoretical framework for dynamical decoupling based on group representation theory, grounded in the finite subgroups of $\SU(d)$, and more broadly of any semisimple Lie group. The central idea is to identify the symmetries of finite groups that are inaccessible to the irreducible components of the relevant operator space. Such symmetries define decoupling groups and thus directly provide pulse sequences capable of suppressing the corresponding interactions. This offers a systematic approach to constructing decoupling protocols for qudit systems, where intuitive geometric methods are generally no longer available.

When applied to $\SU(2)$, our framework unifies the known role of polyhedral point groups in the decoupling of interacting spin systems~\cite{Llor_1991,Llor_1995,Llor_1995bis,Read_2025}. Applied to $\SU(3)$, it yields a much richer set of results. In particular, we recover the shortest known universal decoupling sequence for a qutrit, constructed on $\Delta(27)/\mathbb{Z}_3$~\cite{Tripathi_2024}, and identify finite subgroups that decouple increasingly complex many-body interactions. In particular, we show that the groups $\Sigma(168)$ and $\Sigma(72\times 3)/\mathbb{Z}_3$ decouple arbitrary one- and two-body anisotropic interactions in qutrit ensembles, thereby providing natural analogues for qutrits of the tetrahedral $\mathrm{TEDD}$ sequence introduced for interacting spin systems~\cite{Read_2025}. We also show that arbitrary anisotropic three-body interactions can be decoupled using $\Sigma(360\times 3)/\mathbb{Z}_3$, although the resulting sequence is likely too long for short-term experiments.

A second important result is that the complexity of these sequences can often be substantially reduced by exploiting the symmetries of the Hamiltonian through subgroup factorizations~\cite{Read_2025facto}. This leads, for spin-$1$ systems with a large zero-field splitting, to significantly shorter sequences suitable for realistic experimental setups. In particular, we identify sequences based on $\Sigma(36\times 3)$ and $\Delta(24)$ that eliminate disorder and dipole interactions while maintaining robustness against broad classes of pulse imperfections. We further show that the residual freedom in the orientation of the finite subgroup within $\SU(3)$ can be used to simplify pulse generators and make the resulting protocols more compatible with experimentally accessible transitions. These results are particularly relevant for platforms such as NV center arrays~\cite{Taylor_2008,Zhou_2020_metrology,Zhou_2023,Lukin_2024}, hexagonal boron nitride defects~\cite{Gottscholl_2021,Tarkanyi_2025,Sasaki_2023,Mahajan_2025,Souvik_2025}, and quadrupolar spin systems~\cite{Jerschow_2005,Ashbrook_2014}.

Another potentially fruitful direction, only briefly explored here, is the use of different group representations in dynamical decoupling. As illustrated in Sec.~\ref{SubSection.diff.irrep} for the case of $\SU(2)$, changing the representation of $\SU(d)$ modifies the irrep decomposition of the interaction subspace. For an ensemble of qudits, this additional freedom can be exploited to tailor the decomposition of the interaction space, thereby enabling new dynamical decoupling strategies and new possibilities for Hamiltonian engineering. More broadly, this perspective suggests new routes both for the identification of DD sequences and for Hamiltonian engineering, which we leave for future investigation.

Beyond dynamical decoupling, we have established a direct link between decoupling groups and quantum error correction. We have proven that any subspace whose states transform under the same one-dimensional irreps of a finite subgroup automatically satisfies the Knill-Laflamme conditions provided that this subgroup is a decoupling group for the relevant set of error operators. This transforms the search for QECCs into a symmetry problem identical to that of the search for decoupling sequences: one must identify the symmetries accessible in the physical Hilbert space but inaccessible to the irreps associated with the dominant errors. In this sense, the framework developed here unifies two tasks usually treated separately, namely Hamiltonian cancellation and code space construction.

This connection leads to several explicit constructions of QECCs. For collective spin systems, we obtain families of logical subspaces protected against disorder, dephasing, depolarization, and certain two-body interactions, with concrete implementations based on tetrahedral, dihedral, octahedral, and icosahedral symmetries. In particular, we identify a tetrahedral code in the collective spin-$6$ sector that corrects arbitrary single-spin disorder and depolarization, as well as a smaller code in the collective spin-$2$ sector that corrects dephasing. For qutrit registers, we show that the same $\SU(3)$ subgroups that are useful for decoupling also generate non-trivial QECCs: $\Delta(24)$ and $\Sigma(36\times 3)$ yield small codes correcting dephasing, while $\Sigma(168)$ and $\Sigma(72\times 3)$ support codes correcting arbitrary errors on a single qutrit. We also show that, in the latter case, the trivial irrep of $\Sigma(72\times 3)$ gives rise to a three-dimensional code space, allowing for the encoding of a logical qutrit rather than a single logical qubit.

Our results also clarify the limits of the method. For $\SU(3)$, no finite subgroup among those considered here is inaccessible to all the irreps required for arbitrary qutrit interactions in four- or five-body systems, and in the context of quantum error correction, the same obstacle prevents the construction of codes correcting arbitrary errors on two qutrits within this symmetry-based approach. For $\SU(4)$ and beyond, the formalism extends directly, but progress is currently limited by the incomplete classification of finite subgroups and by the increasing size of the irreps involved~\cite{Amihay_2001,Teixeira_2025}. Nevertheless, our analysis of $\SU(4)$ suggests that decoupling groups and useful codes may still exist, particularly among the subgroups of larger exceptional groups.

More broadly, the framework presented here is not limited to $\SU(d)$. Since it relies solely on complete reducibility and character theory, it should apply equally well to other semisimple Lie groups provided that noise and control transform naturally under their representations. This opens up possible applications beyond spin and qudit systems, for example in continuous-variable platforms with symplectic symmetry~\cite{Simon_1988,Han_1990,Arvind_1995,Fabre_2020,Hasebe_2020,Mattia_2021,Colas_2022}.

Overall, this work provides a unified and systematic approach for constructing both dynamical decoupling sequences and symmetry-protected quantum codes from finite subgroups of Lie groups. We hope that these ideas will be useful not only for quantum error mitigation and error correction, but also for related problems in Hamiltonian engineering and quantum metrology. In particular, the same symmetry-based construction is directly connected to anticoherent states and anticoherent subspaces, which play an important role in quantum metrology~\cite{Zimba_2006,Baguette_2015,Baguette_2017,Bouchard_17,Martin_2020,Serrano_2025,Pereira_2017,Serrano_Ens_stiga_2025}. It would also be interesting to investigate in greater depth how these ideas connect with other representation-theoretic approaches to QECC~\cite{bradshaw_2026,Teixeira_2024,Teixeira_2026,Aydin_2026}. A more in-depth exploration of the interplay between finite-group symmetries, logical gate design, and experimentally realistic control constraints therefore appears to be a particularly promising avenue.

\section{Acknowledgment}

The authors thank Pierre Mathonet and Naïm Zenaïdi for valuable discussions. CR is a Research Fellow of the F.R.S.-FNRS. JM and ESE acknowledge the FWO and the F.R.S.-FNRS for their funding as part of the Excellence of Science program (EOS project 40007526). The numerical calculations and figures in this manuscript were produced using the Julia programming language, in particular the Makie package~\cite{Makie}.

\appendix

\section{Subgroup $\Sigma(36\times 3)\triangleleft\Sigma(72\times 3)$}\label{ap.36}

The finite group $\Sigma(36 \times 3)$ is an exceptional point group of $108$ elements, but we can focus on the quotient group of $36$ elements $\Sigma(36 \times 3)/\mathbb{Z}_3$ if needed, as we explained in Sec.~\ref{Subsec.V.SU3}. It is clear from the form of the Hamiltonians in Eqs.~\eqref{eq.dipsu3} and \eqref{dis} that $\Sigma(36 \times 3)$ is not a symmetry of both Hamiltonians, so we can rule out the use of the quotient group $\Sigma(72\times 3)/\Sigma(36\times 3)$ as a decoupling group. However, as explained in Sec.~\ref{sec.facto}, we can instead exploit a factorization of the decoupling group and write
\begin{equation}
    \Sigma(72\times 3) = \Sigma(36\times 3) S
\end{equation}
with the representatives $S = \qty{\mathds{1},X}$, where the generator $X$ was defined in Eq.~\eqref{mats.}. Clearly, $S\subset \expval{\expval{X}}$, where $\expval{\expval{X}}$ is the cyclic group generated by $X$. If this cyclic group is a symmetry of the Hamiltonian, we can thus use $\Sigma(36\times 3)$ as a decoupling group. At this stage, $\expval{\expval{X}}$ is not a symmetry of $H_{\mathrm{dd}}^{\mathrm{sec},\SU(3)}$. However, we can orient the group $\Sigma(72\times 3)$ such that the subgroup $\expval{\expval{X}}<\Sigma(72\times 3)$, in this new orientation, is a symmetry of the Hamiltonian. In practice, this is done by conjugating the group in $\SU(3)$, 
\begin{equation}
    \Sigma(72\times 3) \mapsto U^{\dagger}\,\Sigma(72\times 3)\,U, \quad U\in\SU(3)
\end{equation}
by choosing $U$ such that $U^{\dagger}XU$ is diagonal. If this is the case, then $\expval{\expval{U^{\dagger}XU}}$ is a symmetry of the Hamiltonian and the group $U^{\dagger}\Sigma(36\times 3)U$, generated by either $U^{\dagger}\qty{C,V}U$ or $U^{\dagger}\qty{E,V}U$, can be used as a decoupling group. 

\par The unitary $U$ is not unique: indeed, we can multiply $U$ from the right by any element of the normalizer of the Cartan subalgebra of $\SU(3)$, denoted $N\in N(3)$. This is because, by definition, the unitary transformation of a diagonal $\SU(3)$ matrix by an element $N\in N(3)$ yields another diagonal matrix. We could thus choose a unitary $U$ and check the sequences obtained by replacing $U$ with $UN$ for all possible elements $N\in N(3)$ and see which one is more convenient to implement for our Hamiltonian. 

\par Since our Hamiltonian is invariant under transformations by diagonal unitaries, two matrices $N$ and $N' = CN$ would yield the same results if $C$ is diagonal. By defining $C(3)$ as the set of diagonal matrices of $\SU(3)$, we can then work with the quotient group $N(3)/C(3)$ to avoid any redundancy in the numerical search. This group is finite and is isomorphic to the Weyl group of $\SU(3)$, which we denote by $W(3)$. In fact, the Weyl group is isomorphic to the permutation group $S_3$, so we need only consider $6$ different matrices $N\in N(3)$, each of which is a representative of a different coset of $N(3)/C(3)$.

\par To benchmark this new sequence, we consider the following Hamiltonian in the rotating frame 
\begin{equation}
    H = \sum_i\delta_i S_z^{(i)} + H_{\mathrm{dd}}^{\mathrm{sec},\SU(3)}.\label{eq.NV}
\end{equation}
We generate $500$ Hamiltonians, where $\delta_i$ and $J_{ij}$ are generated randomly and follow a uniform distribution in the interval $\qty[-\Delta/2, \Delta/2]$ and $\qty[-J/2,J/2]$ respectively. We consider each sequence to be composed of ideal pulses, separated by a waiting time $\tau$, and calculate the distance~\eqref{eq.dist} for a wide region of the parameter space $(\tau \Delta,\tau J)$. An appropriate orientation of the group is represented by a unitary matrix $U$ composed of the eigenvectors of the generator $X$, such that $X = UX'U^{\dagger}$, with $X' = \mathrm{diag}(-i,i,1)$ and
\begin{equation}
   U = \begin{pmatrix}
       \frac{1+\sqrt{3}}{\sqrt{6+2\sqrt{3}}} & \frac{1-\sqrt{3}}{\sqrt{6-2\sqrt{3}}} &0\\ 
       \frac{1}{\sqrt{6+2\sqrt{3}}} & \frac{1}{\sqrt{6-2\sqrt{3}}} &\frac{-\omega^2}{\sqrt{2}} \\ 
        \frac{\omega}{\sqrt{6+2\sqrt{3}}} & \frac{\omega}{\sqrt{6-2\sqrt{3}}} & \frac{1}{\sqrt{2}} ,
   \end{pmatrix}\label{eq.UD}
\end{equation}
where we have defined $\omega \equiv \xi_3 = e^{2\pi i/3}$. Other possible unitary operators $U$ are obtained by multiplying on the right by the elements of the Weyl group; this has the effect of interchanging the columns of the matrix above, hence swapping the elements of $D$. We define $U(\sigma)$ as the matrix $U$ above after a permutation $\sigma$ of its columns: $U(e)\equiv U$ is the unitary obtained after applying the identity permutation and $U(12)$ is the one obtained after exchanging the first and second columns of $U$. It is worth noting that our Hamiltonian is invariant under conjugation by the element of the Weyl group that permutes the first and last columns (the permutation $(13)$); this transformation effectively swaps $\lambda_{1,2}$ and $\lambda_{6,7}$ while changing the sign of $\lambda_3+\sqrt{3}\lambda_8$. For the interaction term, two additional negative signs are added in the tensor product and cancel each other out. For the disorder term, the additional sign is irrelevant because the random variables $\delta_i$ can be either positive or negative. Consequently, it suffices to compare three different orientations, each described by a different representative of the cosets in the quotient group $W(3)/\expval{\expval{(13)}}$.

\par The generators of $\Sigma(36\times 3)$ can be chosen to be either $\qty{C,V}$ or $\qty{E,V}$, which can be expressed in terms of the exponential of their generators (multiplied by the imaginary unit $i$), \textit{e.g.}, $C = e^{-iH_C}$, where the corresponding generators are given by 
\begin{equation}\label{CVE}
    \begin{aligned}
        H_C &= -\frac{2\pi}{3}\mathrm{diag}(0,1,-1) \\ 
        H_V &= -\frac{\pi}{2} \frac{\sqrt{3}}{3}\begin{pmatrix}
            -1 & -1 & -1 \\ 
            -1 & \frac{1}{2} & \frac{1}{2} \\ 
            -1 & \frac{1}{2} & \frac{1}{2}
        \end{pmatrix} \\ 
        H_E &= -\frac{2\pi}{3}\frac{1}{\sqrt{3}}\begin{pmatrix}
            0 & -i & i \\ 
            i &0 &-i \\ 
            -i & i & 0
        \end{pmatrix}
    \end{aligned}.
\end{equation}
In the orientation described by $U$, the generators are given by 
\begin{equation}
    \begin{aligned}
        U^{\dagger}H_CU &= -\frac{2\pi}{3}\begin{pmatrix}
            0 & 0 & -\eta'\omega^2 \\ 
            0 & 0 & -\eta\omega^2  \\ 
             -\eta'\omega &-\eta\omega & 0 
        \end{pmatrix} \\ 
        U^{\dagger}H_VU &= \frac{\pi}{2} \frac{\sqrt{3}}{2}\begin{pmatrix}
            \frac{1+\sqrt{3}}{4} & \frac{-\frac{1}{2}+i}{\sqrt{6}} & -\eta'(\frac{\omega^2}{2}+i\omega) \\
             & \frac{1-\sqrt{3}}{4} & -\eta(\frac{\omega^2}{2}-i\omega) \\ 
             & & -\frac{1}{2}
        \end{pmatrix} \\ 
        U^{\dagger}H_EU &= -\frac{2\pi}{3}\frac{\sqrt{3}}{3}\begin{pmatrix}
            -\frac{3}{2}\eta'^2 & -\frac{3i}{2\sqrt{2}} & \eta'\frac{\sqrt{3}}{2}\qty(-i\omega +1)\\ 
            & \frac{3}{2}\eta^2 & \eta\frac{\sqrt{3}}{2}\qty(i\omega +1) \\ 
            & & \frac{-\sqrt{3}}{2}
        \end{pmatrix}
    \end{aligned}
\end{equation}
where $\eta \equiv \sqrt{\frac{3+\sqrt{3}}{6}}$ and $\eta'=\sqrt{1-\eta^2}$. 
As we saw earlier, these generators $H$ are involved because this requires experimental control of the transitions between all three levels. In particular, the double quantum transition $\ket{-1}\leftrightarrow\ket{1}$ in NV-centers is not an accessible ESR transition, which complicates the implementation of the sequence with simple MW pulses. However, it is still possible to engineer this effective Hamiltonian using quantum optimal control or Floquet engineering, or to use composite pulses~\cite{Lindon_2023}. Since the pulse sequence will ultimately be designed on the Cayley graph of $\Sigma(36\times3)/\mathbb{Z}_3$, we can choose an Eulerian path to ensure that the systematic errors potentially induced by these more complex pulses will be decoupled to first order. This is guaranteed because the dominant control errors can be represented by a Hamiltonian linear in the Gell-Mann matrices, which belong to a $(1,1)$ irrep for which $\Sigma(36\times3)$ is inaccessible, and thus to a decoupling group.

\par
\begin{figure}[h]
    \centering
    \includegraphics[width=\linewidth]{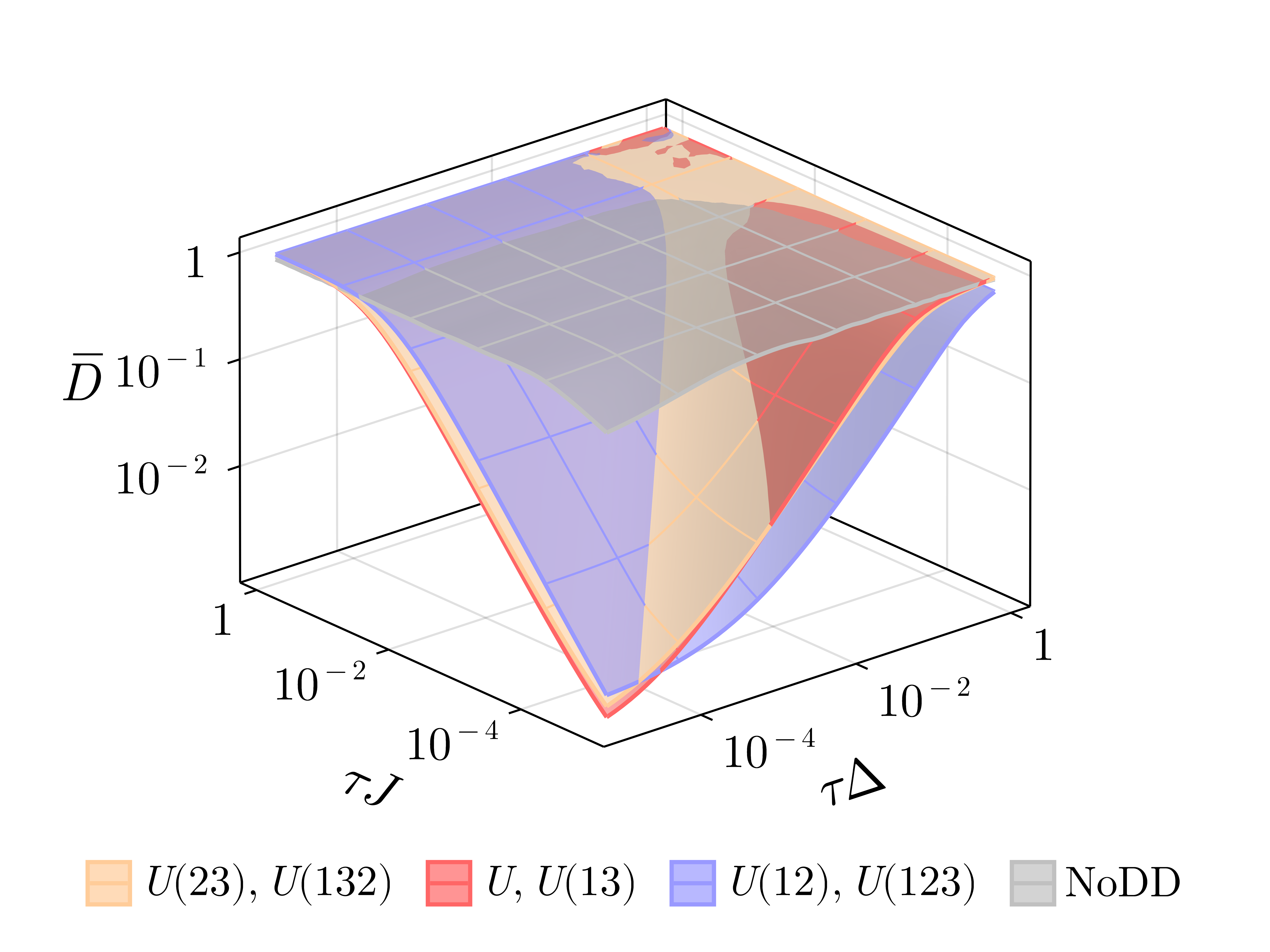}
    \caption{Average distance in the $(\tau \Delta,\tau J)$ parameter space for several orientations of the $\Sigma(36\times 3)/\mathbb{Z}_3$ DD sequence and for 500 randomly generated Hamiltonian (see main text). The "NoDD" curve corresponds to a free evolution of duration equal to total duration of the DD sequence.}
    \label{fig:U36U}
\end{figure}

 Finally, we plot the results in Fig.~\ref{fig:U36U}, using a random Eulerian sequence and the generators $U(\sigma)^{\dagger}\qty{C,V}U(\sigma)$ for all permutations $\sigma$. This demonstrates the first-order decoupling of these particular orientations of the group $\Sigma(36\times 3)$ and identifies a specific orientation, obtained by conjugating the above Hamiltonians by the elements of the Weyl group
\begin{equation}
    U(12) = \begin{pmatrix}
        0 & 1 & 0\\
        1 & 0 & 0\\
        0 & 0 & -1\\
    \end{pmatrix} \quad \mathrm{or}\quad U(123)=\begin{pmatrix}
        0 & 1 & 0\\
        0 & 0 & 1\\
        1 & 0 & 0\\
    \end{pmatrix},
\end{equation}
for which the decoupling properties are enhanced on the disorder term, which is relevant in the case of NV-center ensembles where disorder is the dominant error mechanism~\cite{Zhou_2023,Zhou_2020_metrology,Lukin_2024}. Note that any additional disorder of the form $S_z^2$ will also be decoupled by our sequences, since it can be written as a linear combination of Gell-Mann matrices---\emph{i.e.,} it transforms according to the irrep $(1,1)$. 

\par It should be noted that robustness to finite-duration errors is not guaranteed, even though the sequence is Eulerian. Indeed, the finite duration error Hamiltonian is given by the Hamiltonian~\eqref{eq.NV} transformed unitarily by the propagator of the pulse, \ie,
\begin{equation}
    H_{\mathrm{err}}^P = \frac{1}{\tau}\int_0^{\tau_p}dt\, P^{\dagger}(t)HP(t) , 
\end{equation}
where $P(t)$ is the propagator of the pulse, $\tau_p$ is duration and $H_{\mathrm{err}}^P$ the overall finite duration error. As mentioned in Sec.~\ref{sec.rob}, this unitary transformation generally breaks the symmetry of the Hamiltonian, so that the group factorization used to reduce the sequence from $\Sigma(72\times 3)$ to $\Sigma(36\times 3)$ cannot be used. To ensure finite-duration robustness, one could potentially design the pulses such that the finite duration errors are invariant under the cyclic group $\expval{\expval{X}}$. This requirement translates into a constraint in the numerical search for the propagator $P(t)$. In the case of the Hamiltonian~\eqref{eq.NV}, robustness to finite-duration errors on the disorder is guaranteed because $\Sigma(36\times 3)$ is inaccessible to the $(1,1)$ irrep, so only the symmetry of the interaction term should be considered.

\section{Subgroup $\Delta(24)<\Sigma(168)$}\label{ap.24}

As discussed in Sec.~\ref{method1}, our strategy is to exploit a factorization of a decoupling group $\group$ into a subgroup and a set of coset representatives adapted to the symmetries of the Hamiltonian. In the present case, the relevant inaccessible symmetry is $\Sigma(168)$, introduced in Sec.~\ref{ex.SU3} and generated by the matrices $Y$ and $Z$ defined in Eq.~\eqref{mats.}. We seek a subgroup of $\Sigma(168)$ that can serve as a shorter decoupling group for the Hamiltonians of interest in Sec.~\ref{sec.NV}, namely the secular dipolar interaction~\eqref{eq.dipsu3} and the disorder term~\eqref{dis}.

The exceptional point group $\Sigma(168)$ can be generated by $\qty{Y,Z}$ where $Y^7=\mathds{1}_{3\times 3}$. Since $\Delta(24)$, which is isomorphic to the octahedral point group, has no cyclic subgroup of order $7$, it is evident that $\expval{\expval{Y}}\cap \Delta(24)=\qty{\mathds{1}_{3\times 3}}$, where $\expval{\expval{Y}}$ is the cyclic group generated by $Y$. The product of the two subgroups has 168 elements with no repeated entries. Hence, our exceptional point group can be factorized as 
\begin{equation}
    \Sigma(168) = \Delta(24)\expval{\expval{Y}}.
\end{equation}
Since $Y$ is already diagonal, we should find the realization of $\Delta(24)$ in $\Sigma(168)=\expval{\expval{Y,Z}}$. The opposite problem was addressed in Ref.~\cite{King_2009}, where the authors fixed a specific orientation of $\Delta(24)$ in $\SU(3)$ for the purpose of model building for high-energy physics and identified the similarity transformation that should be applied to the generators of $\Sigma(168)$ to include their specific realization of $\Delta(24)$ as a subgroup. The starting point is a realization of $\Sigma(168)$ generated by 
\begin{equation}
    \begin{aligned}
        A^{[3]} &= \frac{i}{\sqrt{7}} \begin{pmatrix}
        \eta^2 - \eta^5 & \eta - \eta^6 & \eta^4 - \eta^3 \\ 
        \eta - \eta^6 & \eta^4 - \eta^3 & \eta^2 - \eta^5 \\
        \eta^4 - \eta^3 & \eta^2 - \eta^5 & \eta - \eta^6
    \end{pmatrix}, \\ 
    B^{[3]} &= \frac{i}{\sqrt{7}} \begin{pmatrix}
        \eta^3 - \eta^6 & \eta^3 - \eta & \eta - 1\\ 
        \eta^2 - 1 & \eta^6 - \eta^5 & \eta^6 - \eta^2 \\
        \eta^5 - \eta^4 & \eta^4 - 1 & \eta^5 - \eta^3
    \end{pmatrix},
    \end{aligned}
\end{equation}
which satisfy $A^{[3]}B^{[3]}=Y$ and with $\eta \equiv \xi_7 = e^{i2\pi/7}$, and a realization of $\Delta(24)$ generated by
\begin{equation}\begin{aligned}
        M_1 &= \begin{pmatrix}
        0 & 1 & 0\\
        0 & 0 & -1\\
        -1 &0 &0
    \end{pmatrix}, &
    M_2 &= \begin{pmatrix}
        -1&0&0\\
        0&0&-1\\
        0&-1&0
    \end{pmatrix}.
\end{aligned}
\end{equation}
The unitary transformation that must be applied to $A^{[3]}$ and $B^{[3]}$ to include this specific octahedral group as a subgroup corresponds to the unitary $w=w_2w_1$, where 
\begin{equation}\begin{aligned}
        w_1 &= B^{[3]}(A^{[3]}B^{[3]})^2(B^{[3]}A^{[3]})^2 \\
        w_2 &= \begin{pmatrix}
            x & y & z \\
            z & x & y\\
            y & z & x
        \end{pmatrix}
\end{aligned}
\end{equation}
and where the parameters $x$, $y$ and $z$ are given in Eqs.~(3.22)-(3.24) in Ref.~\cite{King_2009},
\begin{equation}
    \begin{aligned}
        x &= \frac{1}{N}\qty(-64-15\eta-56\eta^3-46\eta^4+5\eta^5-27\eta^6) \\ 
        y &= \frac{1}{N}\qty(73\eta+156\eta^2+46\eta^3+12\eta^4+137\eta^5+115\eta^6) \\ 
        z &= \frac{1}{N}\qty(15+35\eta-35\eta^2-23\eta^3+41\eta^4-46\eta^6)
    \end{aligned}
\end{equation}
and $N$ is a normalization factor given by 
\begin{equation}\begin{aligned}
    N ={}&\Bigg[28\bigg(-342125-349668\eta +283769\eta^2\\&+9406\eta^3-501928\eta^4+287955\eta^6\bigg)  \Bigg]^{\frac{1}{3}}
\end{aligned}.
\end{equation}
As a consequence, applying the inverse unitary transformation to $M_1$ and $M_2$, \ie, 
\begin{equation}
    M_1 \mapsto w^{\dagger} M_1w,\quad M_2 \mapsto w^{\dagger} M_2w,
\end{equation}
ensures that $\Delta(24)$ can be used as a decoupling group. As in the case of $\Sigma(36\times 3)$, one can further conjugate the generators by the elements of the quotient group $W(3)/\expval{\expval{(13)}}$ and choose the most effective orientation for the Hamiltonian of interest.

\par
\begin{figure}[t]
    \centering
    \includegraphics[width=\linewidth]{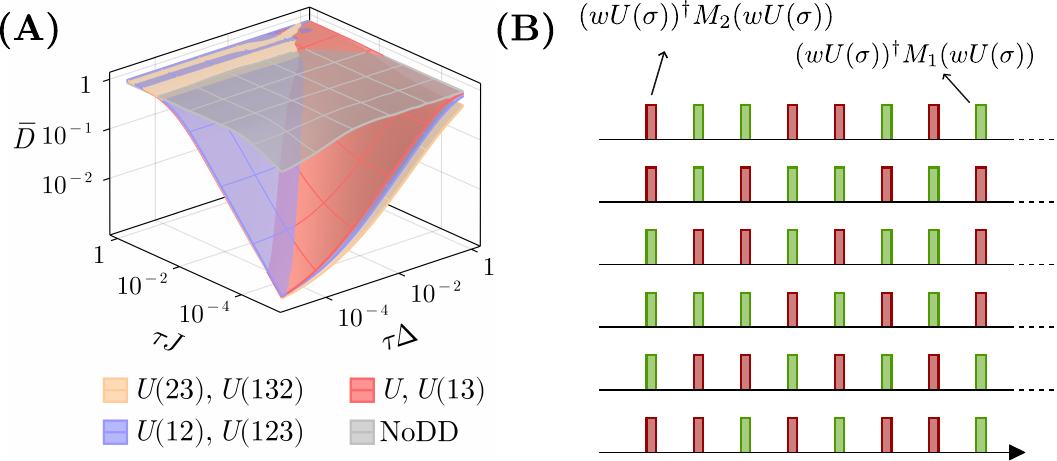}
    \caption{(A) Average distance in the $(\tau \Delta,\tau J)$ parameter space for several orientations of the $\Delta(24)$ DD sequence and for $500$ randomly generated Hamiltonians (see main text). (B) Representation of the octahedral sequence used in the simulation.}
    \label{fig:UO}
\end{figure}

 We now calculate the average distance in the $(\tau\Delta,\tau J)$ parameter space for $200$ randomly generated Hamiltonian of the form~\eqref{eq.NV}, for a random Eulerian path on the Cayley graph of $\Delta(24)$ and choosing the generating set $U(\sigma)w^{\dagger}\qty{M_1,M_2}wU^{\dagger}(\sigma)$ for the different permutations $\sigma$ of the Weyl group $W(3)$. The results are plotted in Fig.~\ref{fig:UO}A, where each orientation is labelled by the corresponding permutation $U(\sigma)$, demonstrating the decoupling properties of these specific orientations of $\Delta(24)$. The most effective orientation in the disorder-dominated regime is obtained by conjugating the generating set by the element 
\begin{equation}
    U(23) = \begin{pmatrix}
        -1 &0 &0\\
        0 & 0 & -1\\
        0 & -1 & 0
    \end{pmatrix}.
\end{equation}
Since $\Delta(24)$ is inaccessible to the irrep $(1,1)$, the resulting sequence (depicted in Fig.~\ref{fig:UO}B) is
robust to systematic errors in the pulses and finite duration errors due to disorder, but not robust to finite duration errors due to the interaction term. Because the group has $24$ elements, the resulting sequence is shorter than the sequence based on the $\Sigma(36\times 3)$ subgroup, which has $36$ elements. However, the analytical form of the generating pulses is far more complicated. 

\section{Subgroup $\Delta(54)\triangleleft\Sigma(72\times 3)$}\label{ap.54}

The finite group $\Sigma(72\times 3)$ can be written as a product of these subgroups and the following set of representatives,
\begin{equation}
    \begin{aligned}
        \Sigma(72\times 3) &= \Delta(27) \qty{\mathds{1}, V,V^2,V^3,X, VX,V^2X,V^3X} ,  \\ 
        \Sigma(72\times 3) &= \Delta(54)\qty{\mathds{1}, X, V, VX}, \\ 
    \end{aligned}
\end{equation}
where $\Delta(27)<\Delta(54)$. We thus require that $\expval{\expval{X}}$ and $\expval{\expval{V}}$ be symmetries of the Hamiltonian. However, the set $\qty{X,V}$ generates $\Sigma(72\times 3)$, so it is impossible to find a Hamiltonian invariant under these cyclic groups but not invariant under $\Sigma(72\times 3)$. Thus, it is not possible to orient this group to use $\Delta(27)$ and $\Delta(54)$ as decoupling groups.

\par Another method would be to find an orientation such that $\Delta(27)=\expval{\expval{C,E}}$ or $\Delta(54)=\expval{\expval{C,E,B}}$ themselves are symmetries of the Hamiltonian. If this is the case, we could design decoupling sequences on the quotient groups 
\begin{equation}
    \begin{aligned}
        \Sigma(72\times 3)/\Delta(27) & \cong Q_8 , \\ 
        \Sigma(72\times 3)/\Delta(54) & \cong K_4\equiv \mathbb{Z}_2\times \mathbb{Z}_2 , \\ 
    \end{aligned}
\end{equation}
where $K_4\equiv \mathbb{Z}_2\times \mathbb{Z}_2$ and $Q_8$ are the Klein four-group and the Quaternion group, respectively. However, it is clear that neither group can be a symmetry of the disorder term, because this symmetry is inaccessible to the irrep $(1,1)$. On the other hand, $H^{\mathrm{sec},\SU(3)}_{\mathrm{dd}}$ has a nice symmetry under the cyclic group $\expval{\expval{B}}$ ---which is the subgroup of the Weyl group generated by the permutation $(13)$---, but it lacks the $\expval{\expval{E}}$ symmetry necessary to generate $\Delta(27)$ and $\Delta(54)$. It is thus impossible to exploit either of these symmetries on the Hamiltonian in its current form.

\par Since there is no symmetry to exploit at this stage , we can resort to a multisymmetrization procedure~\cite{Read_2025facto}; in this protocol, two sequences are nested such that the inner layer performs a first symmetrization and the outer layer performs a second symmetrization on what remains of the Hamiltonian. In the following protocol, the inner layer will have two purposes: it will suppress the disorder term while imposing a certain symmetry on the interaction term $H^{\mathrm{sec},\SU(3)}_{\mathrm{dd}}$. The outer layer will then exploit the additional symmetries provided by the inner layer to provide decoupling on the interaction term.

\par In this case, we can choose an inner sequence that implements a $\expval{\expval{E}}$-symmetrization. This sequence will be composed of three successive applications of the unitary pulse $E$,
\begin{equation}
    \mathcal{S}_{\mathrm{inner}}\, \equiv \, -\,E\, -\,E\, -\,E\label{eq.inner}
\end{equation}
and will act very differently on the disorder and the interaction. First, the disorder term is clearly invariant under unitary transformations by $\mathrm{diag}\qty(1,-1,-1)$ and $\mathrm{diag}\qty(-1,-1,1)$, because it is diagonal. The Abelian group generated by these two elements is isomorphic to the Klein 4-group $K_4$ and is a normal subgroup of the tetrahedron group $\mathrm{T}\cong \Delta(12)$. Because $\Delta(12)$ is inaccessible to $(1,1)$, a decoupling sequence can thus be constructed on the Cayley graph of $\mathrm{T}/K_4 \cong \expval{\expval{E}}$, which results in the sequence~\eqref{eq.inner}. As a consequence, our inner layer will decouple the disorder term. The interaction term will not be immediately decoupled, however. In order to understand what happens to $H^{\mathrm{sec},\SU(3)}_{\mathrm{dd}}$, we first note that the elements $B$ and $E$ generate the Weyl group, which can in fact be written as a product of the two cyclic subgroups generated by these elements,
\begin{equation}
    W(3) \cong \expval{\expval{E}}\expval{\expval{B}}.
\end{equation}
Consequently, we can exploit the symmetry of $H^{\mathrm{sec},\SU(3)}_{\mathrm{dd}}$ under the unitary transformation by $B$; thus, the inner sequence~\eqref{eq.inner} implements the full Weyl symmetrization. The resulting Hamiltonian is therefore invariant under $B$ and $E$, but also under any unitary transformation by the diagonal elements of $\SU(3)$; it thus has $\Delta(54)$ symmetry. Consequently, the outer layer can be chosen to implement a Hamiltonian (or Eulerian) path on the Cayley graph of either of the quotient groups $\Sigma(72\times 3)/\Delta(54)$ and $\Sigma(72\times 3)/\Delta(27)$. Two such sequences are given by 
\begin{equation}\begin{aligned}
         K_4\to \mathcal{S}_{\mathrm{outer}} \equiv& -\,X\,-\,V\,-\,X\,-\,V\,\ , \\
         Q_8\to \mathcal{S}_{\mathrm{outer}} \equiv& -\,X\,-\,X\,-\,X\,-\,V \\
         &\,-\,X\,-\,X\,-\,X\,-\,V \, \ ,
\end{aligned}
\end{equation}
and will ensure that the interaction term is decoupled. We note that any quadratic disorder (terms proportional to $S_z^2$) resulting from crystal strain inhomogeneities or quadrupolar coupling will be symmetrized with the interaction Hamiltonian, as these diagonal operators are also invariant under conjugation by $B$.

\par The nested sequence is then obtained by replacing the free-evolution of the outer sequence by an iteration of the inner sequence, so that the resulting sequence is composed of 12 and 24 pulse intervals, respectively. Note that this sequence is not cyclic, as $(XV)^2\neq \mathds{1}_{3\times 3}$ and $(X^3V)^2\neq \mathds{1}_{3\times 3}$. Cyclicity can be ensured either by changing the last pulse so that it reverses the dynamics generated by the previous pulses or by applying the sequence several times to resolve the identity. We find that $(XV)^4 = \omega \mathds{1}_{3\times 3}$ and $(X^3V)^4=\mathds{1}_{3\times 3}$, so that repeating twice the sequence makes sure that the decoupling protocol effectively implements the identity, up to a global phase.

\par
\begin{figure}[t]
    \centering
    \includegraphics[width=\linewidth]{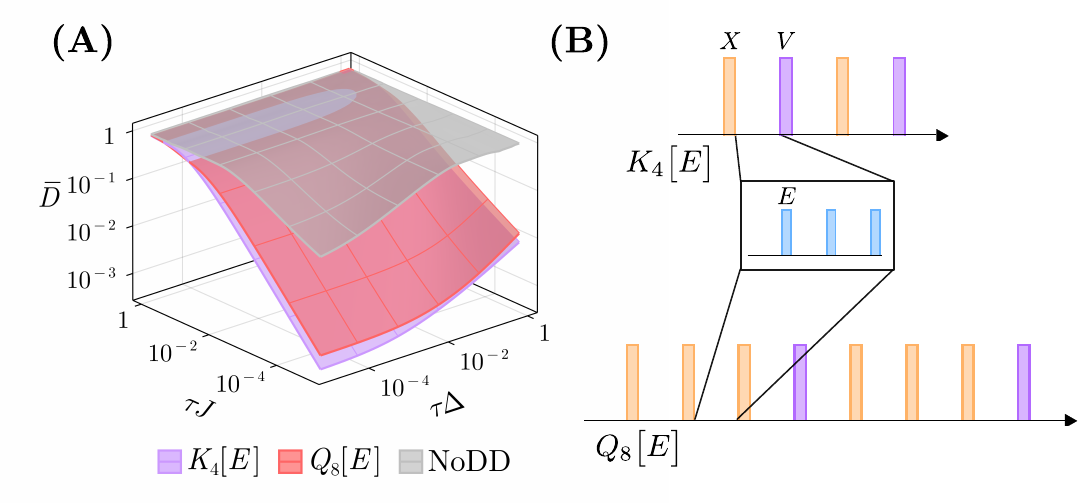}
    \caption{(A) Average distance in the $(\tau \Delta,\tau J)$ parameter space for 100 randomly generated Hamiltonians (see main text) for the two nested sequences $Q_8[E]$ and $K_4[E]$ represented in (B).}
    \label{fig:nested}
\end{figure}

 As a proof of concept, we calculate the Hilbert-Schmidt distance 
\begin{equation}
    D(U,V) = \sqrt{1-\frac{\abs{\mathrm{Tr}\qty[U\,U_{\mathrm{trg}}^{\dagger}]}}{d_S}}
\end{equation}
between the target propagator $U_{\mathrm{trg}}$ (where $U_{\mathrm{trg}}=(XV)^2$ or $U_{\mathrm{trg}}=(X^3V)^2$ depending on the sequence) and the noisy propagator $U$, for a wide regime of parameters $(\tau \Delta,\tau J)$ and $100$ randomly generated Hamiltonians of the form~\eqref{eq.NV} (see Fig.~\ref{fig:nested}A). The results show that the nested sequences (represented in Fig.~\ref{fig:nested}B) perform very well in a disorder-dominated regime, as the associated Hamiltonian is suppressed on the smallest timescale.

\par We note, however, that these nested sequences lack the robustness of most group-based sequences for several reasons. First, they are generally not robust to finite duration pulses, as explained in details in Ref.~\cite{Read_2025facto}. The inner layer will be robust to finite-duration pulses only in the case where the pulse $E$ is implemented by the simple propagator $E(t) = e^{-i\theta(t)H_E}$, which can be done only if $H_E$ is a naturally implementable Hamiltonian. In the context of NV-centers, only two out of the three transitions are accessible to ESR so that the Hamiltonian $H_E$ can only be implemented effectively. The outer-layer sequence will also not generally be robust to finite-duration pulses, as explained in Ref.~\cite{Read_2025facto}. Secondly, choosing an Eulerian path for the outer layer will not guarantee robustness to control errors, because the set of representatives of the quotient groups $\Sigma(72\times 3)/\Delta(27)$ and $\Sigma(72\times 3)/\Delta(54)$ are not finite groups of $\SU(3)$. These limitations diminish the usefulness of our nested sequences in realistic experimental setups.

\section{Orientations of $\Sigma(72\times 3)$ that facilitate experimental implementation}\label{ap.Orient}

\par
\begin{figure}
    \centering
    \includegraphics[width=0.75\linewidth]{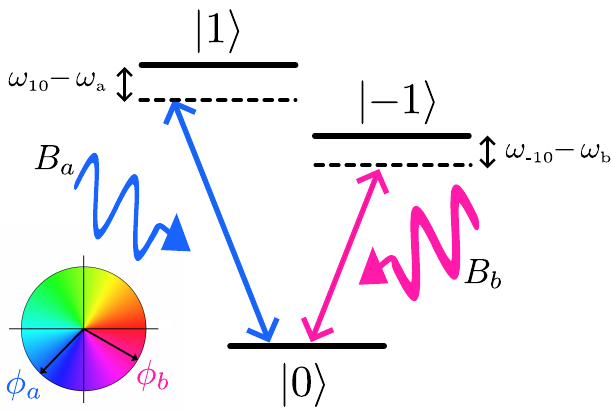}
    \caption{Representation of a double-driving pulse.}
    \label{fig:pulse}
\end{figure}

 We consider the internal Hamiltonian $H_0 = \mathrm{diag}\qty(\omega_{10},0,\omega_{-10})$ and the coupling Hamiltonian $H_{a,b} = -\gamma\vec{B}_{a,b}(t)\cdot \vec{S}$ between the magnetic fields~\eqref{eq.magnfield} and each spin-$1$ in the ensemble. Moving to the rotating frame with respect to the Hamiltonian $\mathrm{diag}\qty(\omega_a,0,\omega_b)$, we find the rotating frame Hamiltonian equal to
    \begin{equation}
    H_{\mathrm{rot}} = \begin{pmatrix}
        \tilde{\omega}_{10} & -\Omega_a e^{-i \phi_a} & 0 \\ 
        -\Omega_a e^{i \phi_a} &\tilde{\omega}_{-10}-\tilde{\omega}_{10} & -\Omega_b e^{-i \phi_b} \\ 
        0 & -\Omega_b e^{i \phi_b} & -\tilde{\omega}_{-10}
    \end{pmatrix}
\end{equation}
where 
\begin{equation}
    \begin{aligned}
        \tilde{\omega}_{10} &\equiv \frac{2}{3}(\omega_{10}-\omega_a) - \frac{1}{3}(\omega_{-10}-\omega_b) , \\ 
        \tilde{\omega}_{-10} &\equiv \frac{2}{3}(\omega_{-10}-\omega_b) - \frac{1}{3}(\omega_{10}-\omega_a) , \\ 
        \Omega_{a,b} &\equiv \frac{\gamma B_{a,b}}{2} .
    \end{aligned}
\end{equation}
We can now directly identify the field parameters that produce the Hamiltonian of interest. An example of such pulse is represented in Fig.~\ref{fig:pulse}, where the frequency of each magnetic field is represented by a colored, double arrow and their amplitude and phase by the thickness and color of the corresponding oscillating field.

\par In Appendix~\ref{ap.36}, we diagonalized $X$ (see Eq.~\eqref{eq.UD}) in order to find a specific orientation of the group $\Sigma(72\times 3)$, for which the generating Hamiltonian of each pulse ($V$ and $X$) was given by 
\begin{equation}\begin{aligned} 
      U^{\dagger}H_VU &= \frac{\pi}{2}\frac{\sqrt{3}}{2}\begin{pmatrix}
            \frac{1+\sqrt{3}}{4} & \frac{-\frac{1}{2}+i}{\sqrt{6}} & -\eta'(\frac{\omega^2}{2}+i\omega) \\
             & \frac{1-\sqrt{3}}{4} & -\eta(\frac{\omega^2}{2}-i\omega) \\ 
             & & -\frac{1}{2}
        \end{pmatrix} , \\ 
    U^{\dagger}H_XU &= -\frac{\pi}{2}\begin{pmatrix}
        -1 & 0 & 0\\ 
        0 & 1 & 0 \\ 
        0 & 0 & 0
    \end{pmatrix} .
\end{aligned}
\end{equation}
Because $UH_XU^{\dagger}$ is already diagonal, we can conjugate it by any of the two families of unitaries 
\begin{equation}
\begin{aligned}
    P(\theta,\phi_1,\phi_2) &= e^{i\theta \lambda_2}e^{i\phi_1\lambda_8}e^{i\phi_2\lambda_3}\\
    P'(\theta,\phi_1,\phi_2) &= e^{i\theta \lambda_7}e^{i\phi_1\lambda_8}e^{i\phi_2\lambda_3}
\end{aligned}
\end{equation}
without populating the forbidden matrix element. By choosing the phases $\phi_1$ and $\phi_2$ for $P$ (resp. $P'$), we can refocus the $[1,3]$ and $[2,3]$ (resp. $[1,3]$ and $[1,2]$) complex numbers of the matrix $U^{\dagger}H_VU$ so their arguments are the same. The non diagonal unitary transformation then transfers the population from one of these matrix elements to the other, and an appropriate value of $\theta$ can be chosen so that the $[1,3]$ matrix element is entirely unpopulated. We now consider the two cases ($P$ and $P'$) separately.

\subsection{Using \texorpdfstring{$P(\theta,\phi_1,\phi_2)$}{Lg}}

Conjugating by $P(\theta,\phi_1,\phi_2)$, we find that the population of the matrix element $[1,3]$ of $U^{\dagger}H_VU$ is entirely transferred to the position $[2,3]$ when 
\begin{equation}
        \theta = \arctan\sqrt{\frac{9+\sqrt{3}}{9-\sqrt{3}}} , \quad
        \phi_2 = -\frac{1}{2}\arccos\frac{3}{\sqrt{13}} ,
\end{equation}
and for any value of $\phi_1$. The generating Hamiltonians are then given by 
\begin{equation}
    \begin{aligned}
        \tilde{H}_V &= \frac{\pi}{2}\begin{pmatrix}
            \frac{1}{3\sqrt{3}} & a & 0 \\ 
            a^*& -\frac{1}{3\sqrt{3}}+\frac{\sqrt{3}}{4}  & be^{i\phi'} \\
            0& b^*e^{-i\phi'} & -\frac{\sqrt{3}}{4}
        \end{pmatrix} , \\
        \tilde{H}_X &= -\frac{\pi}{2}\frac{1}{\sqrt{27}}\begin{pmatrix}
            1 & \sqrt{26} & 0 \\ 
            \sqrt{26} & -1 & 0 \\ 
            0 & 0 & 0
        \end{pmatrix} , 
    \end{aligned}
\end{equation}
where $\phi' = 3\phi_1 + \phi_2$ and 
\begin{equation}
    \begin{aligned}
        a &= -\frac{5}{3}\sqrt{\frac{2}{13}} + \sqrt{\frac{6}{13}}i , \\ 
        b&= -\frac{\sqrt{3}}{2}\frac{\qty(\frac{1}{2}+\sqrt{3}) + i\qty(1+\frac{\sqrt{3}}{2})}{\sqrt{5+2\sqrt{3}}} .
    \end{aligned}
\end{equation}
The remaining parameter to optimize is $\phi'$. Choosing
\begin{equation}
    \phi' = -\arccos\frac{\frac{1}{2}+\sqrt{3}}{\sqrt{5+2\sqrt{3}}}+\phi'',
\end{equation}
we have $be^{i\phi'} = \frac{-\sqrt{3}}{2}e^{i\phi''}$, which simplifies the form of the Hamiltonian considerably, and we can further set $\phi''=0$ for simplicity. 

\subsection{Using \texorpdfstring{$P'(\theta,\phi_1,\phi_2)$}{Lg}}

Conjugating by $P'(\theta,\phi_1,\phi_2)$, we find that the population of $[1,3]$ is entirely transferred to $[1,2]$ when 
\begin{equation}\begin{aligned}
    \theta &= \arctan\sqrt{\frac{9+\sqrt{3}}{5}} ,  \\ 
    3\phi_1-\phi_2 &= \arccos\frac{3}{2\sqrt{5}\sqrt{5+2\sqrt{3}}} .
\end{aligned}
\end{equation}
The generating Hamiltonians are then given by 
\begin{equation}\begin{aligned}
        \tilde{H}_X &= -\frac{\pi}{2}\begin{pmatrix}
        -1 & 0 & 0\\
        0 & \frac{5}{14+\sqrt{3}} & -\frac{9+\sqrt{3}}{14+\sqrt{3}}\\ 
        0 & -\frac{9+\sqrt{3}}{14+\sqrt{3}} & \frac{9+\sqrt{3}}{14+\sqrt{3}}
    \end{pmatrix} , \\ 
    \tilde{H}_V &= \frac{\pi}{2}\frac{\sqrt{3}}{2}\begin{pmatrix}
        \frac{1+\sqrt{3}}{4} & ce^{2i\phi_2} & 0\\
        c^*e^{-2i\phi_2} & \frac{-(17+7\sqrt{3})}{4(14+\sqrt{3})}&d \\ 
        0 & d^*& \frac{-2\sqrt{3}}{14+\sqrt{3}}
    \end{pmatrix} ,
\end{aligned}
\end{equation}
with 
\begin{equation}
\begin{aligned}
c &= \frac{\sqrt{14+\sqrt{3}}}{2\sqrt{6}}\qty(\frac{-1+2i}{\sqrt{5}}),\\
    d &{}= \frac{-\sqrt{3-\sqrt{3}}}{2\sqrt{30}\sqrt{5+2\sqrt{3}}}\bigg(8i+\cos(2\theta)\\&\qquad\qquad\qquad+\sqrt{\frac{15(9+\sqrt{3})}{8}}\sin(2\theta)\bigg).
\end{aligned}
\end{equation}
Choosing $2\phi_2 = \arccos\frac{1}{\sqrt{5}}$, we have $ce^{2i\phi_2} =\frac{\sqrt{14+\sqrt{3}}}{2\sqrt{6}}$. 

\section{Irrep decomposition of the space of operators acting on two qudits}\label{ap:YT}

We give a proof of Eq.~\eqref{eq:quditqudit}, using Young tableaux to compute the irrep decomposition of the tensor product representation $(1,\vec{0}_{d-3},1)^{\otimes 2}$. This is done by representing each irrep by its Young tableau~\cite{Geo:19}, picking one of them and labeling each box of this tableau with a letter, such that the boxes on the first row are labeled with an "a", the ones on the second row with a "b", etc. We also color the boxes accordingly. We then place the colored boxes row by row (first the a's, then the b's, etc) in the second tableau in all possible ways such that the following set of rules is satisfied~:
\begin{enumerate}
    \item They cannot be two boxes of the same color (\textit{e.g.}, two a's) in the same column.
    \item Starting at the box on the top-right of the tableau, and reading the tableau from right to left and top to bottom (we first read the first row, then the second row, etc), there can never be more b's than a's, more c's than b's, etc. 
    \item The boxes of a given row should be pushed as much as possible to the left (there cannot be an empty space on the left side of a given box, expect if this box is in the first column).
    \item A tableau with a column composed of more than $d$ boxes can be discarded.
    \item At the end, a column with exactly $d$ boxes can be discarded (but not the tableau).
\end{enumerate}
\par
\begin{figure*}[t]
    \centering
    \includegraphics[width=0.98\linewidth]{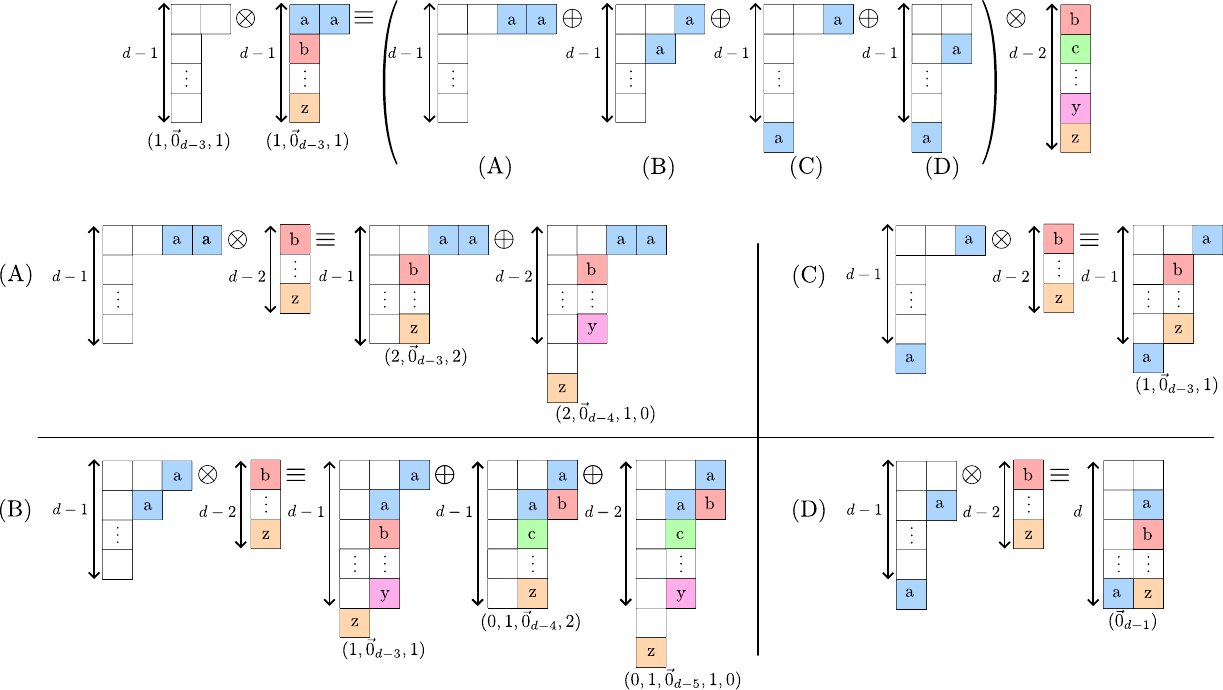}
    \caption{Decomposition of the tensor product representation $(1,\vec{0}_{d-3},1)^{\otimes 2}$ into irreps of $\SU(d)$.}
    \label{fig:YT}
\end{figure*}

 The first step is then to place the boxes labeled with an "a" in all possible ways. There are four possibilities that satisfy the set of rules described above, which are represented in the top panel of Fig.~\ref{fig:YT}. For each of these possibilities, one can keep adding the colored boxes b's, c's, etc, and we consider each case separately. 
\begin{itemize}
    \item[(A)] The following boxes can either go in the first or second column. However, once a box goes to the first column column, the remaining boxes have to go to the first column as well. Since tableaux with more than \textit{d} boxes get discarded, this leaves two option~: either put the remaining $d-2$ boxes in the second column, resulting in $(2,\vec{0}_{d-3},2)$, or the first $d-3$ boxes in the second column and the last one in the first column, resulting in $(2,\vec{0}_{d-4},1,0)$. 
    \item[(B)] The box "b" can go on the first three column. If it goes in the first column, the remaining boxes go there too and the tableau end up being discarded. It will thus go in the second or third column. If it goes in the second column, then the $d-4$ following boxes will go there as well, and the last one ends up in the first column, resulting in $(1,\vec{0}_{d-3},1)$. If we place "b" in the third column, the following $d-4$ boxes go to the second column. The last one then either go to the second column as well, or end up in the first column, resulting in $(0,1,\vec{0}_{d-4},2)$ or $(0,1,\vec{0}_{d-5},1,0)$ respectively. 
    \item[(C)] Once a box ends up in the first column, the tableau gets discarded, so the only way to place all boxes is to stack them in the second column, resulting in $(1,\vec{0}_{d-3},1)$.
    \item[(D)] Again, the only way is to stack them in the second column so that the tableau has only two columns composed of exactly $d$ boxes, which corresponds to the trivial irrep $(\vec{0}_{d-1})$.
\end{itemize}
It is easy to verify that no irrep is missing by calculating the dimensions of the irreps on the right and adding the dimensions, making sure that the result is the dimension of $(1,\vec{0}_{d-3},1)^{\otimes 2}$, that is $d^2-1$.

\section{Elegant bases for our codespaces}
In this Appendix, we calculate particular bases of the codespaces mentioned in Sec.~\ref{Sec7.QECC} that possess more symmetries than a generic state in the codespace. For the first two examples, we use the characters of the one-dimensional $\SU(2)$-irreps of the point groups $\mathrm{D}_2$, $\mathrm{T}$ and $\mathrm{O}$, which we present in the tables~\ref{tab:char.D2}-\ref{tab:char.O} and are labeled as $\rho_i$, $\sigma_i$ and $\tau_i$.

\subsection{Tetrahedral code}\label{Ap.T}

The Hilbert space of a single spin-6 decomposes in irreps of the tetrahedral group as 
\begin{equation}
    \mathcal{H}_S = \sigma_1^{\oplus 2}\oplus \sigma_2 \oplus \sigma_3 \oplus \dots 
\end{equation}
and the codespace is constructed in the subspace $\sigma_1^{\oplus 2}$. In order to choose a nice basis, we first decompose $\mathcal{H}_S$ in irreps of the octahedral group, which gives 
\begin{equation}
    \mathcal{H}_S = \tau_1\oplus \tau_2 \oplus \dots 
\end{equation}
where $\tau_1$ and $\tau_2$ are two one-dimensional irreps of the octahedral point group. Hence, they are also one-dimensional irreps of the tetrahedron; checking the character table of the octahedral point group, we see that the representations $\tau_1$ and $\tau_2$ of the octahedral point group, when restricted to the tetrahedral subgroup, are both isomorphic to $\sigma_1$ (the representations have the same characters). Consequently, the two-dimensional subspace $\tau_1\oplus \tau_2$ of $\Hs_S$ can be identified as $\sigma_1^{\oplus 2}$ and two orthogonal states of the codespace can be identified as the quantum states with octahedral symmetry corresponding to $\tau_1$ and $\tau_2$. The logical zero (resp. one) presented in Fig.~\ref{fig:codespaces_spin}B corresponds to the quantum state living in the one-dimensional subspace $\tau_1$ (resp. $\tau_2$).

\par Consequently, the elements of the octahedral point group with a character that is different in the irreps $\tau_1$ and $\tau_2$, such as the order-4 $\pi/2$ rotations with characters equal to $1$ and $-1$ respectively, can be used to carry out a phase gate in this basis. Depending on the error set, this logical Pauli-$Z$ gate might be fault-tolerant or not. If the goal is to correct arbitrary disorder and depolarization (Sec.~\ref{arb.dis}), the logical gate is fault-tolerant. However, if the error set is that described in Sec.~\ref{twobdy}, it is generally not as the gate will map correctable errors to non-correctable errors.

\subsection{2-Dihedral code}\label{Ap.D}
The Hilbert space of a single spin-2 decomposes in irreps $\rho_i$ of the $\mathrm{D}_2$ group as
\begin{equation}
    \mathcal{H}_S = \rho_1^{\oplus 2}\oplus \rho_2 \oplus \rho_3 \oplus \rho_4 .
\end{equation}
The codespace is constructed on $\rho_1^{\oplus 2}$. We can also decompose the Hilbert space in irreps of the tetrahedral group, which contains $\mathrm{D}_2$ as a subgroup, as 
\begin{equation}
    \mathcal{H}_S = \sigma_2 \oplus \sigma_3\oplus \dots 
\end{equation}
where $\sigma_2$ and $\sigma_3$ are one-dimensional irreps of $\mathrm{T}$, and we can once again check the character table to try and construct the codespace $\rho_1^{\oplus 2}$ using only the one-dimensional irreps of $\mathrm{T}$ restricted to $\mathrm{D}_2$. Once again, checking the character table, it is easy to see that $\sigma_2$ and $\sigma_3$ restricted to $\mathrm{D}_2$  are both isomorphic to $\rho_1$. Hence, the basis of $\rho_1^{\oplus 2}$ can be chosen such that the logical zero (resp. one) is the tetrahedron state living in the one-dimensional subspace $\sigma_2$ (resp. $\sigma_3$).

\par Any element of the $C_3$ classes of the tetrahedron group will thus implement a relative phase $e^{i2\pi/3}$ between $\ket{0}_L$ and $\ket{1_L}$. Furthermore, choosing to apply the rotation around the $z$ axis ensures that this logical gate is fault-tolerant as it will commute with the set of errors.

\begin{table}[h]
    \centering
    \begin{tabular}{c|c|c|c|c}
         \multirow{2}{*}{$\mathrm{D}_2$}& $C_1$ & $C_2$ & $C_2$ & $C_2$ \\
         & 1 & 1 & 1 & 1\\\hline 
        $\rho_1$ & 1 & 1 & 1 & 1\\ 
        $\rho_2$ & 1 & 1 & -1 & -1\\ 
        $\rho_3$ & 1 & -1 & 1 & -1\\ 
        $\rho_4$ & 1 & -1 & -1 & 1\\ 
    \end{tabular}
    \caption{Character table of $\mathrm{D}_2$.}
    \label{tab:char.D2}
\end{table}

\begin{table}[h]
    \centering
    \begin{tabular}{c|c|c|c|c}
         \multirow{2}{*}{$\mathrm{T}$}& $C_1$ & $C_2$ & $C_3$ & $C_3$ \\
         & 1 & 3 & 4 & 4\\\hline 
        $\sigma_1$ & 1 & 1 & 1 & 1\\ 
        $\sigma_2$ & 1 & 1 & $\omega$ & $\omega^2$\\ 
        $\sigma_3$ & 1 & 1 & $\omega^2$ & $\omega$\\ 
        \vdots & \vdots &\vdots &\vdots &\vdots 
    \end{tabular}
    \caption{Character table of $\mathrm{T}$, where we show only the characters of the one-dimensional irreps. The elements of the $\mathrm{D}_2$ subgroup of $\mathrm{T}$ are contained in the first and second column.}
    \label{tab:char.T}
\end{table}

\begin{table}[h]
    \centering
    \begin{tabular}{c|c|c|c|c|c}
         \multirow{2}{*}{$\mathrm{O}$}& $C_1$ & $C_2$ & $C_2$ & $C_3$ & $C_4$\\
         & 1 & 3 & 6 & 8 & 6 \\\hline 
        $\tau_1$ & 1 & 1 & 1 & 1 & 1 \\ 
        $\tau_2$ & 1 & 1 &-1 & 1 & -1 \\ 
        \vdots & \vdots &\vdots &\vdots &\vdots &\vdots
    \end{tabular}
    \caption{Character table of $\mathrm{O}$, where we show only the characters of the one-dimensional irreps. The elements of the tetrahedral subgroup $\mathrm{T}<\mathrm{O}$ are contained in the first, second and fourth column.}
    \label{tab:char.O}
\end{table}

\subsection{\texorpdfstring{$\Sigma(36\times 3)$}{Lg} and \texorpdfstring{$\Sigma(72\times 3)$}{Lg} codes}\label{ap:Sigmas}

\begin{figure}[h]
    \centering
    \includegraphics[width=\linewidth]{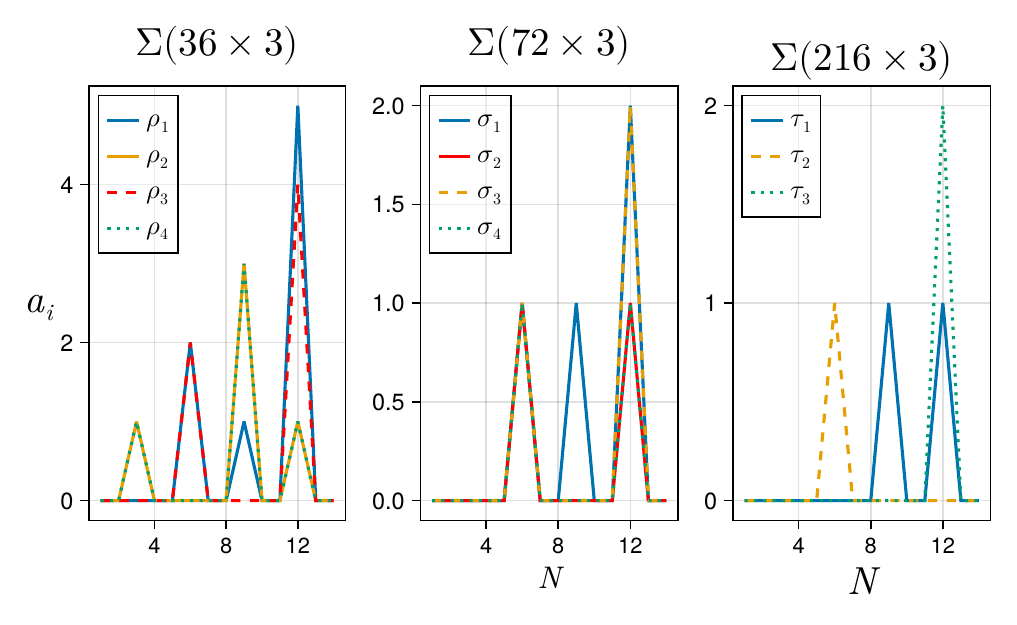}
    \caption{Multiplicity of the one-dimensional irreps of different finite groups in the $\SU(3)$ irrep $(N,0)$, as a function of $N$.}
    \label{fig:app}
\end{figure}

We can follow the same line of reasoning for the codes constructed on the groups $\Sigma(36\times 3)$ and $\Sigma(72\times 3)$, using the chains of subgroups 
\begin{equation}
    \Sigma(36\times 3) \triangleleft \Sigma(72\times 3) \triangleleft \Sigma(216\times 3).
\end{equation}
The character tables of the three groups can be found in Ref.~\cite{Grimus_2010}. We show in Fig.~\ref{fig:app} the multiplicities of the one-dimensional irreps of each of these groups in the $\SU(3)$-irrep $(N,0)$, as a function of $N$. We name the irreps of $\Sigma(36\times 3)$, $\Sigma(72\times 3)$ and $\Sigma(216\times 3)$ $\rho_i$, $\sigma_i$ and $\tau_i$ respectively to avoid any confusion.

\par In the case of $\Sigma(72\times 3)$, we have four non-isomorphic one-dimensional irreps. When restricted on the subgroup $\Sigma(36\times 3)$, we find, by looking at the character tables, that two of them ($\sigma_1$ and $\sigma_3$) will be isomorphic to the trivial irrep of $\Sigma(36\times 3)$ ($\rho_1$). In the same way, we observe that the other two ($\sigma_2$ and $\sigma_4$) will be isomorphic to another one-dimensional irrep of $\Sigma(36\times 3)$ ($\rho_3$). For $N=6$ (where we define our codespace), we see that the four one-dimensional irreps of $\Sigma(72\times 3)$ appear in the decomposition, hence we can construct 4 orthogonal states invariant under $\Sigma(72\times 3)$. On the other hand, the irreps of $\rho_1$ and $\rho_3$ of $\Sigma(36\times 3)$ appear with multiplicity two. It is then straightforward that a basis of $\rho_1^{\oplus 2}$ (resp. $\rho_3^{\oplus 2}$) can be chosen such that the two orthogonal states are symmetric under $\Sigma(72\times 3)$, where one state corresponds to $\sigma_1$ (resp. $\sigma_2$)and the other to $\sigma_3$ (resp. $\sigma_4$). 

\par Similarly, $\Sigma(216\times 3)$ has three non-isomorphic one-dimensional irreps and all of them, when restricted to the subgroup $\Sigma(72\times 3)$, are isomorphic to the trivial irrep of $\Sigma(72\times 3)$ ($\sigma_1$). For $N=12$, the three-dimensional codespace is defined in $\sigma_1^{\oplus 3}$. Since the irreps $\tau_1$ and $\tau_2$ of $\Sigma(216\times 3)$ appear with multiplicity two and one, respectively, we can construct three orthogonal states with the symmetry of $\Sigma(216\times 3)$ which span the codespace. 

\bibliographystyle{quantum}
\bibliography{References}
\end{document}